\documentclass[
preprint,
superscriptaddress,
 amsmath,amssymb,
 aps,
]{revtex4-1}
\usepackage{subcaption}
\usepackage{xcolor}
\usepackage{graphicx}
\usepackage{dcolumn}
\usepackage{bm}
\usepackage{tikz}
\usepackage[utf8]{inputenc}
\usepackage{amsfonts}
\usepackage{chemfig}
\usepackage{mhchem}
\usetikzlibrary{calc,fadings,decorations.markings}
\usetikzlibrary{decorations.pathreplacing,calc,positioning,shapes.multipart,arrows.meta}
\tikzset{two parallel arrows/.style={decorate,decoration={show path construction,
      lineto code={
       \draw [-latex] ($(\tikzinputsegmentfirst)!#1!90:(\tikzinputsegmentlast)$) 
        -- ($(\tikzinputsegmentlast)!#1!-90:(\tikzinputsegmentfirst)$); 
       \draw [latex-] ($(\tikzinputsegmentfirst)!#1!-90:(\tikzinputsegmentlast)$) 
        -- ($(\tikzinputsegmentlast)!#1!90:(\tikzinputsegmentfirst)$); 
      }}},two parallel arrows/.default=2pt}

\begin{document}

\title{Gelation, Clustering and Crowding in the Electrical Double Layer of Ionic Liquids}

\author{Zachary A. H. Goodwin}
\email{zachary.goodwin13@imperial.ac.uk}
\affiliation{Department of Chemistry, Imperial College of London, Molecular Sciences Research Hub, White City Campus, Wood Lane, London W12 0BZ, UK}
\affiliation{Thomas Young Centre for Theory and Simulation of Materials, Imperial College of London, South Kensington Campus, London SW7 2AZ, UK}

\author{Michael McEldrew}
\affiliation{Department of Chemical Engineering, Massachusetts Institute of Technology, Cambridge, MA, USA}

\author{J. Pedro de Souza}
\affiliation{Department of Chemical Engineering, Massachusetts Institute of Technology, Cambridge, MA, USA}

\author{Martin Z. Bazant}
\email{bazant@mit.edu}
\affiliation{Department of Chemical Engineering, Massachusetts Institute of Technology, Cambridge, MA, USA}
\affiliation{Department of Mathematics, Massachusetts Institute of Technology, Cambridge, MA, USA}

\author{Alexei A. Kornyshev}
\email{a.kornyshev@imperial.ac.uk}
\affiliation{Department of Chemistry, Imperial College of London, Molecular Sciences Research Hub, White City Campus, Wood Lane, London W12 0BZ, UK}
\affiliation{Thomas Young Centre for Theory and Simulation of Materials, Imperial College of London, South Kensington Campus, London SW7 2AZ, UK}
\affiliation{Institute of Molecular Science and Engineering, Imperial College of London, South Kensington Campus, London SW7 2AZ, UK}

\date{\today}

\begin{abstract}
Understanding the bulk and interfacial properties of super-concentrated electrolytes, such as ionic liquids (ILs), has attracted significant attention lately for their promising applications in supercapacitors and batteries. Recently, McEldrew \textit{et al.} developed a theory for reversible ion associations in bulk ILs, which accounted for the formation of all possible Cayley tree clusters and a percolating ionic network (gel). Here we adopt and develop this approach to understand the associations of ILs in the electrical double layer at electrified interfaces. With increasing charge of the electrode, the theory predicts a transition from a regime dominated by a gelled or clustered state to a crowding regime dominated by free ions. This transition from gelation to crowding is conceptually similar to the overscreening to crowding transition. 
\end{abstract}

\maketitle

\section{Introduction}

Ionic liquids (ILs) are concentrated electrolytes solely composed of molecular ions, which are often bulky and asymmetric~\cite{Welton1999,Hallett2011,Fedorov2014,Hermann2008}. As such, an IL does not contain electrolysable solvents, and therefore, the electrochemical stability window of ILs is typically larger than dilute aqueous electrolytes~\cite{Fedorov2014}. This property permits larger voltages for operation in supercapacitors, which increases the energy that can be stored in the device~\cite{Kondrat2016,son2020ion}. The promise of IL-based technologies has caused significant interest in understanding ILs, and other concentrated electrolytes, such as water-in-salt electrolytes (WiSEs)~\cite{suo2015,suo2017water,vatamanu2017,Lannelongue2018,chen2020water,mceldrew2018,Han2021WiSE,Groves2021wise,Patsahan2022,Yu2022Energy,Zhang2020wise} and \textcolor{black}{salt-in-ILs (SiILs)~\cite{dokko2013solvate,lewandowski2009ionic,molinari2019general,molinari2019transport,Zhang2018,yoon2013fast,howlett2004high}}. 

Indeed, owing to the high concentration and lack of high dielectric solvent, the energy of electrostatic interactions between nearest neighbours is much larger than thermal energy~\cite{Fedorov2014}. This manifests through decaying oscillations in charge density around an ion in the bulk or as a function of distance from an interface~\cite{Fedorov2008a,Fedorov2008,Georgi2010}, referred to as overscreening~\cite{Bazant2011}, which can give long-ranged electrostatic screening lengths~\cite{Coles2020length,pedroRTILs,pedro2022force,gavish2018solvent,emily2021}. \textcolor{black}{It was shown by Levy \textit{et al.}~\cite{levy2019spin} that these overscreening structures, from molecular dynamics simulations, can be well described by a nearest-neighbour spin-glass model, which suggests these long-ranged structures are built upon short-ranged Coulomb interactions.} Surprisingly, surface force measurements performed in ILs have reported extremely long-ranged \textit{monotonic} decay lengths~\cite{Smith2016,Gebbie2013,Gebbie2015,Gebbie2017rev,smith2017struct,Han2020IL,Jurado2016,Jurado2015,Jurado2017EDL,Mao2019nano}, referred to as underscreening~\cite{UND,underalpha}. Gebbie \textit{et al.}~\cite{Gebbie2013,Gebbie2015} interpreted these results as ILs behaving as ``dilute electrolytes'', with $\ll$ 1\% of ions free and the remaining ions being bound in neutral ion pairs, which do not participate in electrode screening. Whereas, Han \textit{et al.}~\cite{Han2020IL} have proposed that these long-ranged screening lengths arise because of the $\sim 10$~nm domain formation of nano-aggregates in ILs, \textcolor{black}{with indications of these large aggregates being observed in molecular simulations~\cite{Wang2005,Bernardes2011,mceldrew2020correlated,Lopes2006}.} 

These observations motivated many to study the role of ion pair formation in ILs~\cite{Ma2015,Lee2015,Zhang2015,MacFarlane2009,Kirchner2014,adar2017bjerrum,Araque2015}, and Avni \textit{et al.}~\cite{avni2020charge} even accounted for small finite aggregates (triples and quadruplets). \textcolor{black}{Sometimes, only free ions were explicitly treated, with ion pairs being implicitly treated as `voids'}~\cite{goodwin2017mean,Chen2017,goodwin2017underscreening,feng2019free}. This proved to be successful in reproducing the conductivity of ILs based on a Nernst-Einstein relation~\cite{feng2019free}, where 10-20$\%$ of ions were free, with the remaining being bound up in immobile aggregates. A dynamic exchange between these two states occurs, with a small activation energy ($\sim1~k_BT$) for the excitation from `bound' state to the `free' state. Thus the concept of an intrinsic narrow gap `ionic semiconductor', as introduced in Ref.~\citenum{Fedorov2014}, has received substantiation~\cite{feng2019free}. That work showed a dynamic picture of the free ions, but the exact nature of the immobile clusters was not discerned. Moreover, free ion approaches were also able to explain the qualitative changes in differential capacitance as a function of temperature~\cite{Chen2017}, where a `melting' of `frozen' structures is observed to occur as a function of increasing temperature and voltage. However, the fraction of free ions there was inferred from fitting the differential capacitance, and the nature of the `frozen' states only discussed on a qualitative level.

\begin{figure}
    \centering
    \includegraphics[width=0.45\textwidth]{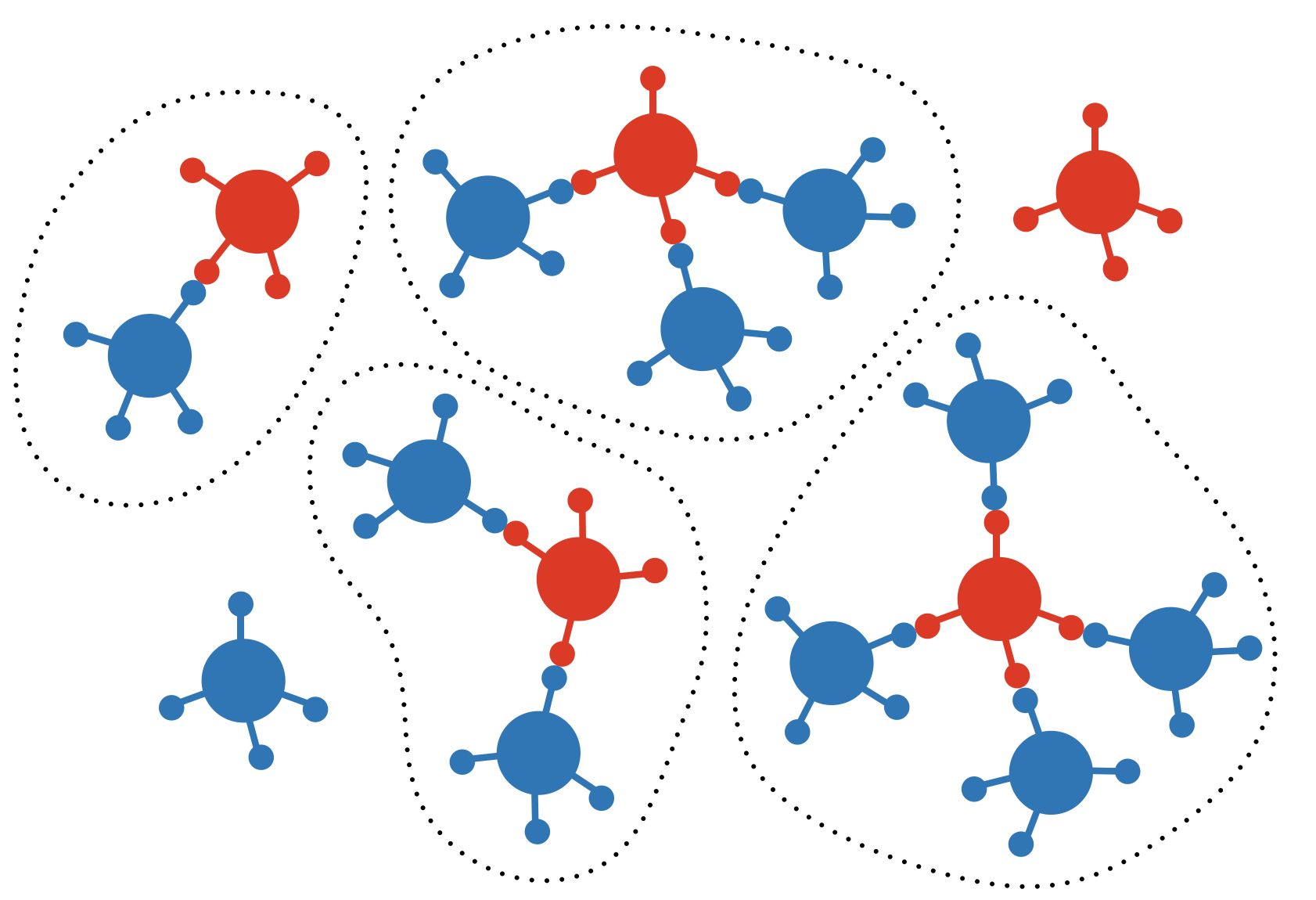}
    \caption{\textbf{Schematic of possible aggregates containing one cation}. Cations are shown in red and anions in blue, where both can form a maximum of 4 associations, as shown by the ``dangling bonds''. A free cation and anion is shown in the top right and bottom left corner, respectively. Dotted lines encircling the different clusters have been used to clearly demark each cluster.}
    \label{fig:into_fig}
\end{figure}

Therefore, in such a concentrated system as ILs, \textcolor{black}{the formation of clusters larger than ion pairs should occur~\cite{Dupont2004,Wang2005,Bernardes2011,Lopes2006}. McEldrew \textit{et al.}~\cite{mceldrew2020theory,mceldrew2020correlated,mceldrew2021wise,mceldrew2021salt} developed a theory, based on the theories of thermoreversible polymers~\cite{flory1942thermodynamics,tanaka1989,tanaka1990thermodynamic,tanaka1994,tanaka1995,ishida1997,tanaka1998,tanaka1999,tanaka2002}, which was able to systematically describe clusters beyond ion pairs, and even a percolating ionic network of infinite size~\cite{borodin2017liquid,choi2018graph,jeon2020modeling}, referred to as the gel.} One of the main parameters of the theory is the number of associations an ion can make, where it is assumed cations can only bond to anions and vice versa, which determines the possible clusters that can be formed. If ions can only form one bond, then ion pairs can only occur. If a cation has four association sites, it can form various clusters; for clusters containing a single cation there can be free cations, neutral ion pairs, and negatively charged triples, quadruplets and quintuplets, as schematically shown in Fig.~\ref{fig:into_fig}. These clusters can be much larger if more cations are involved, which are accounted for in the theory of McEldrew \textit{et al.}~\cite{mceldrew2020theory,mceldrew2020correlated,mceldrew2021wise,mceldrew2021salt}.

The theory of McEldrew \textit{et al.}~\cite{mceldrew2020theory} was applied to bulk ILs~\cite{mceldrew2020correlated}, where a consistent theory of ionic transport was also developed based on vehicular motion of the clusters~\cite{mceldrew2020correlated,france2019}, and other concentrated electrolytes~\cite{mceldrew2021wise,mceldrew2021salt}. While the ionic association theory has given a detailed description of the cluster statistics in the bulk, it has not yet been applied to predict the distributions of clusters near charged interfaces in the electrical double layer (EDL)~\cite{goodwin2021review}. 

Here we develop a theory of aggregation and gelation of ILs in the EDL, and investigate the qualitative behaviour in this system. Before proceeding to the mathematical formalism of this theory, we briefly discuss the principles of the chemical equilibrium which hold in a theory which accounts for thermoreversible associations in the bulk and the EDL of ILs. This discussion will be followed by a short preview of the results, which shall foreshadow the remaining paper.

\section{Chemical Equilibrium Considerations}

\begin{figure}
\begin{center}
\includegraphics[width=0.45\textwidth]{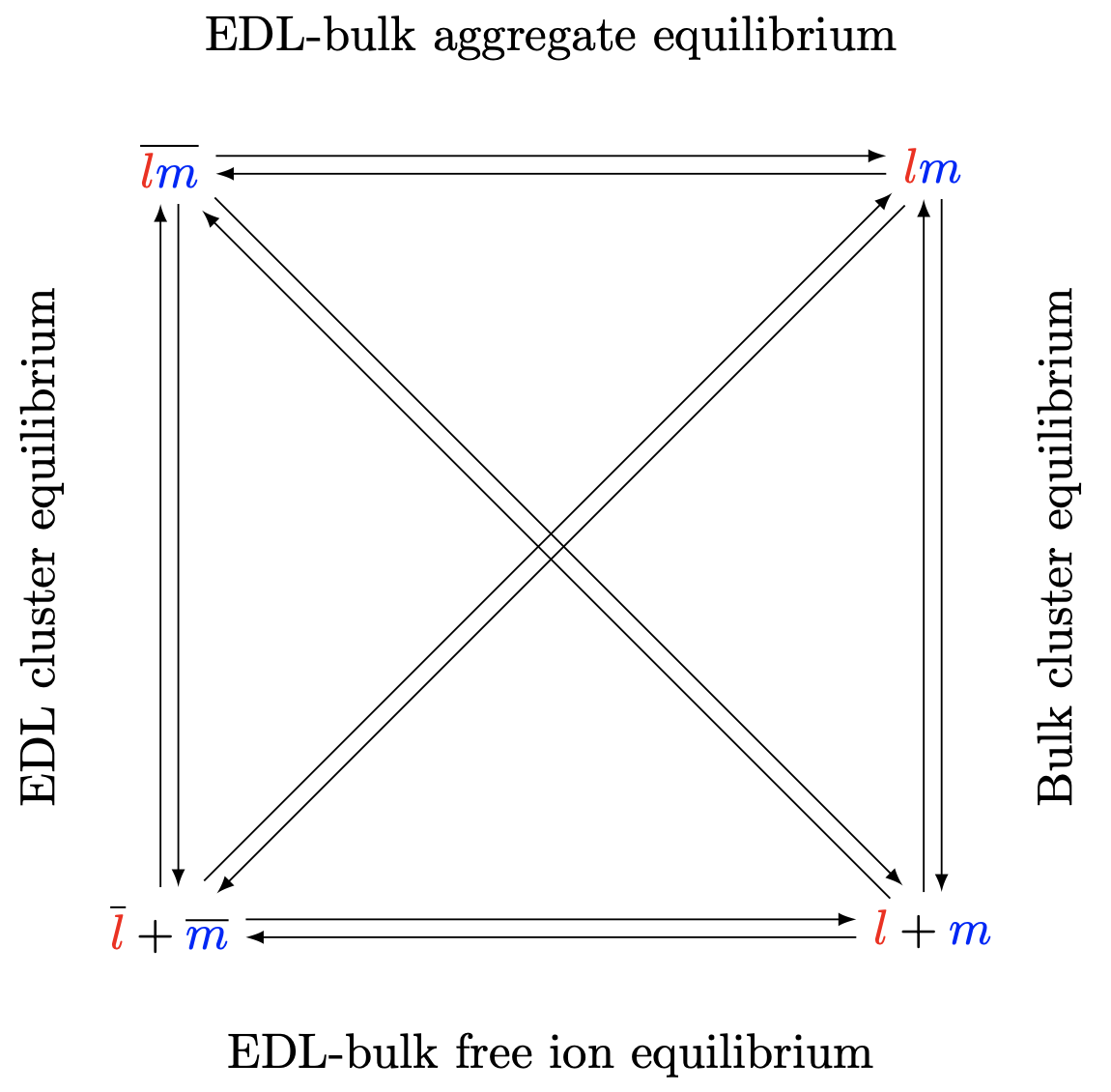}





\end{center}
\caption{\textbf{Schematic depicting all possible chemical equilibria for thermoreverisble associations between ions of an IL, in the bulk and EDL.} Free cations in the bulk are denoted by $l$, and free anions $m$, which can form clusters of rank $lm$. This equilibrium is referred to as the bulk cluster equilibrium. An equivalent equilibrium should hold in the EDL, where barred quantities are used. There is also an equilibrium between the bulk and EDL.}
\label{fig:chem_eq_all}
\end{figure}

One of the first questions which arises when considering clustering in the EDL is: what chemical equilibrium should hold? In the bulk, McEldrew \textit{et al.}~\cite{mceldrew2020theory} established the equilibrium between $l$ free cations and $m$ anions with clusters of rank $lm$ (when there are $l$ cations and $m$ anions associated). This bulk cluster equilibrium, as seen in Fig.~\ref{fig:chem_eq_all}, was shown to agree with the cluster distribution independently computed from molecular simulations for ILs~\cite{mceldrew2020correlated}. To study the EDL, there must be an equilibrium between ions near the interface and the bulk. 

If a ``standard approach" is taken for the EDL-Bulk equilibrium, i.e. where the ions in the bulk are assumed to accumulate in the EDL based on a Poisson-Boltzmann or Poisson-Fermi distribution from the bulk concentrations~\cite{Kornyshev2007,kilic2007a,Bazant2009a,goodwin2021review}, the ions are not permitted to reversibly associate in the EDL~\cite{adar2017bjerrum,goodwin2017mean,goodwin2017underscreening,Chen2017}. Such approaches bring ions to the EDL based on the charge and volume of the ion, but once the species are in the EDL, no further change of the ionic associations can occur. Moreover, with such an approach, it is not clear how the gel phase, if it forms, should be treated, and assumptions about how it will respond to an electrostatic potential will be required. For example, the gel could be assumed to not respond to the potential/field, and it remains as a constant background (charge), similar to solid electrolytes~\cite{Kornyshev1981}. The gel is not a rigid solid~\cite{mceldrew2020theory}, however, so this approach would have short-comings. Alternatively, an approach similar to how solvent is treated in the dilute electrolyte limit could be taken, i.e. the gel can be taken as some constant background dielectric, which would be trivially pushed out of the EDL based on how ions come to the EDL~\cite{Bazant2009a,goodwin2021review}. 

In the bulk, the gel phase is in equilibrium with free ions, and changes to the physical variables (such as temperature) can cause the balance of the equilibrium to shift~\cite{mceldrew2020theory}. Therefore, to model the cluster equilibrium in the EDL, \textit{we enforce that the cluster distribution in the EDL has the same form as in the bulk}, but is altered by the differing volume fractions of cations and anions in the EDL. \textcolor{black}{An equilibrium between the bulk and EDL then needs to be established, which keeps both the bulk and EDL cluster equilibria consistent. Note only one of the four EDL-bulk equilibrium (bulk free ions-EDL free ions; bulk free ions-EDL clusters; bulk clusters-EDL free ions; bulk clusters-EDL clusters) shown in Fig.~\ref{fig:chem_eq_all} needs to be established for the remaining three to also hold.} This approach should naturally show how the gel and all clusters respond to electrostatic fields, \textcolor{black}{as a consequence of all chemical equilibria shown in Fig.~\ref{fig:chem_eq_all} being held consistently.}  

\section{Preview}

The results of this theory shall be briefly summarised, as there are conceptual advances here which are beneficial to bear in mind before the technical parts are introduced. Firstly, the main advancement this paper makes is the account of the \textit{chemical equilibrium between clusters within the EDL}. Consideration of this equilibrium permits a natural way to investigate strongly aggregating and gelating electrolytes in the EDL.

\begin{figure}
    \centering
    \includegraphics[width=0.45\textwidth]{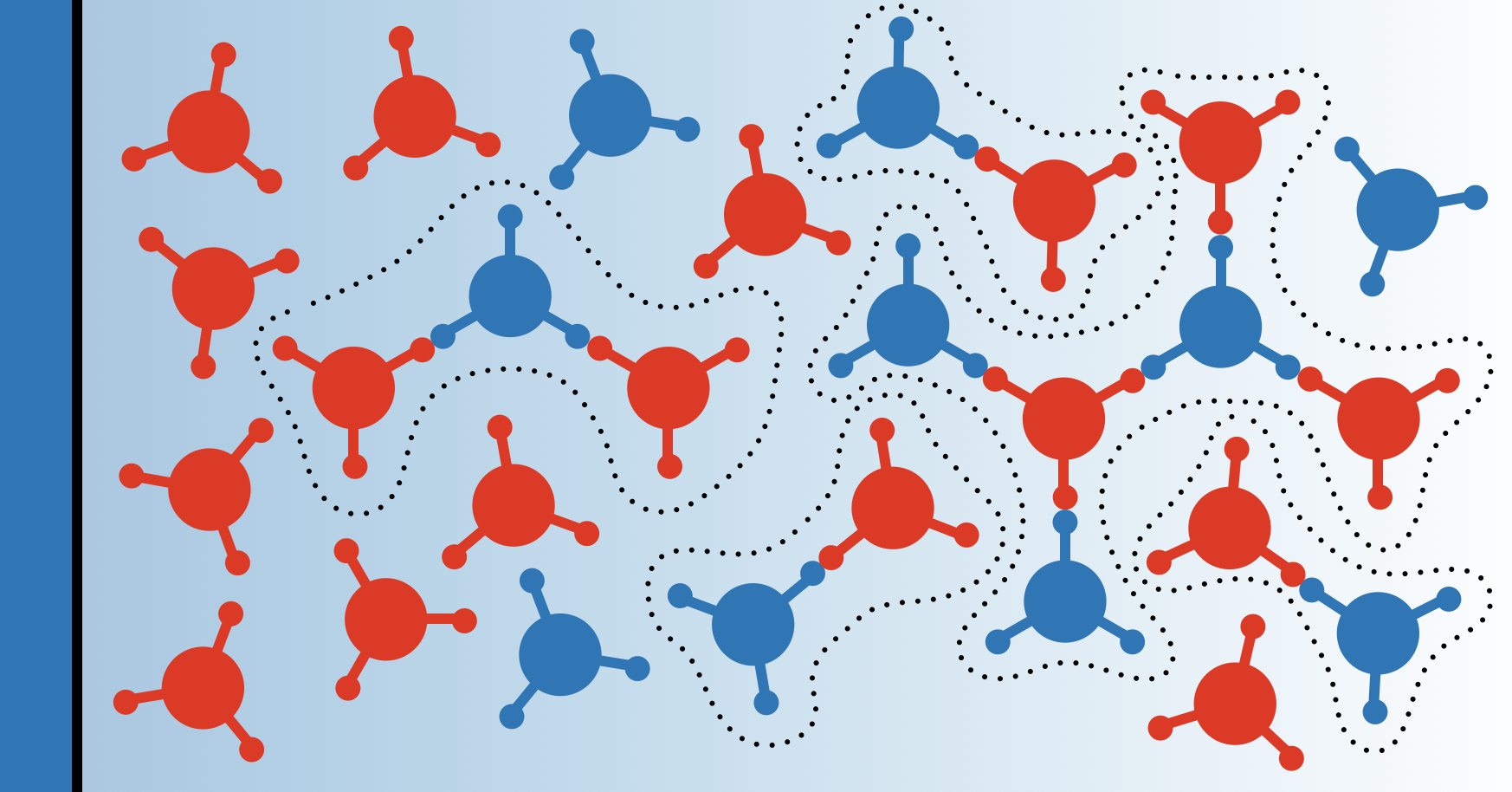}
    \caption{\textbf{Schematic of changes to the clustering in the EDL}. On the left depicted is a negatively charged electrode, which polarized the EDL with a potential distribution decaying to the right. The same notation for the clusters as Fig.~\ref{fig:into_fig} is used, but where each ion can form a maximum of 3 associations.}
    \label{fig:EDL_F}
\end{figure}

This theory predicts the following. For electrolytes which do not contain a gel, the large clusters are broken down / expelled from the EDL in favour of free ions, as schematically shown in Fig.~\ref{fig:EDL_F}. The account of clusters beyond free ions results in a differential capacitance response that is larger than that of just a free ion approach. Overall, this is the expected result for the pre-gel regime, and mainly differs to previous free ion theories (FIT) through quantitative measures.

For electrolytes which contain gel, the physical picture is more complex. It is found that \textit{the gel can screen electrode charge}. This is because the gel becomes charged in an electrostatic potential, with the charge being counter-ion dominated. This arises because the equilibrium between cation-anion associations changes in the EDL, where there are unequal numbers of cations/anions. For an accumulation of anions (cations) in the EDL, one finds that the gel is also negative (positive), because the equilibrium tries to shift back (Le Chatelier's principle) and remove some of the free anions (cations). Moreover, the cations (anions) are removed for positive (negative) potentials, which causes cations (anions) to dissociate from the gel to replenish the free cation (anion) concentration. At sufficiently large potentials, the ratio of cations/anions (anions/cations) in the EDL becomes small enough for the gel to be destroyed. This causes the gel to dissociate and release large aggregates. 

\textcolor{black}{For even larger potentials, these large aggregates are also broken down, and more and more free ions accumulate in the EDL, until the crowding regime is reached. Here crowding refers to an electrolyte composition where practically only free counter-ions exist, which can be achieved locally in the EDL. Gradually, the excluded volume effect causes the thickness of the EDL to increase.} Thus the differential capacitance of the EDL, which is inversely proportional to the EDL thickness, starts decreasing at large potential drops across the EDL. At small voltages, however, the differential capacitance is increasing~\cite{Kornyshev2007,kilic2007a}, and therefore, the capacitance curve will be a typical ``double hump camel'' shape overall~\cite{Kornyshev2007}. Again we find that accounting for the gel/cluster response increases and smooths out the response in comparison to a free ion theory. 

Note that one must not expect the theory to provide the picture with `molecular resolution', explicitly showing overscreening oscillations of charge density and the spatial structure within clusters. Within the large clusters or the gel, alternation of counterions and coions should give rise to such oscillations. The presented theory is able to provide a `\textcolor{black}{coarse}-grained' structure of charge distribution in the EDL, spread over different clusters, the balance between which shifts with the polarization of the electrode. Needless to say, such theory will be thermodynamic, not covering any dynamic aspects of the EDL charging.

\section{Theory}

\subsection{Ionic Liquid Associations}

\begin{figure}
    \centering
    \includegraphics[width=0.4\textwidth]{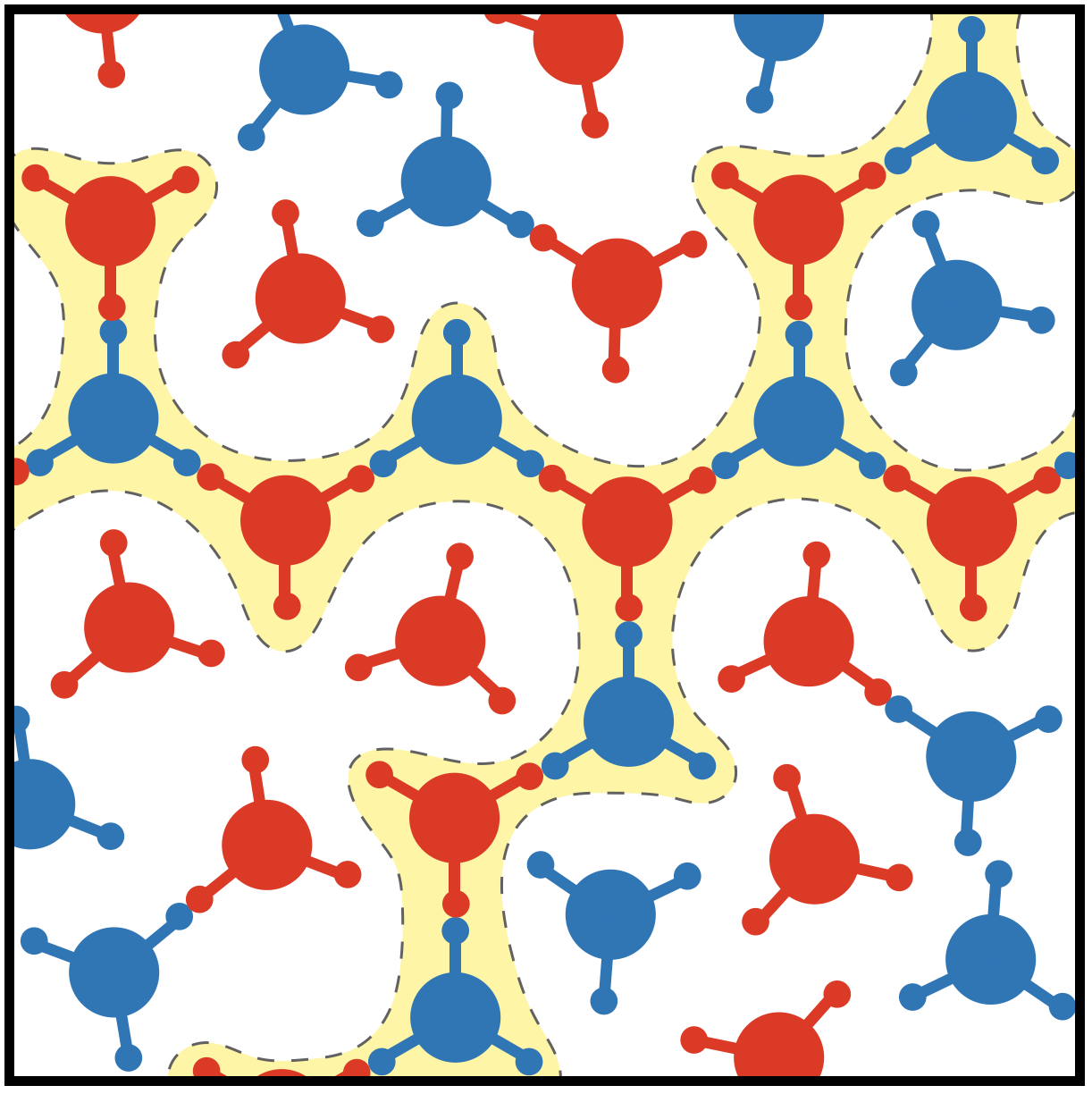}
    \caption{\textcolor{black}{\textbf{Schematic of co-existing sol and gel}. Cations and anions are shown in red and blue, respectively, and both have a functionality of 3. The percolating ionic network, i.e., the gel, is indicated by the enclosed dashed line. The remaining cations and anions are part of the sol. In the absence of the gel, all cations and anions are part of the sol. }}
    \label{fig:SOL_GEL}
\end{figure}

The limit of a symmetric, incompressible IL is employed here, i.e., one where the volumes of cations and anions are the same ($v_+ = v_- = v$), no voids are considered (every lattice site is either occupied by a cation or anion), and the number of associations a cation and anion can form is the same (referred to as functionality, given by $f_+ = f_- = f$). \textcolor{black}{It is assumed that cations can only bind to anions, and anions to cations, i.e. no cation-cation and anion-anion associations are accounted for, which has been shown to be a good approximation for ILs in Ref.~\citenum{mceldrew2020correlated}. No interactions beyond ions which are directly associated are taken into account in determining the extent of associations (in the bulk). The ions are assumed to form Cayley tree clusters, i.e. ones in which there are no intra-cluster loops. Therefore, cations and anions can reversibly form ``branched alternating co-polymers'' of various sizes, from ion pairs to a percolating ionic network. These clusters/gel are schematically shown in Figs.~\ref{fig:into_fig} and \ref{fig:SOL_GEL}.}

\textcolor{black}{Physically, these short-range associations between cations and anions in ILs represent the strong electrostatic correlations (beyond mean-field) between ions of opposite sign. Levy \textit{et al.}~\cite{levy2019spin} showed that the discrete charges of ions with a given packing (e.g. from molecular dynamics simulations) could be well reproduced by a nearest neighbour spin-glass theory that maximizes their alternating-charge ordering, which implies that our short-ranged model of Coulomb correlations should be accurate for these concentrated electrolytes. Moreover, in Ref.~\citenum{mceldrew2020correlated} it was found that IL ions have well defined `hot-spots', where cations (anions) prefer to reside around anions (cations), which also supported these assumptions and permitted the functionality to be determined (from the number of hot-spots).}

\textcolor{black}{ILs are the simplest possible super-concentrated electrolyte, and should serve to clearly investigate the EDL of electrolytes which can form associations beyond ion pairs. For further details about the IL limit, see Ref.~\citenum{mceldrew2020correlated}. }

The electrolyte is treated on the level of a lattice-gas~\cite{mceldrew2020theory,mceldrew2020correlated}. The number of lattice sites, $\Omega$, is given by 
\begin{equation}
\Omega = \sum_{lm}(l + m)N_{lm} + N_{+}^{gel} + N_{-}^{gel},
\end{equation}

\noindent where $N_{lm}$ is the number of clusters of rank $lm$, and $N_{+/-}^{gel}$ is the number of cations/anions in the gel phase. \textcolor{black}{The cations and anions which are not part of the gel, i.e., all of the free cations, free anions and clusters of rank $lm$, are part of the \textit{sol}. This is schematically shown in Fig.~\ref{fig:SOL_GEL}. In the absence of gel, all cations and anions are in the sol. Throughout, free ions and clusters shall not be explicitly referred to as being in the sol, as they are such by definition. Therefore, we do not explicitly include $sol$ superscripts for free ions and clusters for clarity of notation.} Equivalently, dividing by the total number of lattice sites yields
\begin{equation}
1 = \sum_{lm}(l + m)c_{lm} + c_{+}^{gel} +c_{-}^{gel},
\end{equation}

\noindent where $c_{lm} = N_{lm}/\Omega$ is the dimensionless concentration ($\#$ per lattice site) of rank $lm$ clusters, and the dimensionless concentration of cations/anions in the gel is $c_{+/-}^{gel} = N_{+/-}^{gel}/\Omega$, i.e., the volume fraction $c_{+/-}^{gel} = \phi_{+/-}^{gel}$. The volume fraction of a cluster of rank $lm$ is given by $\phi_{lm} = (l + m)c_{lm}$. The volume fraction of cations and anions in the sol phase is 
\begin{equation}
    \phi_+^{sol} = \sum_{lm}lc_{lm},
\end{equation}
\begin{equation}
    \phi_-^{sol} = \sum_{lm}mc_{lm}.
\end{equation}

\noindent The total volume fraction of cations/anions is $\phi_{+/-} = \phi_{+/-}^{sol} + \phi_{+/-}^{gel}$. When there is no gel, the superscript $sol$ shall be dropped for clarity of notation. \textcolor{black}{In Fig.~\ref{fig:SOL_GEL}, the partitioning of the sol and gel is schematically shown.}

The free energy of the cluster equilibrium~\cite{mceldrew2020theory,mceldrew2020correlated} is taken to be
\begin{align}
\beta \mathcal{F} &= \sum_{lm} \left[N_{lm}\ln \left( \phi_{lm} \right)+N_{lm}\Delta_{lm} \right] \nonumber\\
&+ \Delta^{gel}_{+} N^{gel}_{+} + \Delta^{gel}_{-} N^{gel}_-,
\label{eq:Fb}
\end{align}

\noindent where $\beta = 1/k_BT$ is inverse thermal energy, $\Delta_{lm}$ is the free energy of formation of a cluster of rank $lm$ from free cations and anions, and $\Delta_{+/-}^{gel}$ is the free energy change of cations /anions associating with the gel. 

Establishing chemical equilibrium~\cite{mceldrew2020theory,mceldrew2020correlated,mceldrew2021salt}, as shown in the  Supplementary Material (SM), the cluster distribution is expressed as 
\begin{equation}
    c_{lm}=\frac{W_{lm}}{\lambda}  \left(\lambda f\phi_{10}\right)^l \left(\lambda f\phi_{01}\right)^m.
\end{equation}

\noindent where $\lambda = \exp(-\beta \Delta f_{+-})$ is the ionic association constant, \textcolor{black}{with $\Delta f_{+-} = \Delta u_{+-} - T\Delta s_{+-}$ denoting the free energy of an association, determined by the binding energy ($\Delta u_{+-}$) and (configurational) entropy of an association ($\Delta s_{+-}$)}. Here $W_{lm}$ is the number of ways to arrange $l$ cations and $m$ anions in a Cayley tree~\cite{mceldrew2020theory,mceldrew2020correlated}, i.e. the combinatorial contribution \textcolor{black}{to the free energy of formation of a cluster}, which is given by
\begin{align}
    W_{lm}=\frac{(fl-l)!(fm-m)!}{l!m!(fl-l-m+1)!(fm-m-l+1)!}.
\end{align}

The cluster distribution, $c_{lm}$, is expressed in terms of $\phi_{10}$ and $\phi_{01}$, but, in principle, these are unknown quantities. In the bulk, the volume fraction of cations and anions is, however, known. Introducing association probabilities, $p_{ij}$, the probability that an association site of species $i$ is bound to species $j$, permits the volume fraction of free cations to be written as $\phi_{10} = \phi_+(1 - p_{+-})^{f}$ and free anions as $\phi_{01} = \phi_-(1 - p_{-+})^{f}$. 

The association probabilities can be determined through conservation of associations
\begin{align}
    f\phi_+p_{+-} = f\phi_-p_{-+} = \zeta,
    \label{eq:p1}
\end{align}

\noindent and mass action law 
\begin{align}
    \lambda\zeta = \frac{p_{+-}p_{-+}}{(1-p_{+-})(1-p_{-+})},
    \label{eq:p2}
\end{align}

\noindent to give the probability of cations binding to anions
\begin{equation}
    p_{+-} = \dfrac{1 + f\lambda - \sqrt{1 + 2f\lambda + f^2\lambda^2(1 - 2\phi_+)^2}}{2f\lambda\phi_+},
\label{eq:ppm}
\end{equation}

\noindent and anions binding to cations
\begin{equation}
    p_{-+} = \dfrac{1 + f\lambda - \sqrt{1 + 2f\lambda + f^2\lambda^2(1 - 2\phi_-)^2}}{2f\lambda\phi_-}.
\label{eq:pmp}
\end{equation}

\noindent Note that the volume fractions have been explicitly retained in these equations, as these probabilities can be investigated as a function of the volume fraction of cations and anions.

When the probability of a cation (anion) having another cation (anion) connected through an anion (cation), a percolating ionic network can form. This condition is given by $1 = (f-1)^2p^*_{+-}p^*_{-+}$, \textcolor{black}{where the stars are used to denote the critical probabilities for formation of the percolating ionic network}. When the probabilities reach this condition, the volume fractions of cations and anions in the gel and sol must be determined. Flory's treatment of the post-gel regime is employed, in which the volume fraction of free ions can be written equivalently in terms of overall association probabilities, $p_{ij}$, and association probabilities taking into account only the species residing in the sol, $p^{sol}_{ij}$
\begin{align}
    \phi_+(1-p_{+-})^{f}=\phi_+^{sol}(1-p^{sol}_{+-})^{f},
    \label{eq:gel1}
\end{align}
\begin{align}
    \phi_-(1-p_{-+})^{f}=\phi_-^{sol}(1-p^{sol}_{-+})^{f}.
    \label{eq:gel2}
\end{align}

\noindent Using Eqs.~\eqref{eq:gel1}-\eqref{eq:gel2} in addition to Eqs.~\eqref{eq:p1}-\eqref{eq:p2}, with sol-specific quantities, permits the determination of the sol probabilities and volume fractions. The fraction of species, $i$, in the sol is given by $w_i^{sol} = \phi^{sol}_i/\phi_i$. See SM for more details. 

\subsection{Cluster/Gel Composition of ILs}

Having summarised the equations for the cluster equilibrium, we turn to understanding the behaviour of the system with unequal volume fractions of cations and anions. In principle, one cannot arbitrarily control the volume fractions of cations or anions. The charge of an electrode sets the number of \textcolor{black}{excess counter} charges which must be in the EDL, to have overall charge neutrality. These charges shall be distributed in space according to the Poisson equation. However, it is quite useful for conceptual understanding to perform a thought experiment where one can change the volume fraction of cations/anions arbitrarily. Therefore, we shall investigate the composition of the electrolyte as a function of the volume fraction of cations. To start, we shall briefly summarise the composition of the electrolyte in the bulk, i.e. where $\phi_+ = \phi_- = 1/2$, for $\lambda = 1$ and $\lambda = 10$, \textcolor{black}{with $f = 3$}. Then we shall progress onto investigating $\phi_+ \neq \phi_-$.

\begin{figure}
    \centering
    \includegraphics[width=0.45\textwidth]{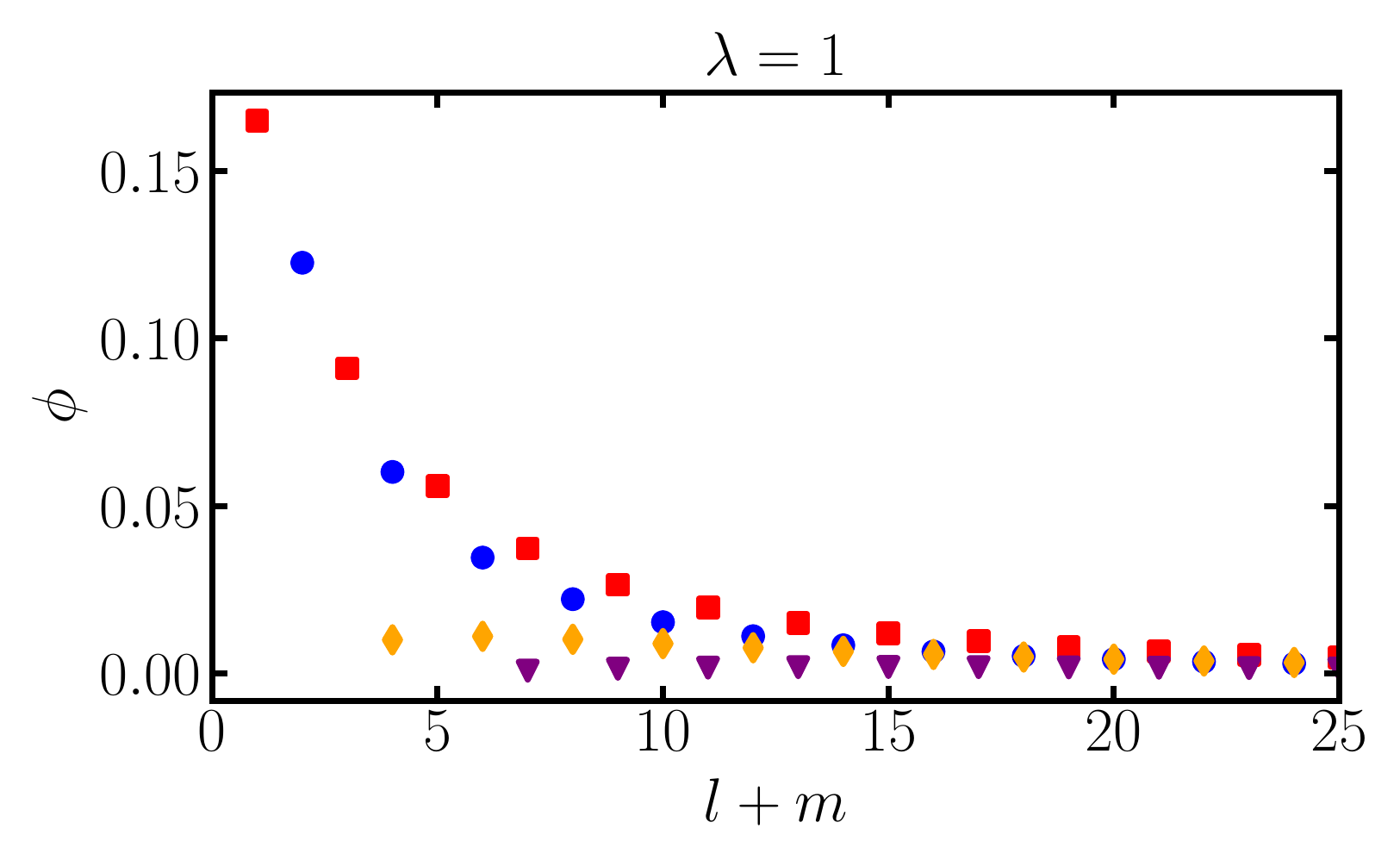}
    \centering
    \includegraphics[width=0.45\textwidth]{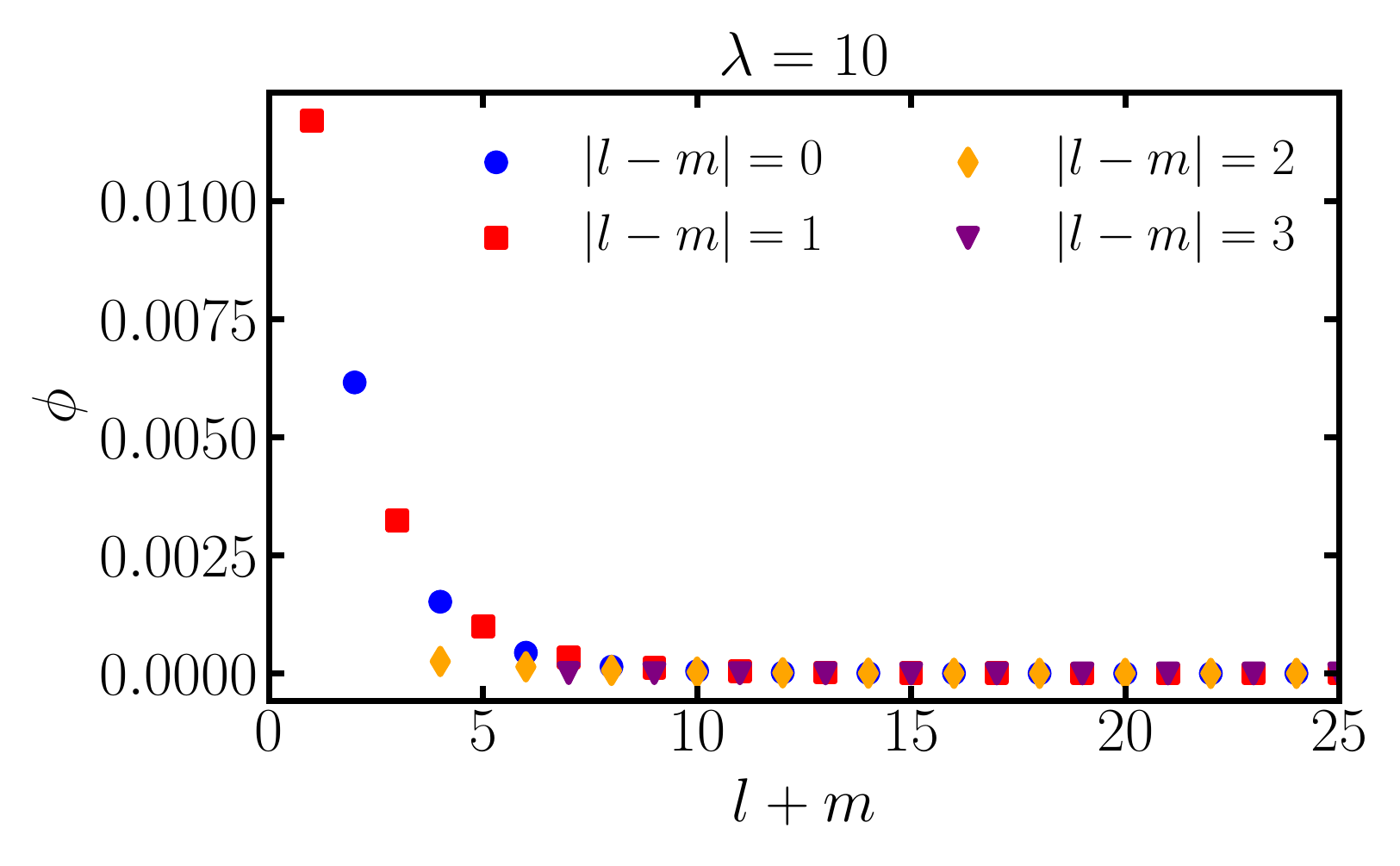}
    \caption{\textbf{The composition of the bulk IL is dominated by neutral and singly charged clusters}. Volume fraction of each rank of cluster, for a given charge $|l-m| = q$ with $q = 0,1,2,3$ and $f=3$, as a function of the size of a cluster $l+m$.}
    \label{fig:cluster_size}
\end{figure}

\subsubsection{Bulk Composition}

\textcolor{black}{In the bulk, the critical value for the association constant for the formation of the gel is given by $\lambda^* = 2(f-1)/[f(f-2)^2]$. For $f=3$, this puts $\lambda^* = 4/3$, which means the chosen association constants are on either side of the gel point. These two different values of $\lambda$ can be considered as an example of those at a high and low temperature, respectively.} 

\textcolor{black}{For larger functionalities, $\lambda^* < 1$, which might be surprising, as this corresponds to a positive free energy of an association, $\Delta f_{+-}$. It was found in Ref.~\citenum{mceldrew2020correlated}, from the temperature dependence of the association constant, that the binding energy is negative, but the entropy of an association is positive, which may result in $\Delta f_{+-} > 0$. The reason why a gel can form with $\lambda < 1$, and why there are still significant associations despite $\Delta f_{+-} \le 0$, can be understood in several ways. The free energy of an association $\Delta f_{+-}$, is not the only contribution to the free energy of formation of clusters of rank $lm$, there are additional entropic contributions (combinatorial entropy and part of the configurational entropy - see Refs.~\citenum{mceldrew2020theory,mceldrew2020correlated} for more details) which drive the system towards a mixture rather than a pure state (of free ions). This has also been found by Lee \textit{et al.}~\cite{lee2014room} when considering ion pair formation in ILs, where $2/3$ of ions were found to be free. This was because the Debye screening length was much shorter than the radius of an ion, which meant the entropy of mixing decided the fraction of free ions, i.e. the IL was effectively treated as an ideal solution. Alternatively, this can be understood from the mass action law, Eq.~\eqref{eq:p2}, which shows that the association probabilities only vanish when $\lambda = 0$. The association constant is, effectively, the equilibrium constant of the formation of an association, which means that when $\lambda = 1$ it is equally favourable to form an association as not to, and therefore, entropy maximisation dictates that a mixture shall form. As the entropy of an association was found to be positive in Ref.~\citenum{mceldrew2020correlated}, the association constant tends to 0 at large temperatures, which is the expected limit of no associations. Note that associations between cations-anions still occur, despite $\Delta f_{+-} > 0$ due to the negative binding energy.}

\textcolor{black}{Having described the extent of associations for different $\lambda$, we turn to understanding the cluster distribution in more detail.} In Fig.~\ref{fig:cluster_size} the volume fraction of each cluster of rank $lm$ is shown as a function of the size of each cluster ($l+m$) for the bulk $\phi_+ = \phi_- = 1/2$ with $f = 3$. This is plotted for various overall charges of clusters, with charges up to $\pm3$ being considered. As it is the symmetric case, the volume fraction of a $+q$ charged cluster is equal to the $-q$ charged clusters. Therefore, these are summed together, and each different symbol represents a $|l-m| = q$, for $q=0,1,2,3$. 

For $\lambda = 1$, see Fig.~\ref{fig:cluster_size}(left), the IL is in the pre-gel regime. There is approximately 16$\%$ of free ions, with the remaining fraction of ions being bound up into clusters. The neutral ion pairs are the next most populous species, followed by singly charged triples, etc. These singly charged and neutral clusters dominate the electrolyte composition (for a symmetric IL in the bulk~\cite{mceldrew2020theory}). A cluster with $|l-m|=2$ can first occur when there are 4 ions in the cluster. The volume fractions of these clusters increases slightly with $l+m$, since for larger $l+m$ not all association sites must be occupied (unlike $l+m=4$ for $f=3$), \textcolor{black}{before decreasing again for large cluster ranks}. A similar story occurs for $|l-m|=3$, where it is found to have even lower volume fractions. 

For $\lambda=10$, see Fig.~\ref{fig:cluster_size}(right), the IL is in the post-gel regime, with most of the ions being bound in the gel. For the remaining clusters, a similar situation can be seen for the populations of clusters. For more information about the bulk cluster composition of an IL, see Ref.~\citenum{mceldrew2020correlated}. 

\subsubsection{``EDL'' Composition}

\begin{figure}
    \centering
    \includegraphics[width=0.45\textwidth]{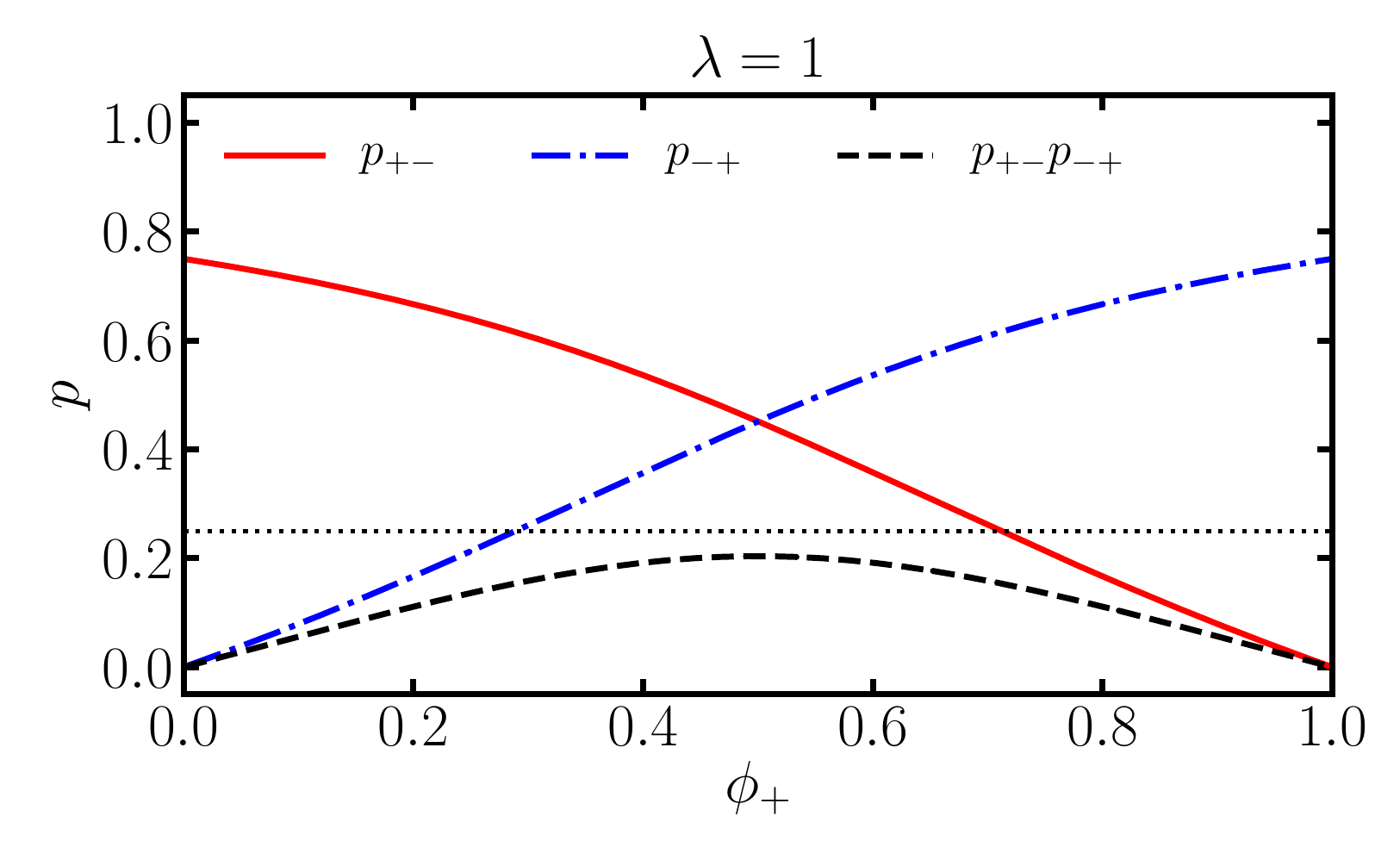}
    \centering
    \includegraphics[width=0.45\textwidth]{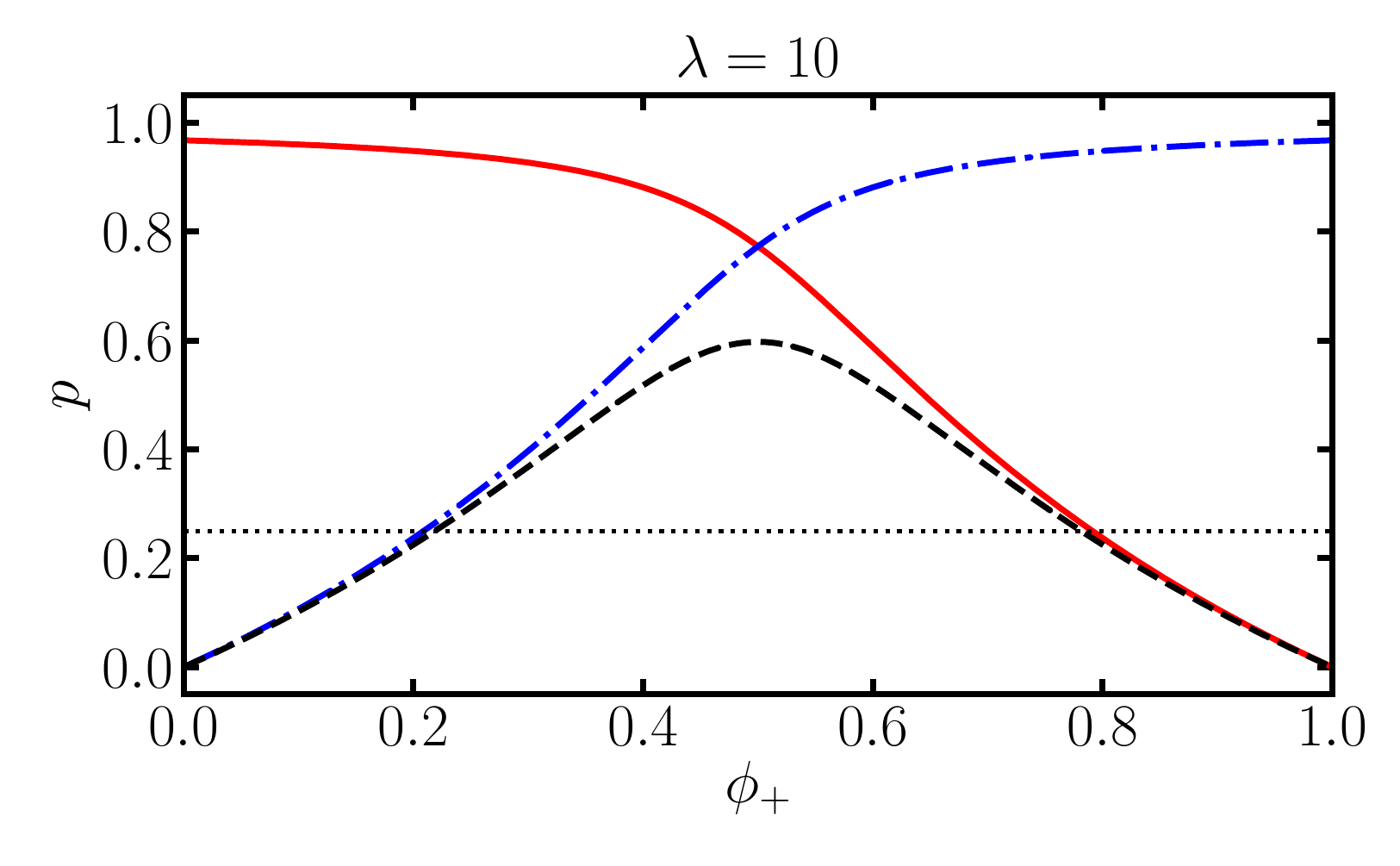}
    %
    %
    \caption{\textbf{The IL becomes less associating as the volume fractions of cations increases/decreases from the bulk value.} Association probabilities, as indicated in the legend, as a function of volume fraction of cations. Two association constants are shown, as indicated in the titles. Both plots are for a functionality of 3, and the horizontal dotted line denotes the critical probability for the onset of a gel, $p^*_{+-}p^*_{-+} = (f-1)^{-1}$.}
    \label{fig:prob_arb}
\end{figure}

In Fig.~\ref{fig:prob_arb}(left), the association probabilities as a function of volume fraction of cations is shown, for $\lambda=1$ 1 with functionality of $f=3$. As the volume fraction of cations increases, the association probability of cations binding to anions ($p_{+-}$) decreases and the association probability of anions binding to cations ($p_{-+}$) increases. The opposite is true for a decreasing volume fraction of cations. This is a consequence of the conservation of associations, which states $p_{+-}/p_{-+} = \phi_-/\phi_+$. At $\phi_+ = 1$, the association probability of cations binding to anions necessarily goes to zero, $p_{+-} = 0$, and $p_{-+}$ reaches a constant value equal to or less than 1. Vice versa for $\phi_+ = 0$. 

For $\lambda = 10$, the association probabilities are sufficiently large to create a percolating ionic network in the bulk, as seen by $p_{+-}p_{+-}$ being larger than the dashed horizontal line in Fig.~\ref{fig:prob_arb}(right). Again, $p_{+-}$ decreases with increasing $\phi_+$, and $p_{-+}$ increases with increasing $\phi_+$. There comes a point, for both small and large $\phi_+$, that the volume fraction of cations (or anions) cannot sustain a percolating ionic network, and $p_{+-}p_{+-}$ drops underneath the threshold.

\begin{figure}
    \centering
    \includegraphics[width=0.45\textwidth]{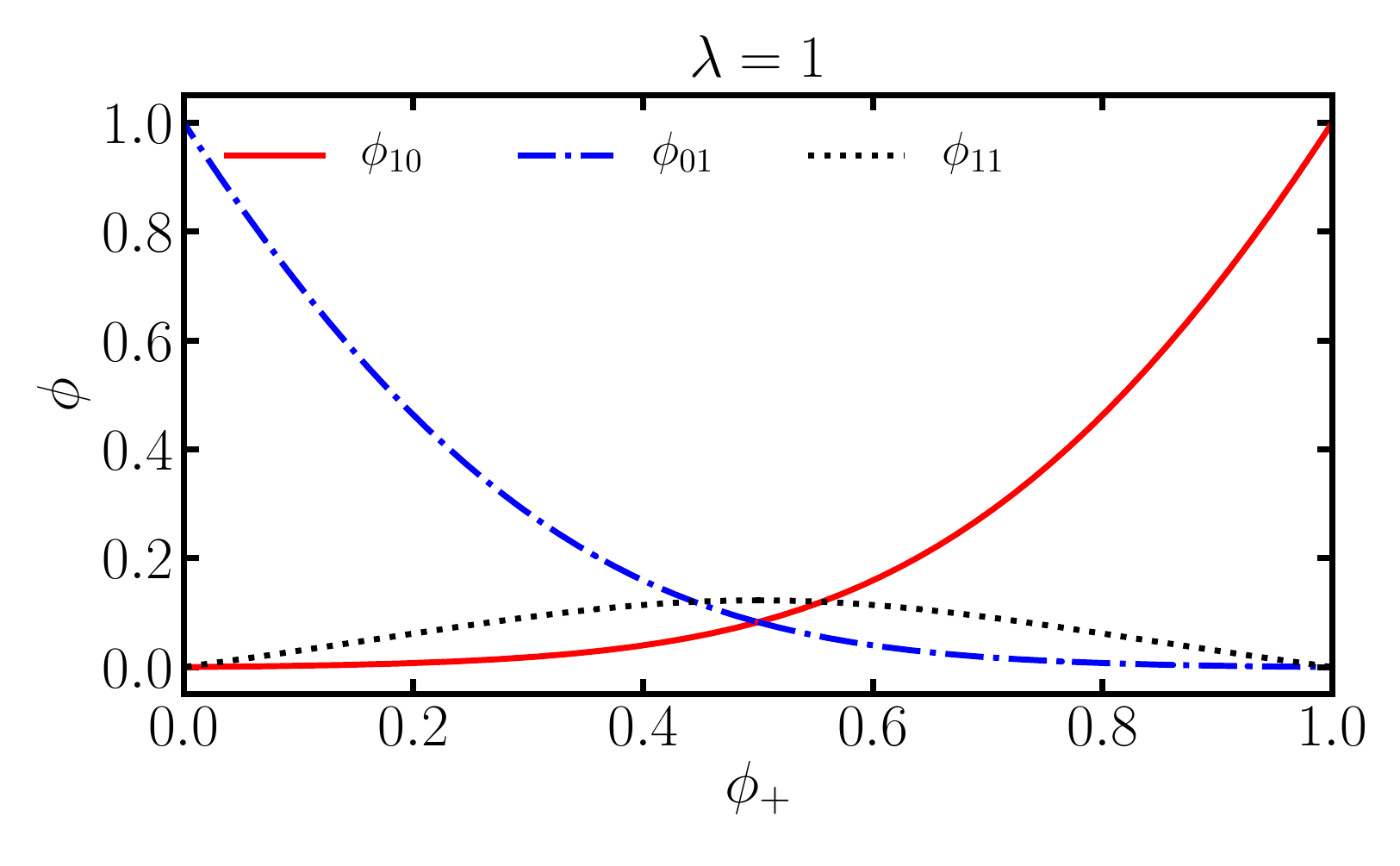}
    \centering
    \includegraphics[width=0.45\textwidth]{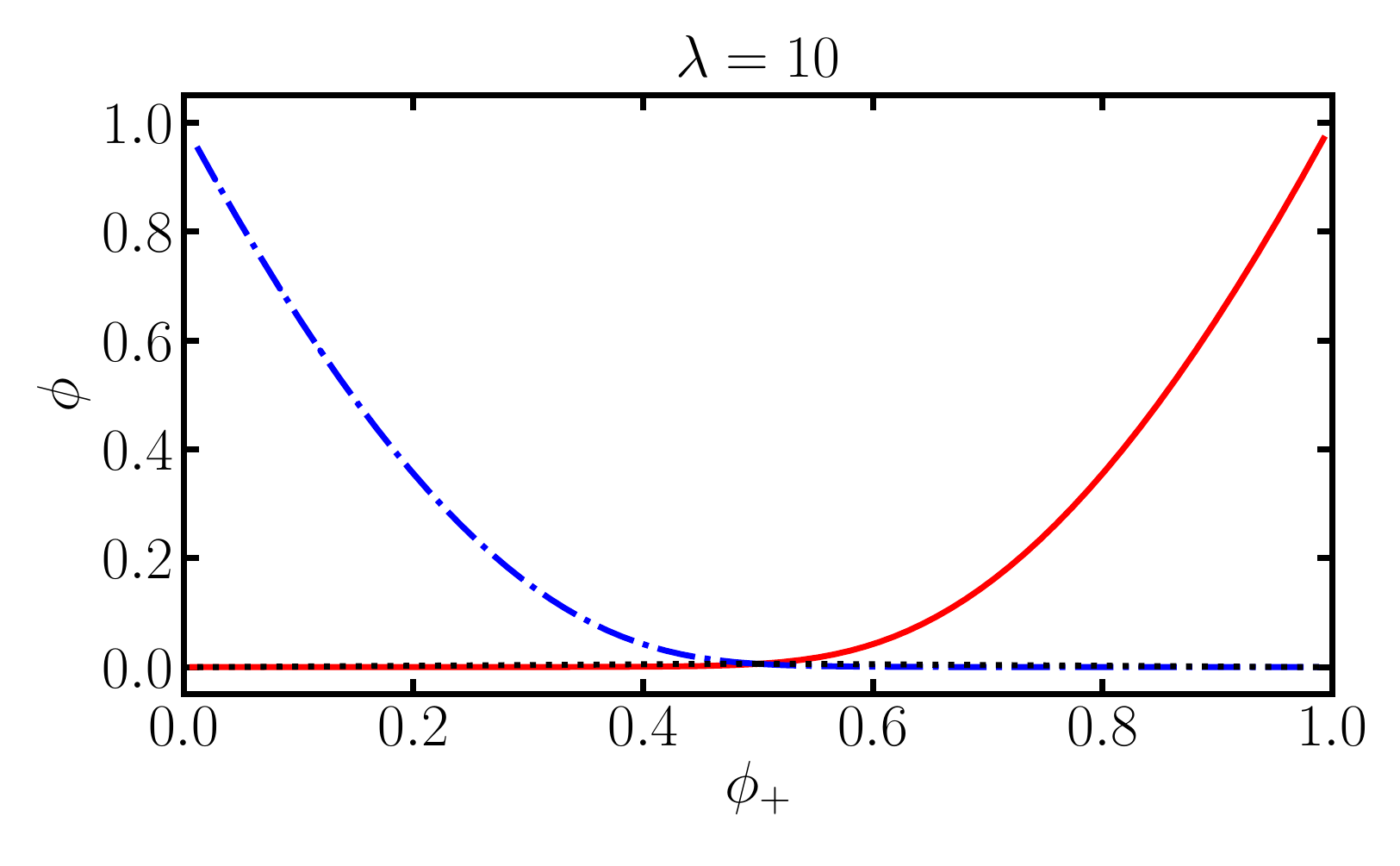}
    \centering
    \includegraphics[width=0.45\textwidth]{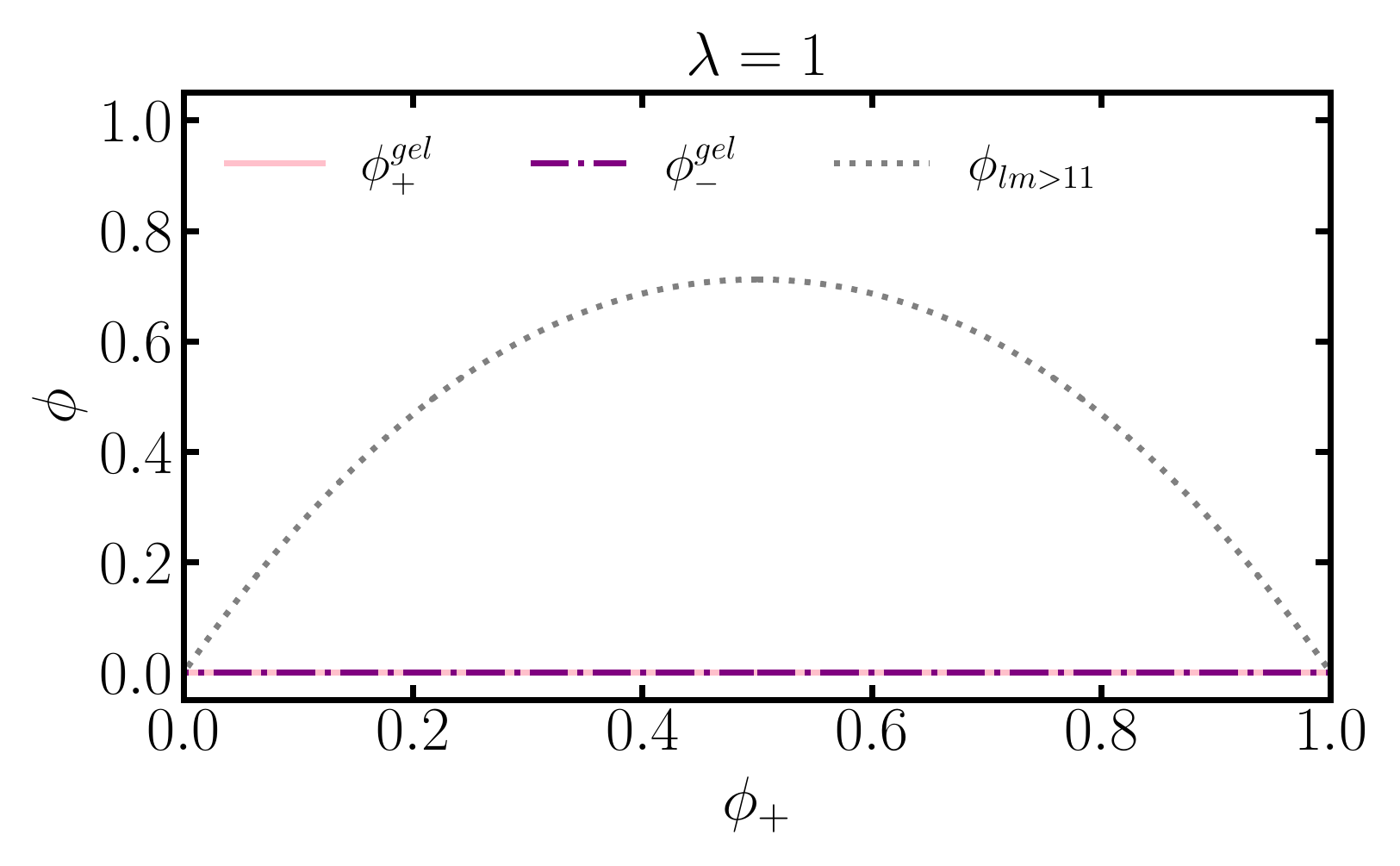}
    \centering
    \includegraphics[width=0.45\textwidth]{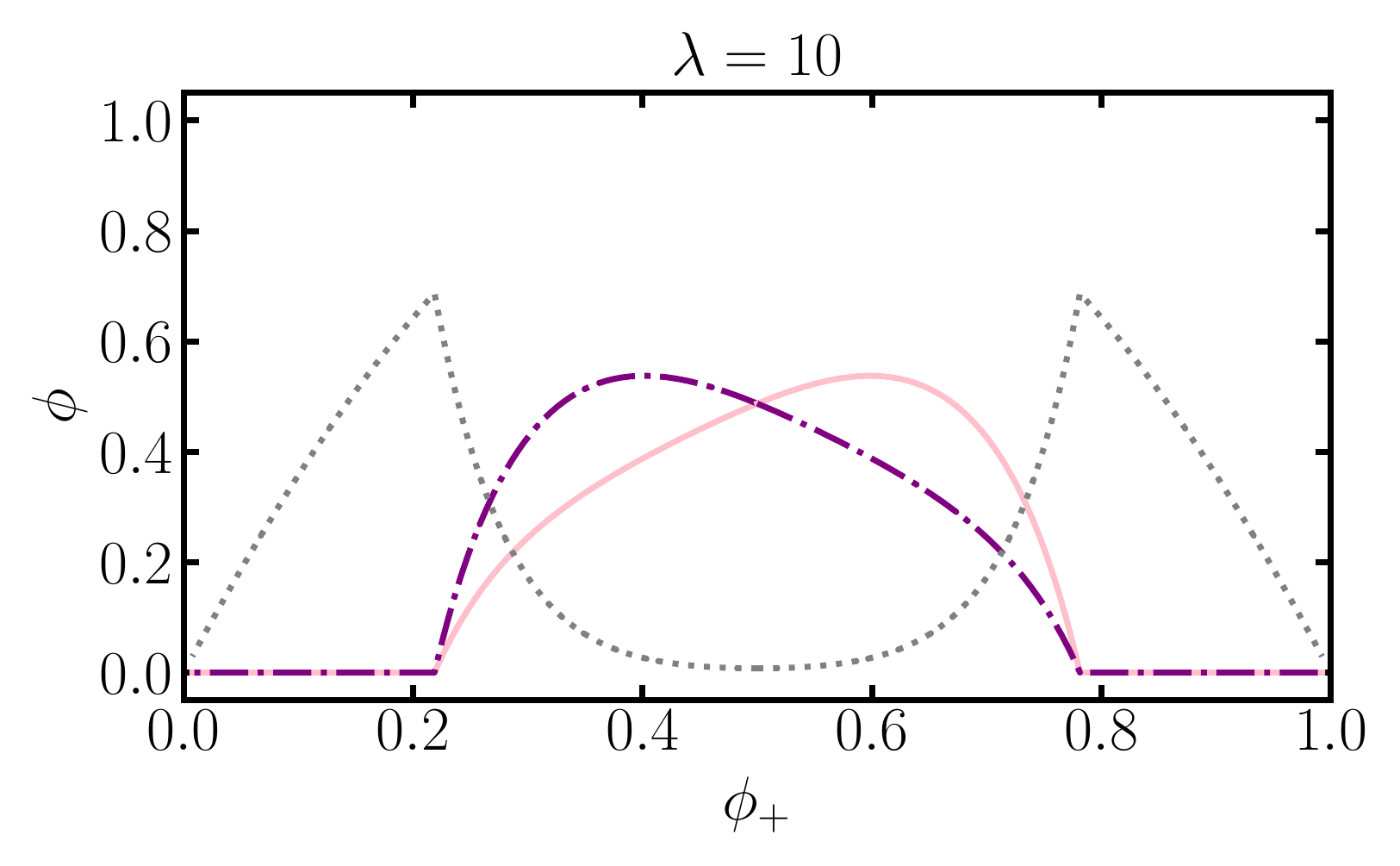}
    %
    %
    \caption{\textbf{Free ions become increasingly dominant and the gel is destroyed for sufficiently charged systems}. Volume fractions of free cations ($\phi_{10}$), free anions  ($\phi_{01}$) and ion pairs  ($\phi_{11}$) as a function of the volume fraction of cations (top panels). Volume fractions of aggregates beyond ion pairs  ($\phi_{lm > 11}$), volume fraction of cations in the gel ($\phi_{+}^{gel}$) and volume fraction of anions in the gel ($\phi_{-}^{gel}$) as a function of volume fraction of cations (bottom panels). Again, all plots are for $f=3$ and the two association constants are shown in the titles.}
    \label{fig:proper_arb}
\end{figure}

Next, how these association probabilities influence the cluster distribution of the electrolyte is shown in Fig.~\ref{fig:proper_arb} as a function of the volume fraction of cations. For $\lambda =1$, the volume fraction of free cations ($\phi_{10}$) increases with the volume fraction of cations ($\phi_+$), and the volume fraction of free anions ($\phi_{01}$) decreases with increasing $\phi_+$. At $\phi_+ = 1$, $\phi_{10} = 1$ and $\phi_{01} = 0$, and vice versa for $\phi_+ = 0$. This is a reflection of the changing association probabilities and volume fractions, as just discussed. The volume fraction of ion pairs ($\phi_{11}$) and aggregates beyond ion pairs ($\phi_{lm > 11}$) decreases for increasing and decreasing volume fractions of cations (relative to its bulk value). This is because of the average association probability, $(p_{+-} + p_{-+})/2$, decreases as the volume fractions of cations is changed from the bulk value. 

For $\lambda = 10$, the volume fraction of free cations and free anions behave in an analogous manner to $\lambda = 1$, and the volume fraction of ion pairs is always extremely small. The aggregates beyond ion pairs behave qualitatively differently, however, with a highly non-monotonic dependence on $\phi_+$. To understand this, the behaviour of the gel must first be described. In the bulk, the system is significantly gelled. As the volume fraction of cations increases ($\phi_+$), the volume fraction of cations in the gel ($\phi_+^{gel}$) also increases and the volume fraction of anions in the gel ($\phi_-^{gel}$) decreases. This is because the prevalence of cations shifts the free ion-gel equilibrium to accommodate more cations. Therefore, \textit{the gel becomes charged and decreases in total volume fraction}, until it reaches the critical volume fraction where it can no longer be sustained. This decreasing volume fraction of gel causes a dramatic increase in the volume fraction of aggregates beyond ion pairs, as closer to the gel point larger and larger clusters dissociate from the gel. When the critical volume fraction to sustain a gel is reached, there is no longer gel to create large clusters, and the decreasing average association probability causes a dramatic decrease in the volume fraction of these larger clusters. 

Overall, the phenomenology described in this thought experiment is exactly what is expected for the behaviour of the clusters in the EDL: a transition from gelation/aggregation to crowding. Therefore, what is required is a connection between the volume fractions of cations/anions and the electrostatic potential, which can be linked to the spatial distribution of charges through the Poisson equation. 

Inspecting the equations for $p_{\pm\mp}$ and $\phi_{10/01}$ demonstrates that, for a given volume fraction of cations, the volume fractions of free anions and free cations are not independent, but constrained by the associations. Therefore, if EDL-bulk equilibrium was established with ``standard approaches", such as Boltzmann or Fermi functions for free cations and free anions, the system of equations (the above equations for the concentrations of each species and the Poisson equation) becomes over-determined and the cluster distribution is no longer consistent. Therefore, an alternative approach to investigate the EDL properties in a consistent way is sought. In the SM, this is shown explicitly from the free energy. Moreover, alternative approaches to make the cluster distribution consistent everywhere is outlined. 

It is interesting to note that this observation also applies to approaches, such as Refs.~\citenum{goodwin2017underscreening,Chen2017}, where the associations were only treated through the explicit treatment of the remaining free ions. In the bulk, the number of free ions was determined by the binding free energy of cations and anions. The bulk concentration of free cations and free anions was then utilised in a Poisson-Fermi theory for the EDL~\cite{goodwin2017mean,goodwin2017underscreening,adar2017bjerrum,avni2020charge}. This essentially treated the associations as irreversible in the EDL, but reversible in the bulk, which is also how Ref.~\citenum{Ma2015} investigated strongly associating ILs within a sophisticated classical density functional theory. For thermoreversible associations in the EDL, the number of associations based on the concentration of cations and anions must also be determined in the EDL. This deficiency in such approaches~\cite{goodwin2017mean,goodwin2017underscreening,adar2017bjerrum,avni2020charge} does not, however, qualitatively change the predictions of those theories in terms of the free ions.

\subsection{Boltzmann Closure of Free Ions}

To close the system of equations without over-determining, a single relationship is required to connect the electrostatic potential with the volume fractions. To achieve this, the following closure relationship is taken based on the ratio of free cations and free anions in the EDL 
\begin{equation}
     \dfrac{\phi_{10}}{\phi_{01}}e^{-2\alpha u} = \dfrac{\bar{\phi}_{10}}{\bar{\phi}_{01}} = \dfrac{\bar{\phi}_+(1 - \bar{p}_{+-})^f}{\bar{\phi}_-(1 - \bar{p}_{-+})^f},
    \label{eq:ansatz}
\end{equation}

\noindent where the bar has been used to indicate quantities in the EDL, and $u$ is the electrostatic potential in units of thermal volts (which is given by $1/\beta e$, with $e$ denoting the elementary charge, i.e. the magnitude of an ion studied here). The factor $\alpha$ is a parameter which describes additional correlations beyond mean-field electrostatics (defined such that $0 \leq \alpha \leq 1$), as introduced in Ref.~\citenum{goodwin2017mean}, \textcolor{black}{and shown to work well in Ref.~\citenum{BBKGK}.} Since $\lambda$ accounts for short-ranged attractive interactions between cations-anions, only the short-ranged repulsion between cations-cations/anions-anions are considered in $\alpha$. This prevents double counting the short-ranged attraction between cations-anions. Moreover, $\alpha$ is not assumed to be based on the free ion fractions, but the total ion fractions, since the gel can also become charged. This parameter acts to smooth-out the response of ions to the electrostatic potential, as Poisson-Boltzmann (or Fermi) is well known to overestimate the response~\cite{Bazant2009a,goodwin2021review}. This is analogous to dressed ion theory, where the charge of an ion is rescaled to smaller values because of correlations between ions~\cite{Gebbie2017rev}. 

Equation~\eqref{eq:ansatz} assumes that free ions behave in a Boltzmann way, whilst being consistent with the cluster distribution and incompressibility constraint. This expression can be used to solve for the local volume fraction of cations or anions within the EDL, which then determines the local free cation and free anion volume fractions. These can be used in the cluster distribution
\begin{equation}
    \bar{c}_{lm}=\frac{W_{lm}}{\lambda}  \left(\lambda f\bar{\phi}_{10}\right)^l \left(\lambda f\bar{\phi}_{01}\right)^m,
\end{equation}

\noindent to find the concentrations of all clusters. Moreover, when the free cation/anion volume fractions are known, so are the association probabilities. This means that the gel can also be treated in a consistent way. 

For $\lambda \ll 1$, the association probabilities tend to zero $p_{+-} = p_{-+} \approx 0$, which means $\bar{\phi}_+/\bar{\phi}_- \approx e^{-2\alpha u}$. Using the incomprehensibility condition, this can be solved to give $\bar{\phi}_+ - \bar{\phi}_- = -\tanh \alpha u$ for the charge density, which is the expected result when there are no associations~\cite{Kornyshev2007,kilic2007a}.

\subsubsection{Linear Response}

When there are associations, the approach cannot generally be solved analytically, but at linear response some insight can be gained. Taking a linear expansion of Eq.~\eqref{eq:ansatz} and introducing a symmetric perturbation of free ions, $\bar{\phi}_{10} = \phi_{10} + \delta \bar{\phi}_f$ and $\bar{\phi}_{01} = \phi_{01} - \delta \bar{\phi}_f$, yields $\bar{c}_{10} = c_{10}(1 - \alpha u)$ for the free cation concentration and $\bar{c}_{01} = c_{01}(1 + \alpha u)$ for the free anion concentration. Using this in the cluster distribution yields
\begin{align}
\bar{c}_{lm} = c_{lm}[1 - (l - m)\alpha u].
\end{align}

Introducing the EDL volume fraction and probabilities as the bulk perturbed by a small value, $\bar{\phi}_{\pm} = 1/2 \pm \delta \bar{\phi}$ and $\bar{p}_{\pm\mp} = p \pm \delta \bar{p}$, respectively, the volume fraction of cations changes by
\begin{equation}
    \delta \bar{\phi} = \dfrac{-(1-p)\alpha u}{2[1 + (f-1)p]}.
\end{equation}

\noindent where $p = (1 + f\lambda - \sqrt{1 + 2f\lambda})/f\lambda$ is the association probability in the bulk. 

Therefore, the Poisson equation takes the form
\begin{equation}
\nabla^2u = \dfrac{e^2\beta \alpha u}{v\epsilon_0\epsilon}\sum_{lm} (l-m)^2 c_{lm} = \dfrac{e^2\beta\alpha u}{v\epsilon_0\epsilon}\dfrac{(1 - p)}{\left[1 + (f-1)p\right]},
\end{equation}

\noindent where $\epsilon_0$ and $\epsilon$ are the permittivity of free space, and relative permittivity and $v$ is the volume of an ion (i.e. a lattice site). The screening length
\begin{equation}
    \ell = \dfrac{1}{\sqrt{\alpha}\kappa}\sqrt{\dfrac{1 + (f-1)p}{(1 - p)}}
    \label{eq:debye}
\end{equation}

\noindent is based on the ionic strength of the cluster distribution, with $\kappa = \sqrt{v\epsilon_0\epsilon/e^2\beta}$. When there are no associations, i.e. $p=0$, the Debye length is recovered. In the opposite limit the association probability tends to 1, and extremely large screening lengths can emerge. This expression for the ionic strength in terms of $p$ was previously derived in Ref.~\citenum{mceldrew2020correlated} from modifying the expression for the weight average degree of aggregation. Therefore, the Boltzmann closure of free ions appears to be well justified at linear response in the pre-gel regime.

Gebbie \textit{et al.}~\cite{Gebbie2013,Gebbie2015} have suggested that ILs screen the field in a Boltzmann way with few free ions. Thus, this closure relation could also be assumed to hold in the post-gel regime. This would mean the screening length in Eq.~\ref{eq:debye} also holds in the post-gel regime, because it can be derived from the total volume fractions and probabilities. The contribution to this screening length from the clusters and gel can be derived. Again, a symmetric perturbation for the change in volume fractions of sol ($\delta \bar{\phi}^{sol}_+ = - \delta \bar{\phi}^{sol}_- = \delta \bar{\phi}^{sol}$) and sol probabilities ($\delta \bar{p}^{sol}_{+-} = - \delta \bar{p}^{sol}_{-+} = \delta \bar{p}^{sol}$) is introduced. In the system of equations, there are two unknowns, $\delta \bar{\phi}^{sol}$ and $\delta \bar{p}^{sol}$, and therefore, two equations are required. One equation of which is Flory's treatment of the post-gel regime; the second equation is the conservation of associations in the sol. Solving these equations, whilst using the previous results for $\delta \bar{\phi}$ and $\delta \bar{p}$, gives
\begin{equation}
    \delta \bar{\phi}^{sol} = -\dfrac{w^{sol}(1 - p^{sol})\alpha u}{2[1 + (f-1)p^{sol}]},
\end{equation}

\noindent which is an analogous expression for the sol quantities. Therefore, the change in the volume fraction of the gel is then given by
\begin{equation}
     \delta \bar{\phi}^{gel} = -\dfrac{\alpha u}{2}\left\{\dfrac{(1 - p)}{1 + (f-1)p} - \dfrac{w^{sol}(1 - p^{sol})}{1 + (f-1)p^{sol}}\right\}.
\end{equation}

\noindent These results can be used to decompose the total screening length into the contributions from the sol and the gel.

\begin{figure}
    \centering
    \includegraphics[width=0.45\textwidth]{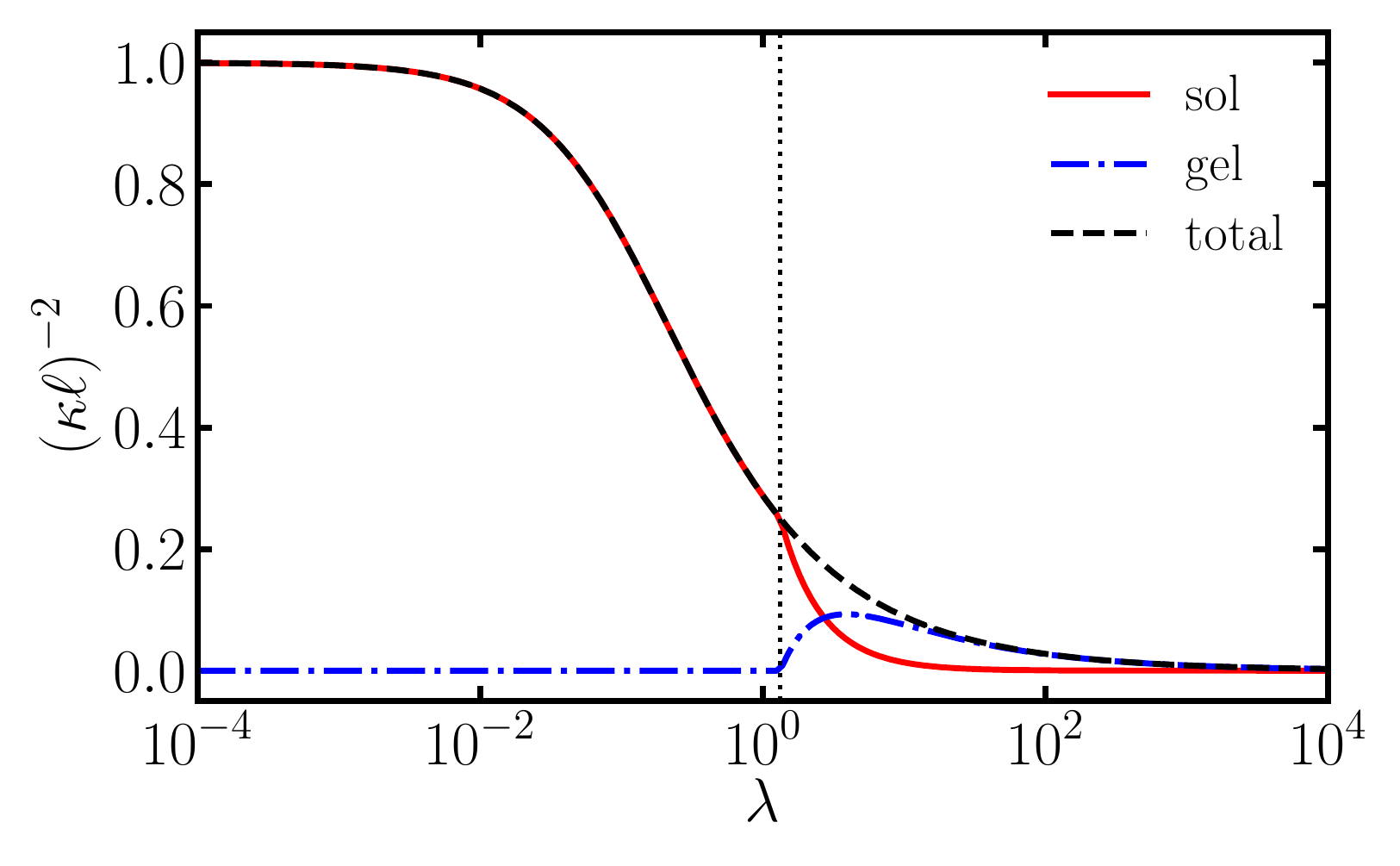}
    \caption{\textbf{The ionic strength decreases with association constant, but in the post-gel regime the gel dominates the ionic strength}. Dimensionless ionic strength, given by $1/(\kappa \ell)^2$, as a function of association constant, $\lambda$, for $f=3$. Note the x-axis is on a logarithmic scale. The $a$ parameter is set to 0.}
    \label{fig:ionic}
\end{figure}

In Fig.~\ref{fig:ionic} the dimensionless ionic strength, $(\kappa \ell)^{-2} = \sum_{lm}(l-m)^2c_{lm}$ (taking $\alpha = 1$), is shown as a function of the association constant. For $\lambda < \lambda^*$\textcolor{black}{, where the star is used to denote the critical association constant for the formation of the gel,} the ionic strength is determined by the clusters. For an association constant just above the critical value, both clusters in the sol and the gel contribute to the screening at linear response. For $\lambda \gg \lambda^*$, practically only the gel contributes to the screening at linear response. This indicates that approaches based on free ions~\cite{goodwin2017underscreening,Chen2017}, which follow the sol contribution closely~\cite{mceldrew2020correlated}, underestimate the ability of the electrolyte to screen. 

\subsubsection{Non-Linear Response}

Generally, the differential equation which needs to be solved is given by
\begin{equation}
\nabla^2u = -\kappa^2\left[\sum_{lm} (l-m) \bar{c}_{lm} + \bar{c}_+^{gel} - \bar{c}_-^{gel}\right] = -\kappa^2\left[\bar{\phi}_+ -\bar{\phi}_- \right].
\label{eq:Poisson_NL}
\end{equation}

\noindent A constant charge boundary condition at the interface is taken and a zero electrostatic potential in the bulk is used. Numerical results for the solution to this non-linear differential equation shall be shown in the next section. \textcolor{black}{Note a Stern layer is not considered here for clarity of discussing the new results.} \textcolor{black}{In the SM, a step-by-step guide of how to implement the equations to obtain a numerical solution is given.}

\section{Results}

\subsection{EDL Structure}

\begin{figure}
    \centering
    \includegraphics[width=0.45\textwidth]{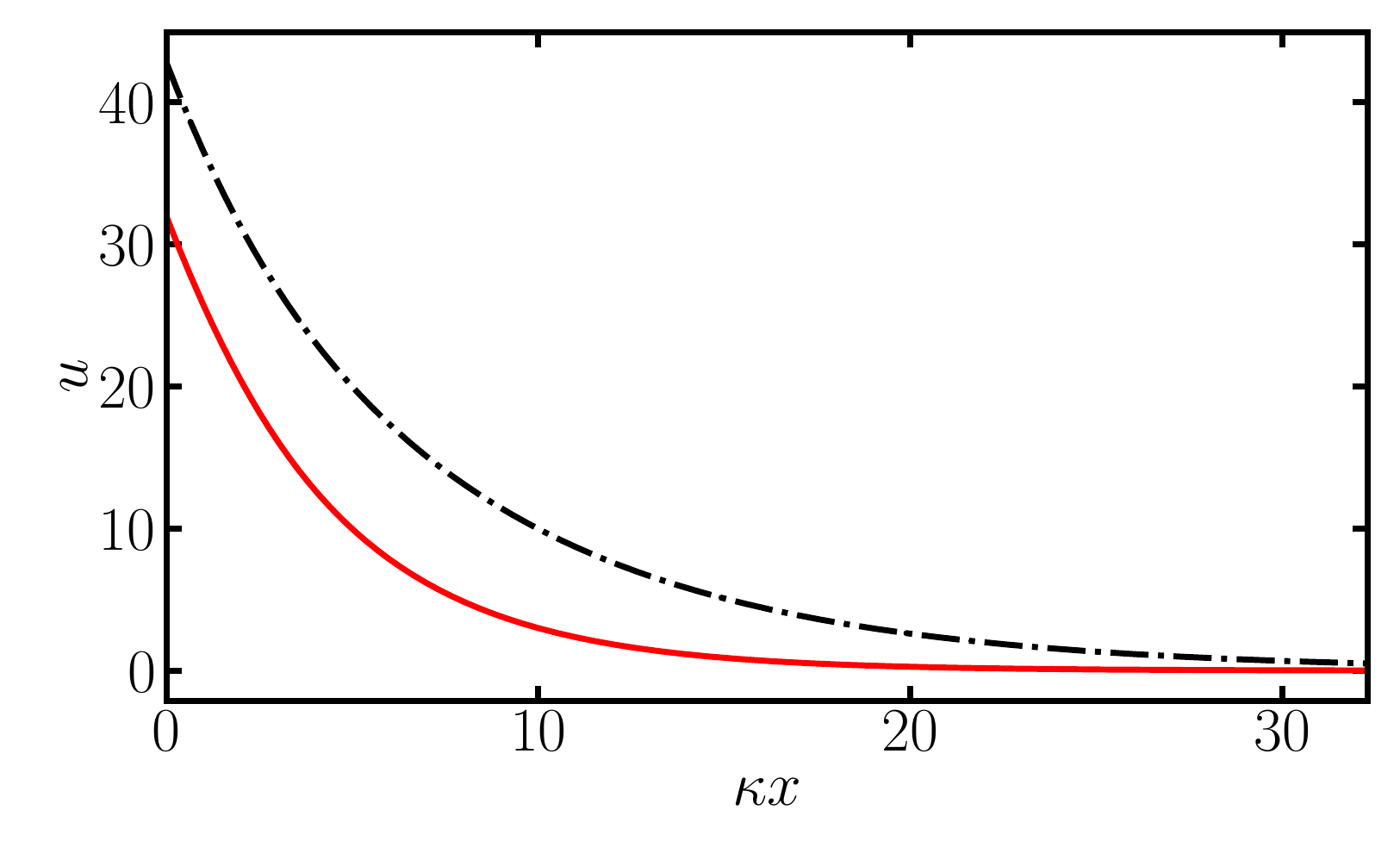}
    \centering
    \includegraphics[width=0.45\textwidth]{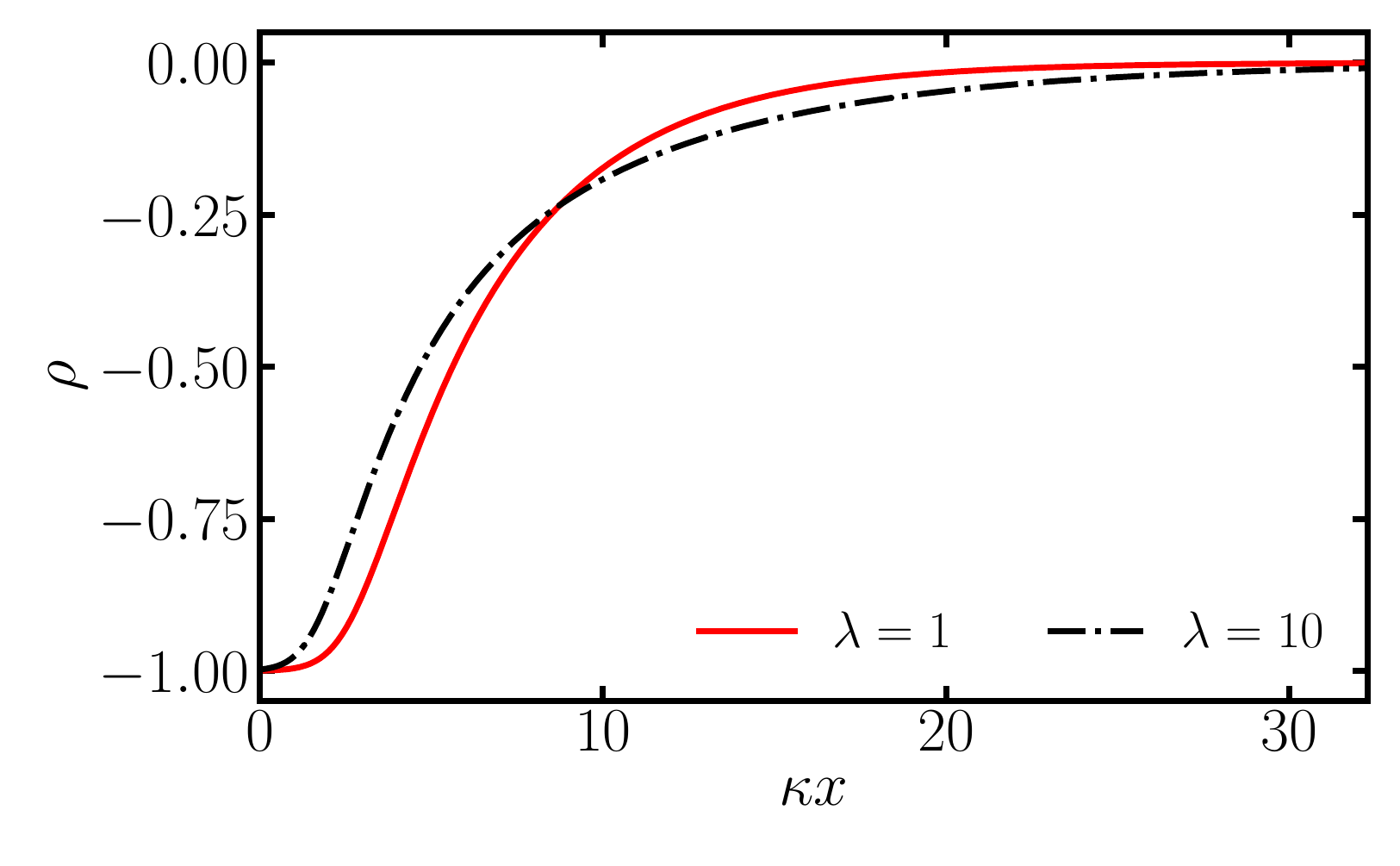}
    \caption{\textbf{Monotonic and decay of electrostatic potential and charge density}. Electrostatic potential in units of thermal volts (left panel) and charge density in units of charge per lattice site (right panel) in the EDL as a function from the interface for a surface charge density of 0.02 Cm$^{-2}$. All calculations are for $f=3$ and $\alpha = 0.2$ at the two association constants shown. For $T=300$~K, $v = 1$~nm$^3$, $\epsilon = 5$, the inverse screening length is $\kappa \approx 0.85~\textrm{\AA}$, which means the EDL structure is of a similar length scale to the size of an ion.}
    \label{fig:EDL_PHI_CD}
\end{figure}

To start, the description of the non-linear solution to Eq.~\eqref{eq:Poisson_NL} in terms of the electrostatic potential and charge density shall be outlined. The results for these are shown in Fig.~\ref{fig:EDL_PHI_CD} as a function of distance from an interface which has a charge of 0.02~Cm$^{-2}$ for the studied association constants. The electrostatic potential (in units of thermal volts) and charge density (in units of the charge per unit lattice site) are both found to monotonically decay from the interface, as expected \textcolor{black}{from a local density approximation.}

For $\lambda = 1$, the potential and charge density decay to the bulk values within $\sim30/\kappa$. The crowding regime, \textcolor{black}{defined by $c_{01/10} \approx 1$ and $c_{10/01} \approx 0$, which has to be taken here because an incompressible IL can only accumulate charge density not number density in the EDL,} can clearly be seen, since the dimensionless charge density reaches $-$1 at the interface. For $\lambda=10$, the electrostatic potential and charge density decay further from the interface, owing to the larger screening lengths from more associations. The crowding regime is also reached near the interface, as can be seen from the charge density reaching $-$1. These results are expected, and we can now turn to decomposing the charge density into the different contributions to understand the response of the IL further.

\begin{figure}
    \centering
    \includegraphics[width=0.45\textwidth]{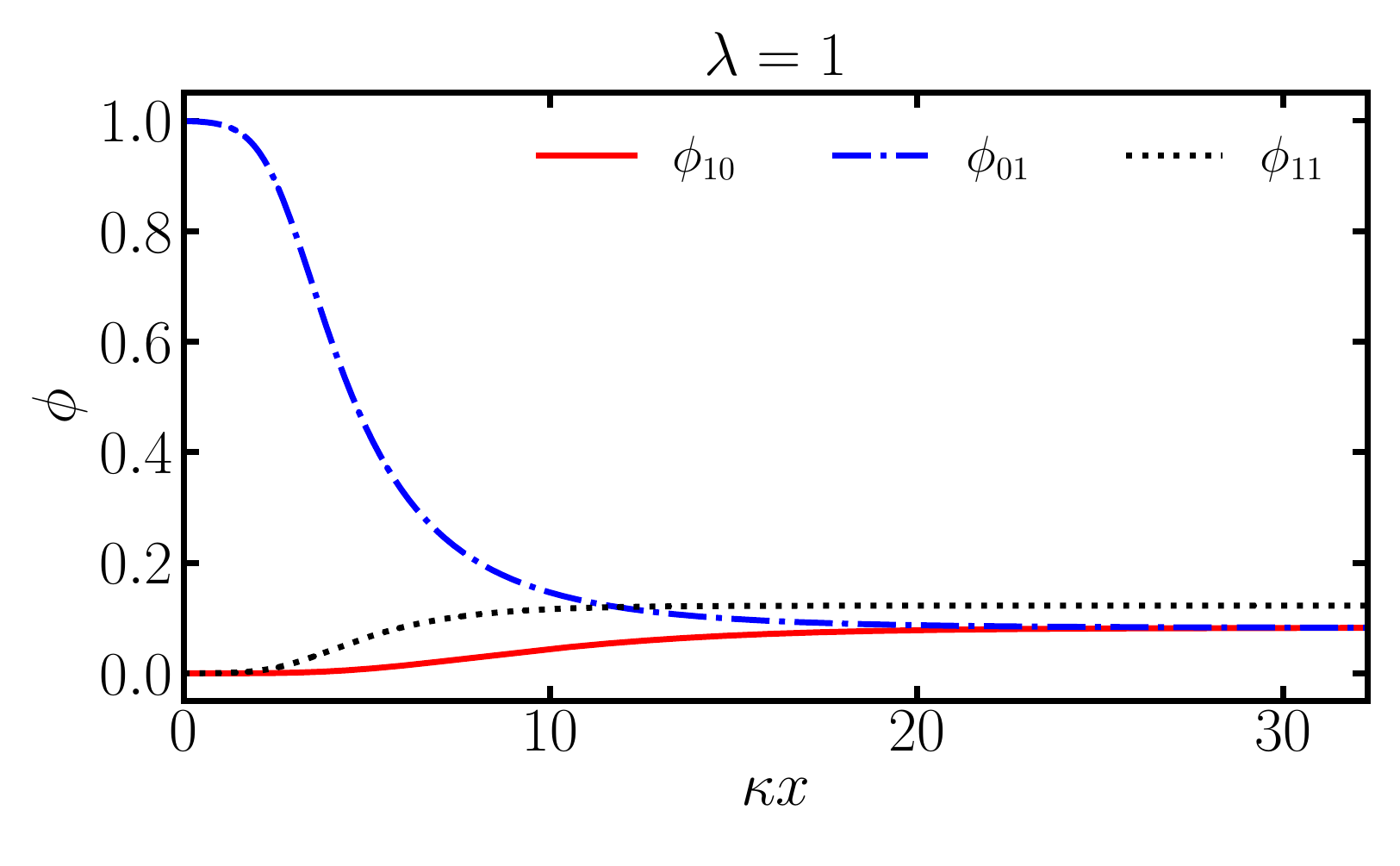}
    \centering
    \includegraphics[width=0.45\textwidth]{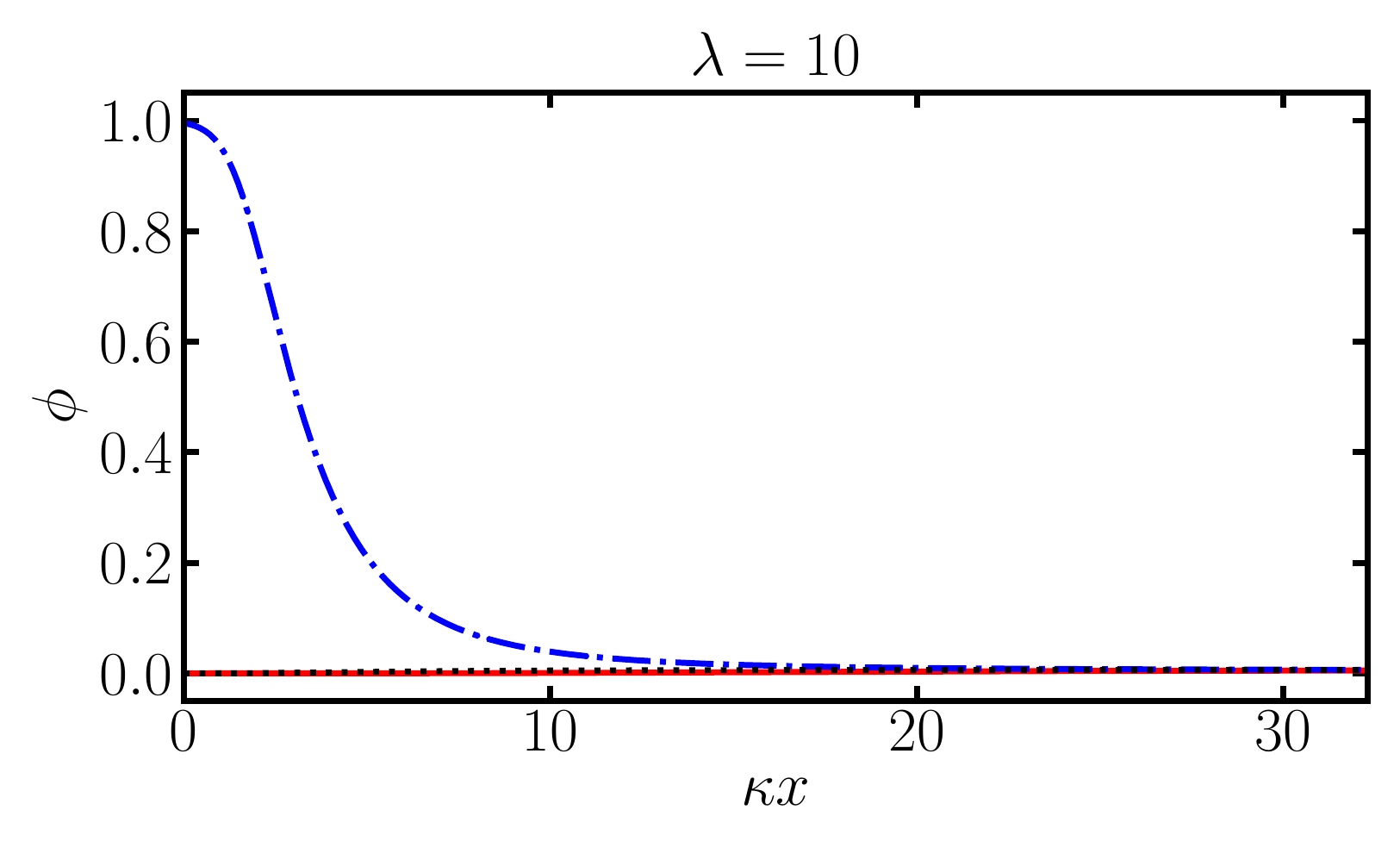}
    \centering
    \includegraphics[width=0.45\textwidth]{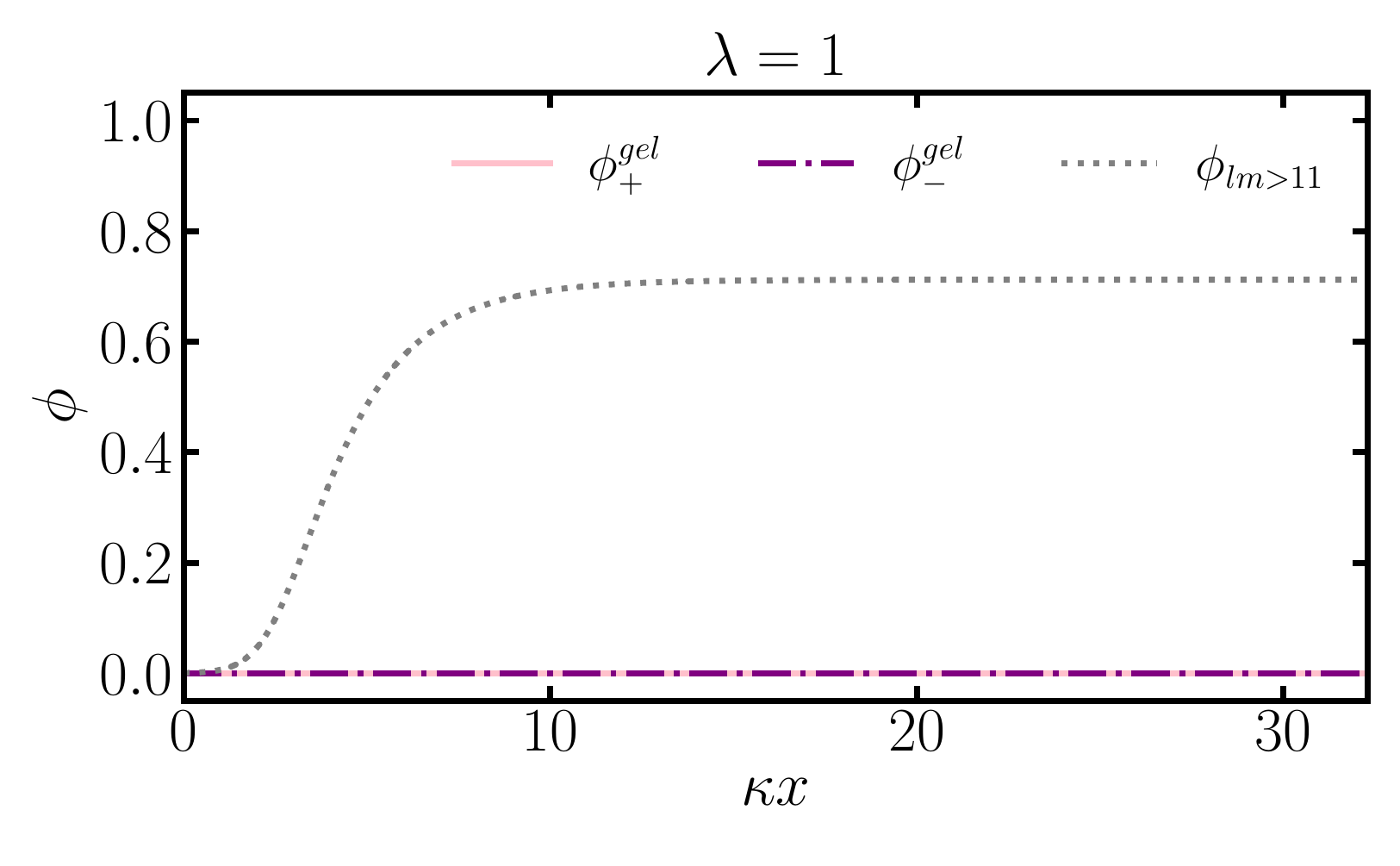}
    \centering
    \includegraphics[width=0.45\textwidth]{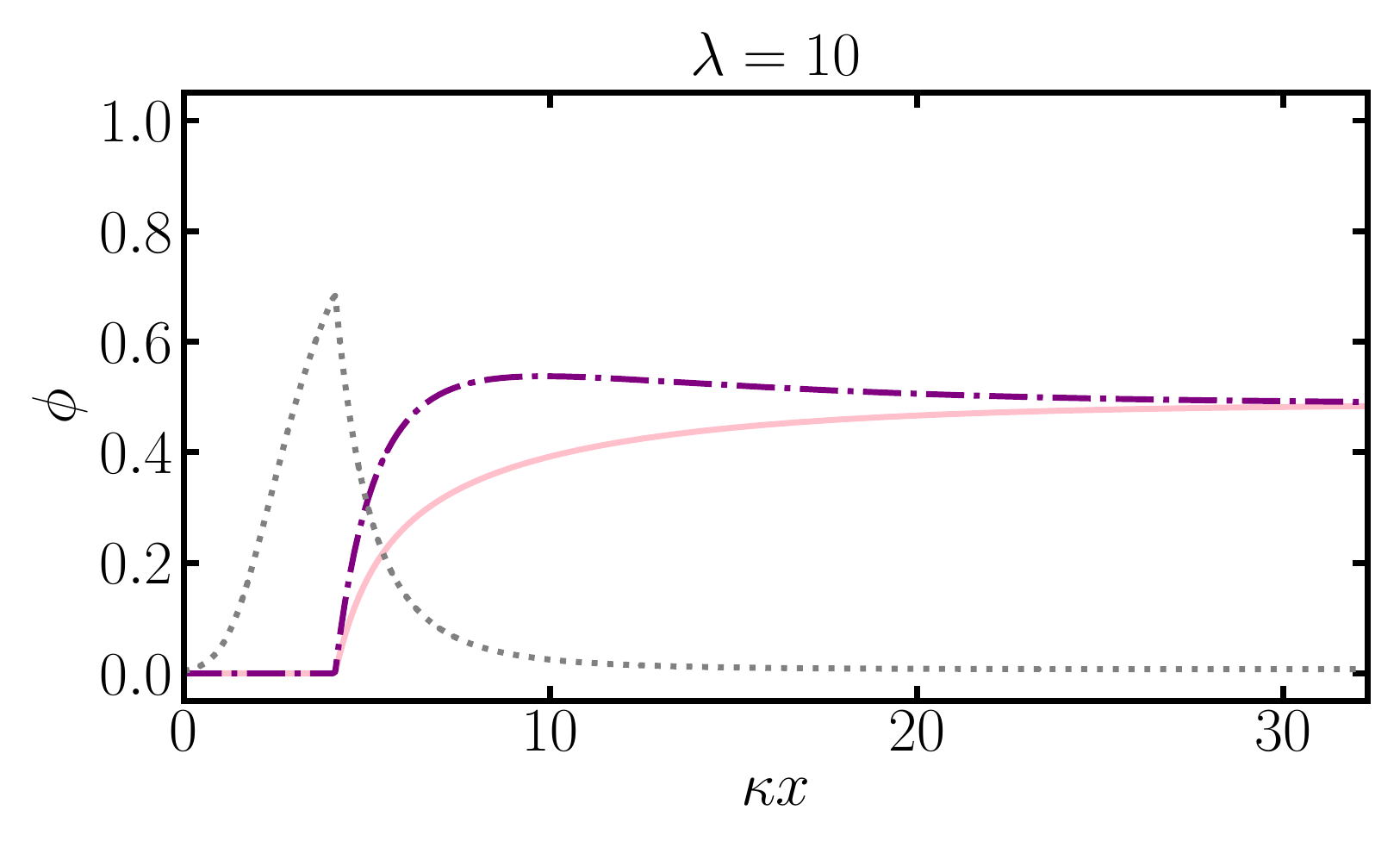}
    %
    %
    \caption{\textbf{For strongly associating ILs, the gel can screen the electrode charge, but only far from the interface, as crowding of free ions near the interface dominates}. Volume fractions in the EDL as a function from the interface for a surface charge density of 0.02 Cm$^{-2}$. Top panels show the volume fraction of free cations ($\phi_{10}$), free anions ($\phi_{01}$) and ion pairs ($\phi_{11}$). The bottom panels show the volume fractions of finite aggregates beyond ion pairs, and the volume fraction of cations in the gel and volume fraction of anions in the gel. These calculations were performed with the same parameters as Fig.~\ref{fig:EDL_PHI_CD}.}
        \label{fig:proper_EDL}
\end{figure}

In Fig.~\ref{fig:proper_EDL}, \textcolor{black}{the charge density of Fig.~\ref{fig:EDL_PHI_CD} is decomposed into the free ions, ion pairs, clusters beyond ion pairs, and cations and anions in the gel.} For $\lambda = 1$, free anions substantially increase close to the interface, where the crowding regime is reached. The free cations are expelled from the EDL rapidly. Both ion pairs and finite aggregates beyond ion pairs are depleted at length scales between the crowding regime of free anions and the expulsion of free cations. This is because it is more favourable to have the free ions in the EDL than clusters.

For $\lambda = 10$, the structure of the EDL is more complicated. At short distances from the positively polarised electrode, the free anions again reach the crowding regime. The free cations and ion pairs are practically zero in the bulk, and do not appear to change much in the EDL. For increasing distances from the interface, $\phi_{lm > 11}$ initially increases, as the number of cations/anions starts to equalise, which permits the large aggregates to be created. This occurs until the gel point is reached, where the large clusters combine. For further distances, there is a significant volume fraction of gel, but the number of cations and anions in the gel are not equal. There are more anions in the gel than cations, which means the gel screens the field. In fact, in this regime, the field is practically only screened by the gel, since the volume fraction of free anions is much smaller. Again, the gel can screen the electrostatic potential because of the shift in the equilibrium between how the free ions associate to the gel when there are unequal numbers of cations/anions.

\subsection{Differential Capacitance}

The differential capacitance, $C$, can be calculated numerically within this theory through
\begin{equation}
    \dfrac{C}{C_0} = \dfrac{d \tilde{\sigma}}{d u_0},
\end{equation}

\noindent where $\tilde{\sigma}$ is the dimensionless surface charge density, $C_0 = \epsilon_0\epsilon\kappa$ is the Debye capacitance without associations, and $u_0$ is the dimensionless potential drop across the entire EDL. When $\lambda \ll 1$, i.e. all free ions, and $\alpha=1$ the differential capacitance numerically calculated within the presented theory exactly reproduces that of Refs.~\citenum{Kornyshev2007,kilic2007a}.

\begin{figure}
    \centering
    \includegraphics[width=0.45\textwidth]{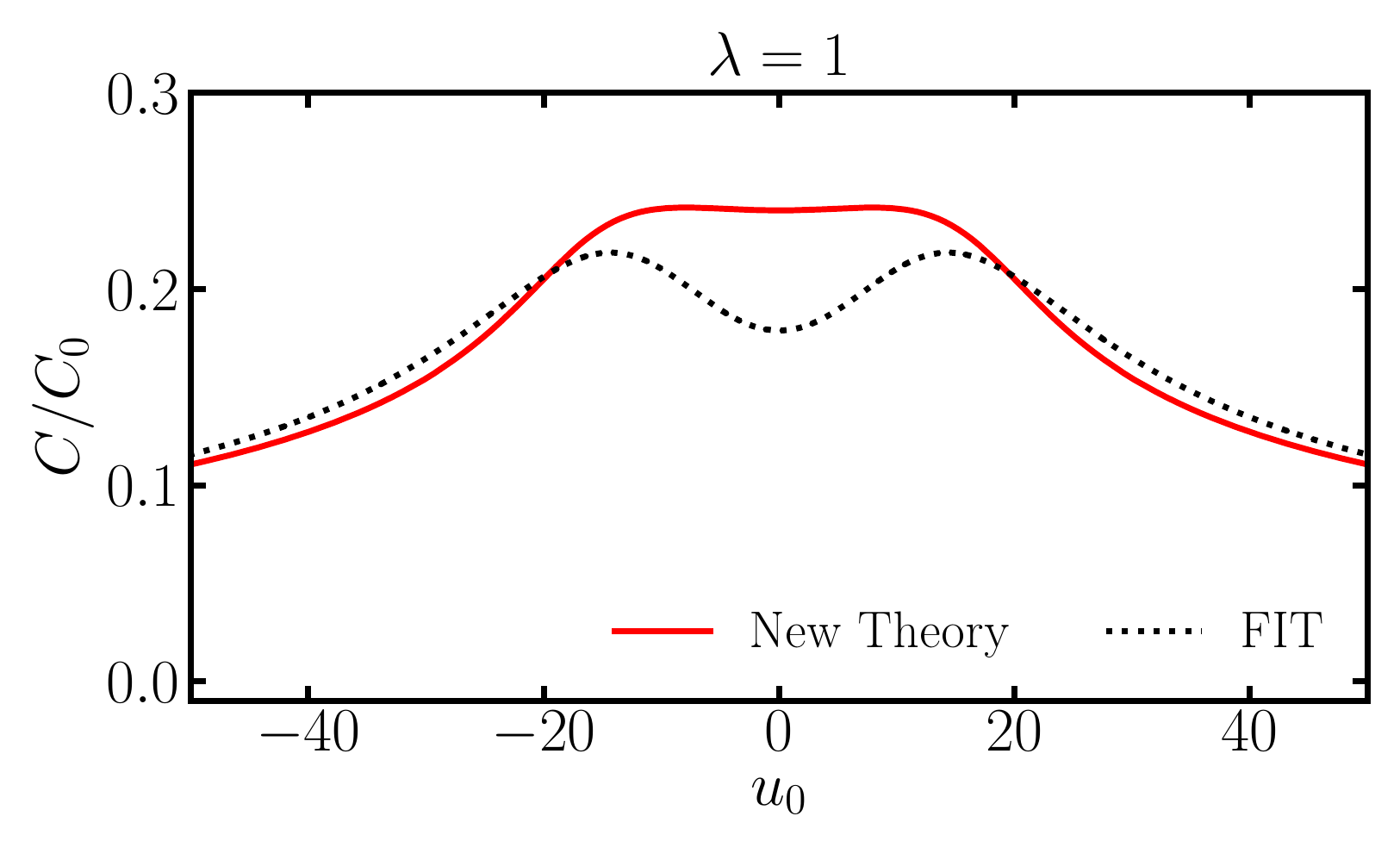}
    \centering
    \includegraphics[width=0.45\textwidth]{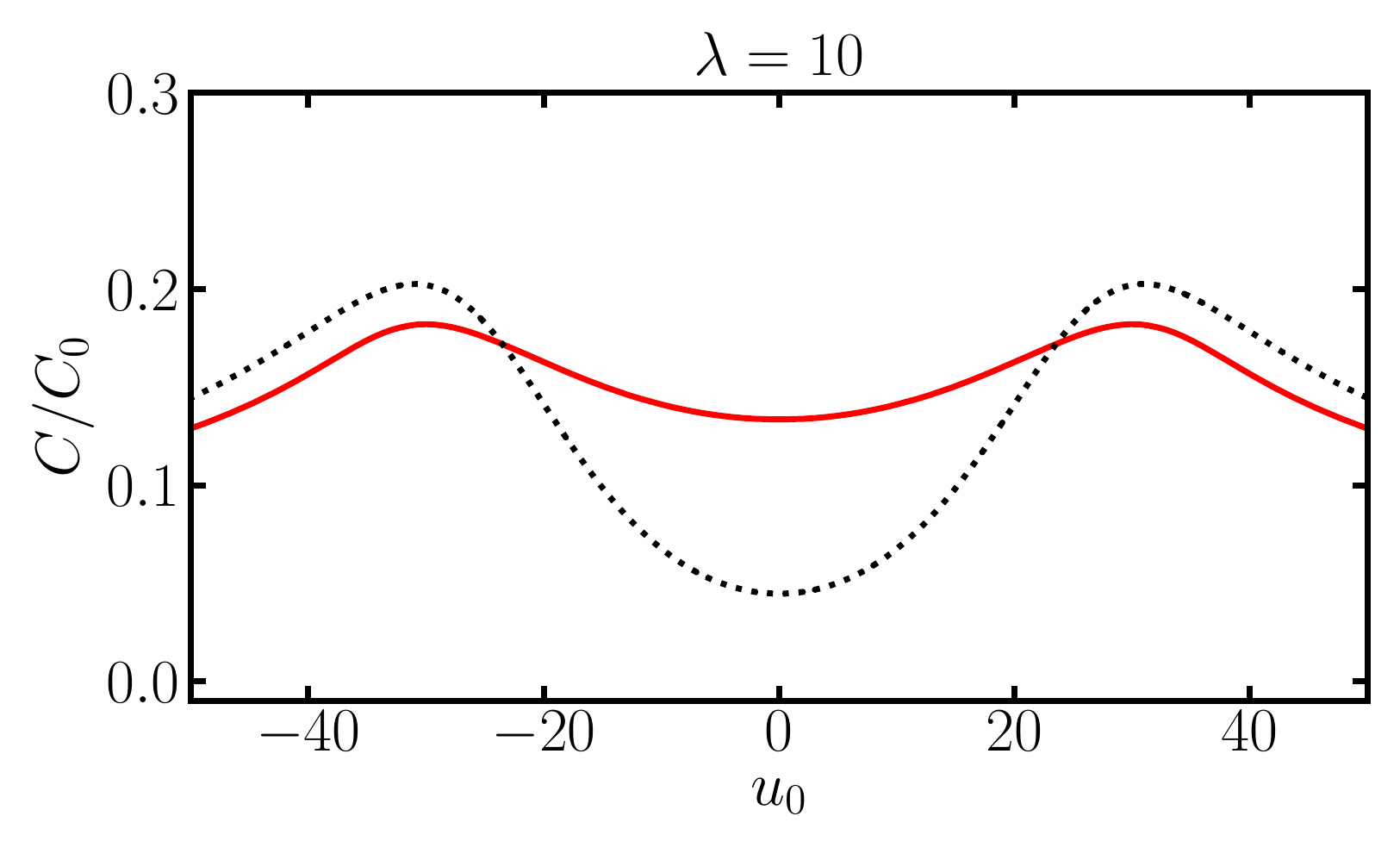}
    \caption{\textbf{Account of clusters beyond free ions in the EDL causes the capacitance at zero charge to increase, and the curves become more bell-like}. Differential capacitance in units of Debye capacitance without associations, $C_0 = \epsilon_0\epsilon\kappa \approx 75-100~\mu$Fcm$^{-2}$, as a function of voltage in units of thermal volts (25.6~mV at room temperature). The solid (red) line corresponds to the differential capacitance computed numerically for the theory presented here. The dotted (black) line corresponds to the expression in the free ion theory (FIT) of Ref.~\citenum{Chen2017} with free ion fractions, $\gamma$, of 0.16 and 0.01 for $\lambda = 1$ and $\lambda = 10$, respectively. All curves are plotted for $\alpha = 0.2$.}
    \label{fig:DC}
\end{figure}

In Fig.~\ref{fig:DC} the numerical differential capacitance as a function of potential is plotted (solid line) alongside the analytical expression derived in Ref.~\citenum{goodwin2017mean} 
\begin{equation}
    \dfrac{C}{C_0} = \sqrt{\alpha\gamma}\dfrac{\cosh \left(\alpha u_0/2\right)}{1 + 2\gamma\sinh^2 \left(\alpha u_0/2\right)}\sqrt{\dfrac{2\gamma\sinh^2 \left(\alpha u_0/2\right)}{\ln\left\{1 + 2\gamma\sinh^2 \left(\alpha u_0/2\right)\right\}}},
    \label{eq:cap_free}
\end{equation}

\noindent where $\gamma$ is the free ion fraction based on a similar free ion fraction (dotted line) to the cluster distribution; where the additional factor of $\sqrt{\alpha\gamma}$ comes from the definition of $\kappa$ taken here. Note that Ref.~\citenum{Chen2017} is based on just the free ions in an IL, \textcolor{black}{referred to as a free ion theory (FIT)}, where there are only thermoreversible associations in the bulk.

For $\lambda = 1$, the free ion fraction is approximately 0.16, which is used in Eq.~\eqref{eq:cap_free} of the FIT. Since the free ion fraction is significantly smaller than 1/3, a clear ``camel"~\cite{Kornyshev2007} shaped differential capacitance curve is obtained from  Eq.~\eqref{eq:cap_free}, i.e. initially the differential capacitance increases before a maximum is reached, after which the differential capacitance is governed by universal charge conservation laws~\cite{Kornyshev2007}. In contrast, the numerical solution for the theory presented here only has a slight ``camel" shape, with the differential capacitance at zero charge also being larger than the FIT.

For $\lambda = 10$, the free ion fraction is found to be approximately 0.01. Again, the FIT has a significant U-shape near the potential of zero charge, which goes through a maximum before the crowding regime is reached and the differential capacitance decreases again \textcolor{black}{(note that the differential capacitance follows $C/C_0 \propto 1/\sqrt{u_0}$ for $u_0 > 40$, which is where $|\rho| = 1$ reaches in Fig.~\ref{fig:EDL_PHI_CD}, and therefore, the employed definition of the crowding regime is reasonable)}. The new theory presented here, in contrast, only has a slight ``camel"-shape which is stretched out over the potential range, and again has a larger differential capacitance at zero charge.

Overall, there are two features of the theory presented here against that of Ref.~\citenum{Chen2017}: (1) - The screening length obtained here is always smaller than Eq.~\eqref{eq:cap_free} because larger clusters are explicitly accounted for. Therefore, the capacitance at zero charge is always larger. (2) - The break-down of clusters and gel causes the crowding regime to be reached at smaller potentials than that of Eq.~\eqref{eq:cap_free}, which makes the ``camel" shape less pronounced. 

The two association constants investigated here can be considered to be at two different temperatures for the same IL, since they are related to the free energy of an association through $\lambda = \exp\{-\beta\Delta f_{+-}\}$. Therefore, larger temperatures correspond to smaller $\lambda$. As found in Ref.~\citenum{Chen2017}, we also observe a transition from a camel to bell shape with increasing temperature, owing to the dissociation of ions. Moreover, the capacitance at zero charge increases with decreasing associations because there are more charge carriers to screen the potential.

\subsubsection{Comparison to Experiments}

In Ref.~\citenum{mceldrew2020correlated}, it was found that both the cation and anion in 1-Ethyl-3-methylimidazolium bis(trifluoromethylsulfonyl) imide [Emim][TFSI] have a functionality of 4, and the volumes of each ion are practically the same. This is perhaps as close as one can get to the symmetric IL, which the new theory is applicable for. ILs are typically composed of ions with very different volumes and \textcolor{black}{unequal} functionalities~\cite{mceldrew2020correlated}. \textcolor{black}{Further development of the presented formalism must be achieved to make predictions of asymmetric ILs.}

\begin{figure}
    \centering
    \includegraphics[width=0.45\textwidth]{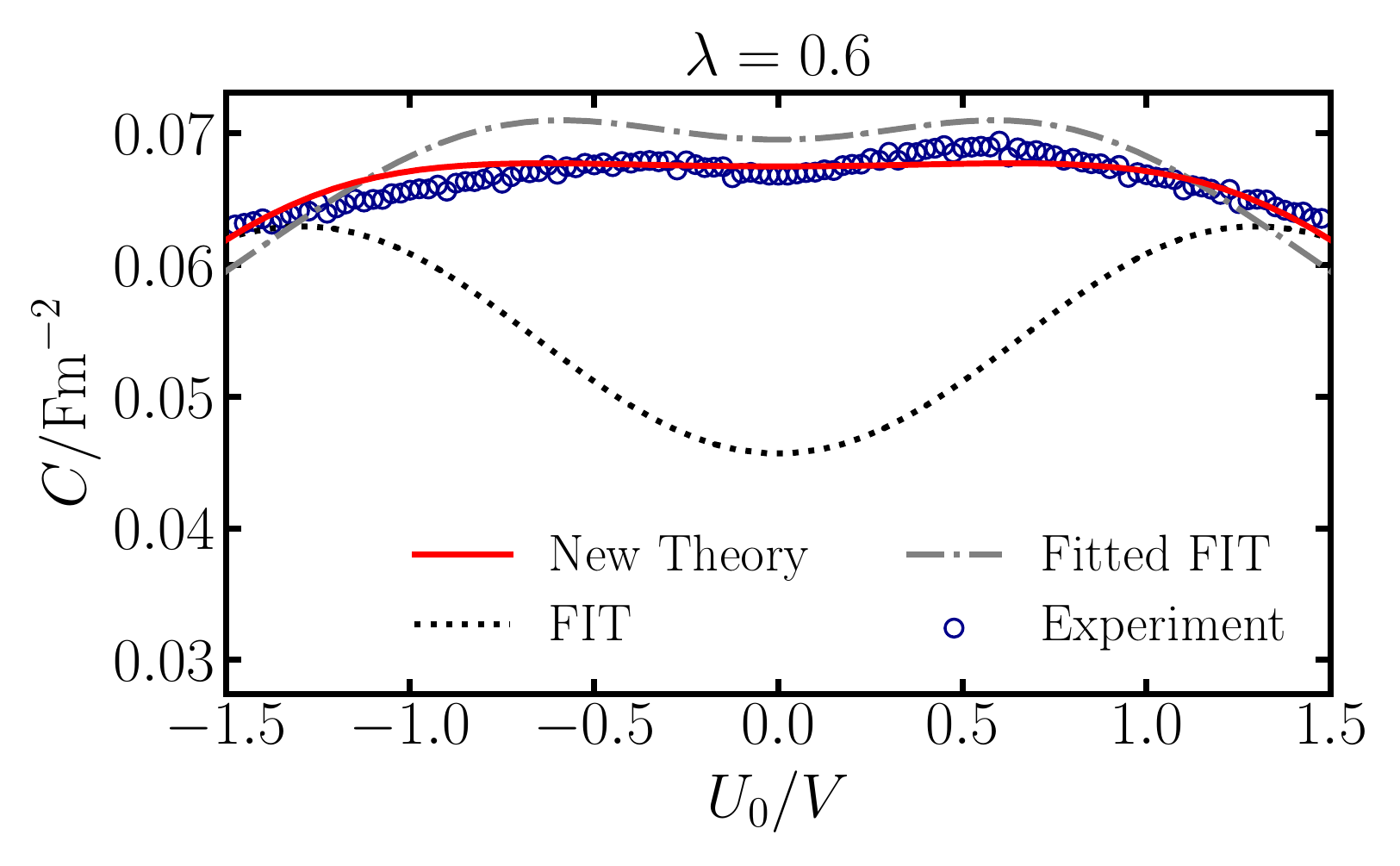}
    \caption{\textbf{Better agreement between experimental differential capacitance and the new theory compared to the free ion theory (FIT), using an independently determined free ion fraction}. Differential capacitance capacitance as a function of potential drop across the EDL. The presented theory is shown by the solid red line and the theory of Eq.~\ref{eq:cap_free} is shown by the dotted black line, using the independently determined free ion fraction. A fitted $\gamma=0.27$ for the FIT is also shown. The IL is taken to have $f=4$ and an association constant of $\lambda = 0.6$ ($\gamma = 0.12$), with T$=298~$K, maximal concentration of 7.76~M, $\alpha = 0.07$~\cite{Monchai2018} and dielectric constant of $\epsilon=1$ (used as a parameter to reduce the capacitance at zero charge). The experimental values for [Emim][TFSI] are shown by open circles, and have been reproduced from Ref.~\citenum{Monchai2018}.}
    \label{fig:DC_EXP}
\end{figure}

In Ref.~\citenum{Monchai2018}, Jitvisate and Seddon reported the experimental differential capacitance curves of various ILs, including [Emim][TFSI]. The data from Ref.~\citenum{Monchai2018} \textcolor{black}{for [Emim][TFSI]} has been reproduced in Fig.~\ref{fig:DC_EXP}. A slight camel shape is found for [Emim][TFSI], with the wings of the capacitance curve at large voltages being approximately symmetric. It was noted in Ref.~\citenum{Monchai2018} that the differential capacitance curve could be approximately fitted with Eq.~\eqref{eq:cap_free} using $\alpha \approx 0.07$ and $\gamma \approx 0.27$. If $\gamma$ is interpreted as the free ion fraction, there would have been slightly less than $1/3$ of free ions. This was also the conclusion of Ref.~\citenum{Chen2017}, based on fitting the differential capacitance curve obtained from molecular dynamics simulations. However, in Ref.~\citenum{Feng2014} the fraction of free ions of [Emim][TFSI] was found to be only $\sim$0.15 at room temperature, which is approximately half of that based on Ref.~\citenum{Chen2017}. 

\textcolor{black}{In Fig.~\ref{fig:DC_EXP} the differential capacitance curve from the new theory and the free ion theory (FIT), i.e., Eq.~\eqref{eq:cap_free}, is shown. The new theory can reproduce the differential capacitance curve reasonably well with $\lambda = 0.6$, as seen in Fig.~\ref{fig:DC_EXP}. For $\lambda=0.6$, the free ion fraction is 0.12, which is close to that of Ref.~\citenum{Feng2014}. For the FIT with a free ion fraction of $\gamma=0.12$, the differential capacitance curve cannot reproduce experimental data well, being too ``camel'' shaped. The FIT with the fitted value of $\gamma = 0.27$ from Jitvisate and Seddon~\cite{Monchai2018} is also shown for comparison. Therefore, the new theory appears to be able to rationalise the results which were not in quantitative agreement with the theory before.}

\textcolor{black}{There are still a couple of free parameters, other than the free ion fraction which was independently taken from Ref.~\citenum{Feng2014}, evaluated to obtain the differential capacitance curve in Fig.~\ref{fig:DC_EXP}. One of these parameters is $\alpha = 0.07$, which was the value fitted by Jitvisate and Seddon~\cite{Monchai2018}. Using this parameter, and setting the dielectric constant to the bulk value (for [Emim][TFSI] this is $\epsilon=12$~\cite{Monchai2018}) obtains a capacitance at zero charge which is too large. Therefore, to obtain a similar capacitance at zero charge, the dielectric constant is set to $\epsilon=1$. As such value is unphysical (at least the polarisability of electronic degrees of freedom ions would contribute the value of 2), such fitting of $\epsilon$ demonstrates a short-coming of the simple, local theory presented here, and motivates further improvement. Note, to permit a transparent comparison with Jitvisate and Seddon~\cite{Monchai2018}, a Stern layer has not been included, which would have reduced the capacitance at zero charge.}

\section{Discussion}

\subsection{Underscreening}

\textcolor{black}{The presented theory has} implications for the interpretation of the experiments in Refs.~\citenum{Smith2016,Gebbie2013,Gebbie2015,Gebbie2017rev,smith2017struct,Han2020IL,Jurado2016,Jurado2015}. Inverting the expression for the screening length in terms of the association probability, we obtain
\begin{equation}
    p = \dfrac{(\ell\kappa)^2 - 1}{f-1+(\ell\kappa)^2}.
    \label{eq:debye_prob}
\end{equation}

\noindent \textcolor{black}{In experiments~\cite{Smith2016}, the screening length multiplied by the inverse Debye length is $\ell\kappa \approx 100$ for ILs}. Using this value yields an association probability of 0.9997 for a functionality of 3. This association probability produces $\sim 3\times10^{-11}$ free ions, \textit{with the rest being gel}, which corresponds to a $\lambda$ in excess of $1\times10^6$ and $\Delta f_{+-} \approx -14/\beta$. At large distances from the charged interface, we have found that it is \textit{actually the gel which screens the long-tail in electrode charge, not the free ions}. This could suggest that ILs are not dilute electrolytes, but a gel which can screen electrode charge. In fact, Jurado \textit{et al.}~\cite{Jurado2015,Jurado2016} found 1-Hexyl-3-methyl-imidazolium ethylsulfate forms lamellar-like structures on charged mica surfaces, which are conceptually similar to the gel, albeit with more order. 

Despite this, the capacitance at zero charge is (in this theory) still fixed by the screening length $\ell$, which means the underscreening paradox remains~\cite{UND}. Further development of the presented theory, such as through taking into account the spatial structure of clusters/gel or accounting for electrostatics beyond that investigated here~\cite{pedroRTILs}, could gain further insight into these puzzling experiments.

\subsection{Overscreening}

\textcolor{black}{Ma \textit{et al.}~\cite{Ma2015} performed sophisticated classical density functional calculations of a coarse-grained IL with varying degrees of ion pairing, up to that suggested by Gebbie \textit{et al.}~\cite{Gebbie2013}. It was noted that the charged density was somewhat insensitive to the extent of ion pairing, which suggested a link between ion pairing and overscreening. This was further shown by Avni \textit{et al.}~\cite{avni2020charge}, where a link between clusters (ion pairs, triplets and quadruplets) and the linearised overscreening theory of Bazant-Storey-Kornyshev (BSK)~\cite{Bazant2011} occurs in the long-wavelength limit. The BSK theory is known to underestimate overscreening in ILs, however, and the interpretation of it can be subtle~\cite{de2020continuum,Chao2020,Leedynamics2015}, with some suggesting that short-range cation-anion repulsion occurs. The alternating ``co-polymer'' structure of the gel/large clusters is conceptually similar to extended overscreening structures (bulk or in the EDL). If one accounts for the spatial structure of larger clusters (than ion pairs, triplets and quadruplets~\cite{avni2020charge}) or the gel, extended overscreening could potentially be obtained.}

Here, for large association constants, we found that the EDL structure has two regimes: one where electrode charge is screened by free ions, and another where it is screened by the gel. The free ions accumulate near the interface, with the gel persisting towards the bulk. Conceptually, this could be considered similar to the overscreening-crowding transition~\cite{Bazant2011}. 

\textcolor{black}{The reader should also bear in mind that the EDL theory presented here is a simple, local density approximation, which cannot explicitly account for overscreening or the internal charge distribution of clusters. However, while a spatial map between ions in clusters, assumed to be Cayley trees here, and charge density does not yet exist, Cayley trees have well defined shells around an ion in a cluster. One can loosely interpret the number of shells around an ion with the distance from that ion, which can be utilised to further establish the link between associations and overscreening.}

Let us consider a cation at a `central' lattice site and examine the probability of an anion being at any of the lattice sites $n$ shells away from the central cation, \textcolor{black}{similar to the pair correlation function between cations and anions}. The anion can be present on that lattice site either by being linked to the central cation as part of the same cluster, $p$, or by random, $p_-^r$. There is also a possibility that a cation could be present on that site by random, $p_+^r$. Assuming that unassociated species are uncorrelated, $p_-^r = p_+^r = p^r$, and taking into account the incompressibility condition, we obtain $p^r = (1 - p)/2$. Thus, the total probability of observing an anion one shell away is $(1+p)/2$. For the next shell, the probability that an anion is there via a link to the central cation is exactly zero, as we require alternating associations. However, there is some probability that an anion (or cation) is there by chance. In this case, the random probability of an anion is $(1 - p^2)/2$, which is in fact the total probability of seeing an anion two shells away. Similarly, the analysis for three shells away yields an anion probability of $(1+p^3)/2$. Thus, we can generalise the probability of seeing an anion $n$ shells away from a central cation to
\begin{equation}
    \mathcal{P}_{+-} = \dfrac{1 + p^n(-1)^{n+1}}{2}.
\end{equation}

\begin{figure}
    \centering
    \includegraphics[width=0.45\textwidth]{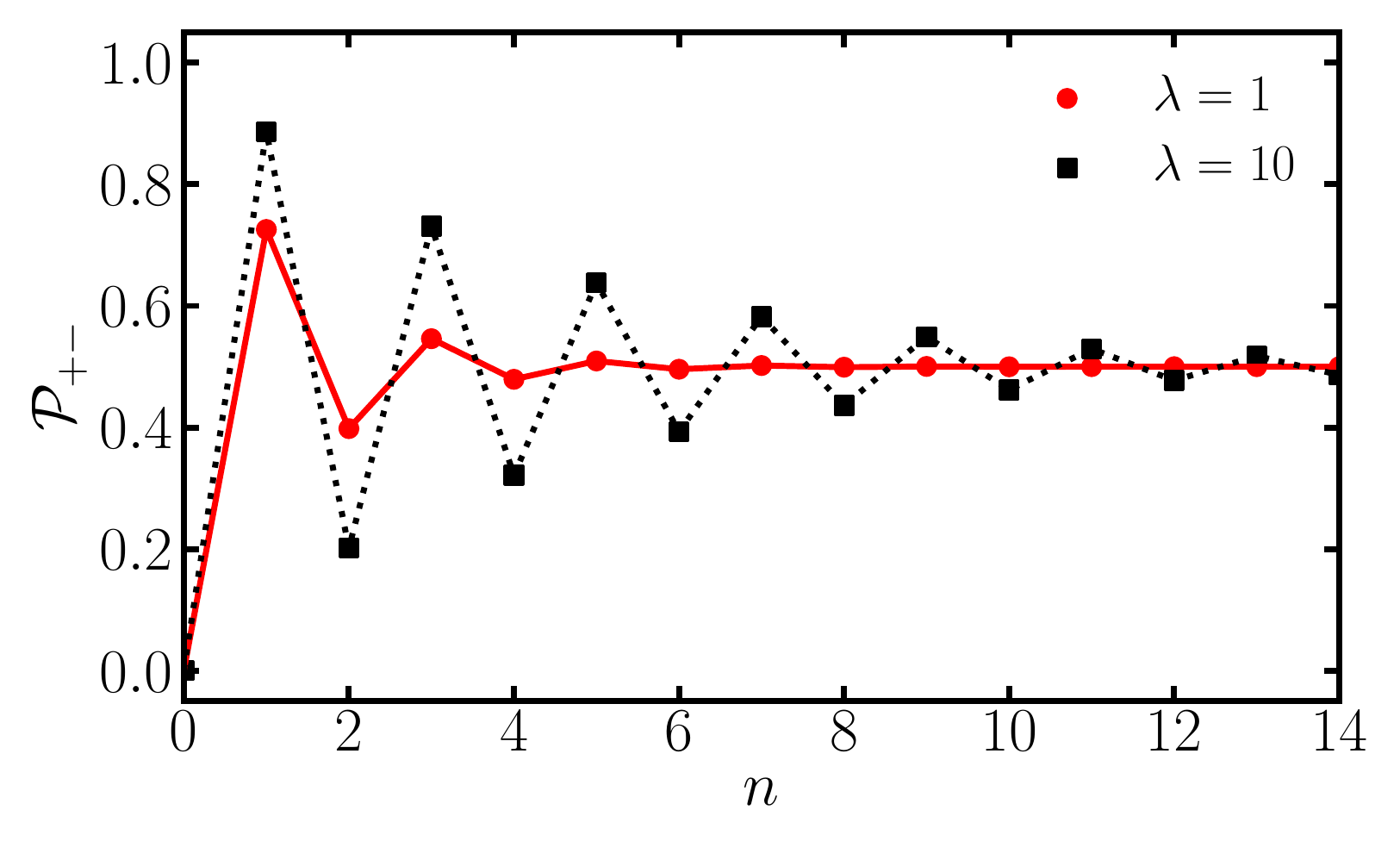}
    \caption{\textbf{Pronounced overscreening could occur as a result of the associations between ions}. The probability of an anion $n$ shells away from a central cation. The functionality was taken to be 3.}
    \label{fig:over}
\end{figure}

In Fig.~\ref{fig:over}, $\mathcal{P}_{+-}$ is plotted as a function of $n$, where we see that the correlation probabilities display pronounced overscreening when $\lambda=10$, but only modest overscreening for $\lambda=1$. Therefore, pronounced overscreening could be more consistent with the gel than clusters. While the developed theory does not spatially resolve the clusters/gel, further development for the spatial distribution of ions in clusters/gel could shed light on this connection.

In this spirit, one issue of the theory (which presumably needs to be dealt with before this can be achieved) is how can extremely large clusters screen the field if it decays on a significantly shorter length scale than the clusters? This is a well known issue for theories of ILs with simple, local density approximations when no associations are included. Taking into account associations increases the screening length relative to the Debye length, but the clusters can become extremely large in size. Therefore, the issue can remain \textcolor{black}{for finite clusters, and perhaps a finite cut-off needs to be introduced for clusters which can contribute to the screening of electrode charge, as is the case of the conductivity of ions in ILs~\cite{mceldrew2020correlated,france2019}}. For the gel, it is conceptually reasonable to think of its composition as locally changing, just as an iceberg melts from its surface. For clusters, their individual structures cannot locally vary as they are discrete. \textcolor{black}{In any case, the theory should work for a large number of free ions and a small number of free ions, when there is significant gel, and should act as an interpolation scheme between these two regimes, with the largest error occurring only right at the gel-point where there are significant numbers of very large clusters.} 

\textcolor{black}{In addition to the above issue, is the applicability of a local density approximation when there are substantial variations of the charge density, on a smaller length scale than the screening length, i.e. the fact there is internal structure to the clusters. The presented theory cannot capture these short-scale variations, with it only being able to deal with mean-field volume fractions of clusters of different ranks and the gel. By developing theories which more accurately account for spatial correlations from electrostatics and non-electrostatic interactions, and which also account for the internal structures of clusters, the overscreening structures and screening lengths observed in surface-force experiments could, perhaps, be more accurately described.}

\subsection{Assumptions of the Theory}

The developed theory has some shortcomings, as just described, but how well do we expect it to handle the averaged volume fractions of each cluster? There could potentially be a number of issues with the developed theory at a charged interface which could cause it to break-down. For example: 
\begin{enumerate}
    \item Do Cayley tree clusters exist near an interface?
    \item Do specific interactions with the interface dominate the compact layer?
    \item Do ion pairs and aggregates rotate in the electrostatic fields?
    \item Does the association free energy depend on the electrostatic field? 
\end{enumerate}

\noindent For each of these questions, we shall state what is expected to happen, and how to confirm/falsify these expectations. Point 1 is the major question to be answered, with 2 and 3 being minor if 1 is found to hold reasonably. In the description of 1, some details of how to test the cluster distribution shall also be given. Overall, it is expected that the cluster distribution in the EDL does not exactly hold, but we believe it is a sufficiently good approximation to understand the qualitative behaviour.

\subsubsection{Cayley trees and cluster distribution in the EDL}

We have shown that clusters are destroyed in favour of free ions at large electrostatic potentials, and therefore, the assumption of Cayley tree clusters is not important at large fields. What needs to be tested, however, is the low-voltage regime of overscreening. Close to the interface, the presence of the interface could block association sites of the ions and drive the ions to form intra-cluster loops. The assumption of Cayley trees applying in the EDL and the assumed cluster distribution can only be confirmed from molecular simulations. 

\textcolor{black}{In bulk ILs, McEldrew \textit{et al.}~\cite{mceldrew2020correlated} found that the spatial distribution function (SDF) of cations (anions) around anions (cations) have well defined ``hot spots'' (for where the concentration is over twice the bulk value), where the ions of opposite sign prefer to reside. When the cations and anions reside in each others hot spots, an association was defined~\cite{mceldrew2020correlated}. This highly directional interaction between cations and anions is a consequence of the complicated shapes of the ions in ILs, and is fundamental to the formation of Cayley tree clusters~\cite{choi2018graph}, which ILs have been found to form in the bulk~\cite{mceldrew2020correlated}. Spherical cut-offs to define associations or kinetic criteria~\cite{feng2019free} are not sufficient to test this approximation. Moreover, Lennard-Jones hard sphere approximation to coarse-graining cannot be used, as the full atomistic details are required to capture these complex interactions. To see if Cayley tree clusters hold in the EDL, the association criteria of McEldrew \textit{et al.}~\cite{mceldrew2020correlated} must be utilised to calculate the clusters in the EDL. This cluster distribution can be compared against the presented theory to test how well the assumed cluster distribution holds. Note that in the simulations, to obtain the volume fraction of each cluster, the clusters will have to be assigned to bins. The size of these bins should be larger than the average size of clusters, which might become problematic right near the gel point when the cluster sizes become very large.}

\textcolor{black}{The gel will also be evident from using the association criteria of McEldrew \textit{et al.}~\cite{mceldrew2020correlated} as a percolating ionic network throughout the simulation. The theory predicts that the gel continuously exists, but changes composition as a function from the interface. Inspection of the percolating ionic network should be able to reveal if there is an equivalence between overscreening and gel screening. Moreover, testing the connection between the associations and overscreening in the bulk would also be interesting.}

\subsubsection{Specific interactions with the interface}

\textcolor{black}{The ions can interact with the interface `specifically' through non-electrostatic interactions, which often causes one type of ion to accumulate at the interface without an applied voltage, owing to the specific interaction of one type of ion being more favourable than the other. These specific interactions with the surface are not accounted for in the presented theory, and will further contribute to breaking the assumed cluster distribution. Therefore, the presented theory might only work well for the diffuse part of the double layer, not the compact layer of ions which are in contact with the interface.}

\textcolor{black}{In a similar spirit to the presented theory, several theories to describe the voltage dependence of the compact layer with clusters were developed by Damaskin-Frumkin-Parsons~\cite{Damaskin1974,Parson1975}. In those theories, the equilibrium between the free states and clustered states were self-consistently determined in an electrostatic potential, which gave rise to a voltage-dependence of the differential capacitance of the compact layer. To account for specific interactions with the interface, one approach could be to have an additional equilibrium between the compact layer, described by the theories of Damaskin-Frumkin-Parsons~\cite{Damaskin1974,Parson1975}, and the diffuse part of the EDL.}

\textcolor{black}{Again, molecular dynamics simulations can be utilised to test if specific interactions cause a break-down of the cluster distribution, using the cluster criteria of McEldrew \textit{et al.}~\cite{mceldrew2020correlated}.}

\subsubsection{Orientation of clusters}

As overscreening is a spatially ordered structure, it is expected that clusters do orient in an electrostatic field. However, we do not expect them to behave as fluctuating Langevin dipoles. It has been shown in Refs.~\citenum{mceldrew2020correlated,Zhang2015} that the lifetime of associations is of the order of $\sim$1-10~ps, which is presumably too short for the rotation of an ion pair or cluster. Therefore, the orientation of clusters must be a weaker contribution than fluctuating Langevin dipoles. In Ref.~\citenum{yufan2020}, ion pairs (which were not permitted to dissociate in the EDL) were treated as Langevin dipoles, where an additional bump in concentrations of ion pairs \textcolor{black}{and a higher propensity for a ``bell''-shaped differential capacitance curve because of dielectric saturation was found}. \textcolor{black}{While ion pairs might not behave as fluctuating Langevin dipoles, the large clusters and gel could have vibrational modes which contribute to the dielectric screening of the electrolyte. This could be introduced in a similar way to Ref.~\citenum{yufan2020}, as some effective dielectric constant which is saturated to lower values in the crowding regime.} Thus, it is expected that this issue is not substantial for the averaged volume fractions, but could cause a breakdown of the assumed cluster distribution, as outlined in the SM. \textcolor{black}{The issue of orientation of clusters can again be confirmed by molecular simulations, provided the cluster distribution can be calculated as described in the previous point.}

\subsubsection{Field dependent binding}

Related to the previous points, if an ion pair is oriented along the electrostatic field decay, the ion pair will presumably be stretched. This could be captured through an electrostatic field dependent binding energy, and therefore, association constant. We would expect the binding energy to decrease with increasing electrostatic fields. This will cause the associations to break-down more than shown here. The presented theory could be considered to overestimate the extent of clusters in the EDL, then. \textcolor{black}{Again, this issue can be confirmed from computing the cluster distribution in the EDL and comparing it against the presented theory.}

\section{Conclusion}

In summary, we have developed a theory for the EDL in ILs which accounts for thermoreversible associations, based on that of McEldrew \textit{et al.} in bulk ILs. The developed theory is constructed from assuming the bulk cluster distribution applies in the EDL, but with modulated volume fractions of cations and anions, which is coupled to the Poisson equation through a Boltzmann closure relation of the free ions. This theory was shown to recover the expected linear response behaviour, and the large potential limits. Therefore, it should serve as a qualitative description for the EDL of an IL with thermoreversible associations. For strongly associating ILs, the free ions crowd near the interface, but far from the interface the gel screens the electrode charge. The differential capacitance was found to have a larger capacitance at zero charge than free ion theories, from the fact that the clusters can also screen, and the transition to ``camel" shape occurs at smaller free ion fractions than models based on free ions. 

\textcolor{black}{ILs provided an extremely useful test ground to develop the theory presented here, since ILs are the simplest possible super-concentrated electrolytes. The aggregation and gelation theory of McEldrew \textit{et al.} has also been developed for water-in-salt electrolytes and salt-in-ionic liquids, so far. Development of the EDL theory for these systems would be relevant for engineering the interfacial behaviour of such electrolytes, which is crucial for their performance in energy storage technologies. Moreover, investigating the consequences of this theory for the electrokinetic behaviour could potentially yield new predictions to confirm the presented theory. }

%
%
%
%
%
%
%
%

\section{Supplementary Material}

In the Supplementary Material, further details of the bulk cluster distribution is given, the bulk-electrical double layer equilibrium is established and the approximations required are outlined, the role of gel terms in the closure relation is outlined, and finally a step-by-step guide of how implement the equations numerically is given.

\section{Acknowledgements}

All authors would like to acknowledge the Imperial College-MIT seed fund. ZAHG was supported through a studentship in the Centre for Doctoral Training on Theory and Simulation of Materials at Imperial College London funded by the EPSRC (EP/L015579/1) and from the Thomas Young Centre under grant number TYC-101. JPD acknowledges support from the National Science Foundation Graduate Research Fellowship under award number \#1122374. MM and MZB acknowledge support from a Amar G. Bose Research Grant. AAK would like to acknowledge the research grant by the Leverhulme Trust (RPG-2016- 223). 

\begin{table}[ht]
\centering
\begin{tabular}{llll}
\hline
\hline
$N_{lm}$         & Number of $lm$ clusters    & $N^{gel}_i$     & Number of species $i$ in gel \\
$f_i$              & Functionality of species $i$ & $v_i$         & Volume of species $i$\\
$\Omega$     & Number of lattice sites      &$\beta$ & Inverse thermal energy\\
 $c_{lm}$       &  Concentration of cluster &  $\phi_{lm}$    & Volume fraction of an $lm$ cluster \\
$\phi_i$         & Total volume fraction of species $i$ &  $c^{gel}_{i}$    & Concentration of species $i$ in gel \\
$\phi^{sol}_i$   & Volume fraction of species $i$ in sol & $\phi^{gel}_i$   & Volume fraction of species $i$ in gel\\
$w^{sol}_i$     & Fraction of species in the sol  & $w^{gel}_i$     & Fraction of species in the gel\\
$\Delta_{lm}$  & Free energy of formation of a rank & $\Delta^{gel}_{i}$ & Free energy change of species $i$ \\
& formation of a rank $lm$ cluster&& formation of an lm cluster\\
$\Delta f_{+-}$ & Free energy of an association & $W_{lm}$      & Combinatorial enumeration \\
$\Delta u_{+-}$ & Association free energy  & $\Delta s_{+-}$      & Configurational entropy of a cluster\\
$\lambda$  & Association constant  & $\zeta$         & Number of anion-cation associations \\ 
$p_{ij}$        & Association probabilities & $p^{sol}_{ij}$  & Association probabilities in the sol\\
\hline
$u$ & Volts in units of thermal volts & $\alpha$ & Short-range correlation parameter \\ 
$\kappa$ & Inverse Debye length & $\ell$ & Screening length\\
$e$ & Elementary charge & $\epsilon\epsilon_0$ & Dielectric Constant \\ $\delta \bar{\phi}$ & Perturbation in volume fractions & $\delta \bar{p}$ & Perturbation in probabilities \\
$\delta \bar{\phi}^{sol}$ & Perturbation in sol volume fractions & $\delta \bar{p}^{sol}$ & Perturbation in sol probabilities \\
$C$ & Differential capacitance & $C_0$ & Capacitance at zero charge \\
$\tilde{\sigma}$ & Dimensionless surface charge density & $u_0$ & Potential drop across EDL\\
\hline
\hline
\end{tabular}
\caption{List of variables and parameters. Top panel - quantities used for the cluster distribution. A $*$ is used to denote values at the gel point. Bottom panel - quantities used in the EDL. A bar is used over any of the variables in the top panel if they are in the EDL.}
\label{tab:my_label}
\end{table}

\bibliography{main.bib}

\begin{thebibliography}{94}%
\makeatletter
\providecommand \@ifxundefined [1]{%
 \@ifx{#1\undefined}
}%
\providecommand \@ifnum [1]{%
 \ifnum #1\expandafter \@firstoftwo
 \else \expandafter \@secondoftwo
 \fi
}%
\providecommand \@ifx [1]{%
 \ifx #1\expandafter \@firstoftwo
 \else \expandafter \@secondoftwo
 \fi
}%
\providecommand \natexlab [1]{#1}%
\providecommand \enquote  [1]{``#1''}%
\providecommand \bibnamefont  [1]{#1}%
\providecommand \bibfnamefont [1]{#1}%
\providecommand \citenamefont [1]{#1}%
\providecommand \href@noop [0]{\@secondoftwo}%
\providecommand \href [0]{\begingroup \@sanitize@url \@href}%
\providecommand \@href[1]{\@@startlink{#1}\@@href}%
\providecommand \@@href[1]{\endgroup#1\@@endlink}%
\providecommand \@sanitize@url [0]{\catcode `\\12\catcode `\$12\catcode
  `\&12\catcode `\#12\catcode `\^12\catcode `\_12\catcode `\%12\relax}%
\providecommand \@@startlink[1]{}%
\providecommand \@@endlink[0]{}%
\providecommand \url  [0]{\begingroup\@sanitize@url \@url }%
\providecommand \@url [1]{\endgroup\@href {#1}{\urlprefix }}%
\providecommand \urlprefix  [0]{URL }%
\providecommand \Eprint [0]{\href }%
\providecommand \doibase [0]{http://dx.doi.org/}%
\providecommand \selectlanguage [0]{\@gobble}%
\providecommand \bibinfo  [0]{\@secondoftwo}%
\providecommand \bibfield  [0]{\@secondoftwo}%
\providecommand \translation [1]{[#1]}%
\providecommand \BibitemOpen [0]{}%
\providecommand \bibitemStop [0]{}%
\providecommand \bibitemNoStop [0]{.\EOS\space}%
\providecommand \EOS [0]{\spacefactor3000\relax}%
\providecommand \BibitemShut  [1]{\csname bibitem#1\endcsname}%
\let\auto@bib@innerbib\@empty
\bibitem [{\citenamefont {Welton}(1999)}]{Welton1999}%
  \BibitemOpen
  \bibfield  {author} {\bibinfo {author} {\bibfnamefont {T.}~\bibnamefont
  {Welton}},\ }\href@noop {} {\bibfield  {journal} {\bibinfo  {journal} {Chem.
  Rev.}\ }\textbf {\bibinfo {volume} {99}},\ \bibinfo {pages} {2071} (\bibinfo
  {year} {1999})}\BibitemShut {NoStop}%
\bibitem [{\citenamefont {Hallett}\ and\ \citenamefont
  {Welton}(2011)}]{Hallett2011}%
  \BibitemOpen
  \bibfield  {author} {\bibinfo {author} {\bibfnamefont {J.~P.}\ \bibnamefont
  {Hallett}}\ and\ \bibinfo {author} {\bibfnamefont {T.}~\bibnamefont
  {Welton}},\ }\href@noop {} {\bibfield  {journal} {\bibinfo  {journal} {Chem.
  Rev.}\ }\textbf {\bibinfo {volume} {111}},\ \bibinfo {pages} {3508} (\bibinfo
  {year} {2011})}\BibitemShut {NoStop}%
\bibitem [{\citenamefont {Fedorov}\ and\ \citenamefont
  {Kornyshev}(2014)}]{Fedorov2014}%
  \BibitemOpen
  \bibfield  {author} {\bibinfo {author} {\bibfnamefont {M.~V.}\ \bibnamefont
  {Fedorov}}\ and\ \bibinfo {author} {\bibfnamefont {A.~A.}\ \bibnamefont
  {Kornyshev}},\ }\href {\doibase 10.1021/cr400374x} {\bibfield  {journal}
  {\bibinfo  {journal} {Chem. Rev.}\ }\textbf {\bibinfo {volume} {114}},\
  \bibinfo {pages} {2978} (\bibinfo {year} {2014})}\BibitemShut {NoStop}%
\bibitem [{\citenamefont {Weing\"{a}rtner}(2008)}]{Hermann2008}%
  \BibitemOpen
  \bibfield  {author} {\bibinfo {author} {\bibfnamefont {H.}~\bibnamefont
  {Weing\"{a}rtner}},\ }\href@noop {} {\bibfield  {journal} {\bibinfo
  {journal} {Angew. Chem. Int. Ed.}\ }\textbf {\bibinfo {volume} {47}},\
  \bibinfo {pages} {654} (\bibinfo {year} {2008})}\BibitemShut {NoStop}%
\bibitem [{\citenamefont {Kondrat}\ and\ \citenamefont
  {Kornyshev}(2016)}]{Kondrat2016}%
  \BibitemOpen
  \bibfield  {author} {\bibinfo {author} {\bibfnamefont {S.}~\bibnamefont
  {Kondrat}}\ and\ \bibinfo {author} {\bibfnamefont {A.~A.}\ \bibnamefont
  {Kornyshev}},\ }\href@noop {} {\bibfield  {journal} {\bibinfo  {journal}
  {Nanoscale Horiz.}\ }\textbf {\bibinfo {volume} {1}},\ \bibinfo {pages} {45}
  (\bibinfo {year} {2016})}\BibitemShut {NoStop}%
\bibitem [{\citenamefont {Son}\ and\ \citenamefont {Wang}(2020)}]{son2020ion}%
  \BibitemOpen
  \bibfield  {author} {\bibinfo {author} {\bibfnamefont {C.~Y.}\ \bibnamefont
  {Son}}\ and\ \bibinfo {author} {\bibfnamefont {Z.-G.}\ \bibnamefont {Wang}},\
  }\href@noop {} {\bibfield  {journal} {\bibinfo  {journal} {J. Chem. Phys.}\
  }\textbf {\bibinfo {volume} {153}},\ \bibinfo {pages} {100903} (\bibinfo
  {year} {2020})}\BibitemShut {NoStop}%
\bibitem [{\citenamefont {Suo}\ \emph {et~al.}(2015)\citenamefont {Suo},
  \citenamefont {Borodin}, \citenamefont {Gao}, \citenamefont {Olguin},
  \citenamefont {Ho}, \citenamefont {Fan}, \citenamefont {Luo}, \citenamefont
  {Wang},\ and\ \citenamefont {Xu}}]{suo2015}%
  \BibitemOpen
  \bibfield  {author} {\bibinfo {author} {\bibfnamefont {L.}~\bibnamefont
  {Suo}}, \bibinfo {author} {\bibfnamefont {O.}~\bibnamefont {Borodin}},
  \bibinfo {author} {\bibfnamefont {T.}~\bibnamefont {Gao}}, \bibinfo {author}
  {\bibfnamefont {M.}~\bibnamefont {Olguin}}, \bibinfo {author} {\bibfnamefont
  {J.}~\bibnamefont {Ho}}, \bibinfo {author} {\bibfnamefont {X.}~\bibnamefont
  {Fan}}, \bibinfo {author} {\bibfnamefont {C.}~\bibnamefont {Luo}}, \bibinfo
  {author} {\bibfnamefont {C.}~\bibnamefont {Wang}}, \ and\ \bibinfo {author}
  {\bibfnamefont {K.}~\bibnamefont {Xu}},\ }\href {\doibase
  10.1126/science.aab1595} {\bibfield  {journal} {\bibinfo  {journal}
  {Science}\ }\textbf {\bibinfo {volume} {350}},\ \bibinfo {pages} {938}
  (\bibinfo {year} {2015})}\BibitemShut {NoStop}%
\bibitem [{\citenamefont {Suo}\ \emph {et~al.}(2017)\citenamefont {Suo},
  \citenamefont {Borodin}, \citenamefont {Wang}, \citenamefont {Rong},
  \citenamefont {Sun}, \citenamefont {Fan}, \citenamefont {Xu}, \citenamefont
  {Schroeder}, \citenamefont {Cresce}, \citenamefont {Wang} \emph
  {et~al.}}]{suo2017water}%
  \BibitemOpen
  \bibfield  {author} {\bibinfo {author} {\bibfnamefont {L.}~\bibnamefont
  {Suo}}, \bibinfo {author} {\bibfnamefont {O.}~\bibnamefont {Borodin}},
  \bibinfo {author} {\bibfnamefont {Y.}~\bibnamefont {Wang}}, \bibinfo {author}
  {\bibfnamefont {X.}~\bibnamefont {Rong}}, \bibinfo {author} {\bibfnamefont
  {W.}~\bibnamefont {Sun}}, \bibinfo {author} {\bibfnamefont {X.}~\bibnamefont
  {Fan}}, \bibinfo {author} {\bibfnamefont {S.}~\bibnamefont {Xu}}, \bibinfo
  {author} {\bibfnamefont {M.~A.}\ \bibnamefont {Schroeder}}, \bibinfo {author}
  {\bibfnamefont {A.~V.}\ \bibnamefont {Cresce}}, \bibinfo {author}
  {\bibfnamefont {F.}~\bibnamefont {Wang}},  \emph {et~al.},\ }\href@noop {}
  {\bibfield  {journal} {\bibinfo  {journal} {Adv. Energy Mater.}\ }\textbf
  {\bibinfo {volume} {7}},\ \bibinfo {pages} {1701189} (\bibinfo {year}
  {2017})}\BibitemShut {NoStop}%
\bibitem [{\citenamefont {Vatamanu}\ and\ \citenamefont
  {Borodin}(2017)}]{vatamanu2017}%
  \BibitemOpen
  \bibfield  {author} {\bibinfo {author} {\bibfnamefont {J.}~\bibnamefont
  {Vatamanu}}\ and\ \bibinfo {author} {\bibfnamefont {O.}~\bibnamefont
  {Borodin}},\ }\href@noop {} {\bibfield  {journal} {\bibinfo  {journal} {J.
  Phys. Chem. Lett.}\ }\textbf {\bibinfo {volume} {8}},\ \bibinfo {pages}
  {4362} (\bibinfo {year} {2017})}\BibitemShut {NoStop}%
\bibitem [{\citenamefont {Lannelongue}\ \emph {et~al.}(2018)\citenamefont
  {Lannelongue}, \citenamefont {Bouchal}, \citenamefont {Mourad}, \citenamefont
  {Bodin}, \citenamefont {Olarte}, \citenamefont {le~Vot}, \citenamefont
  {Favier},\ and\ \citenamefont {Fontaine}}]{Lannelongue2018}%
  \BibitemOpen
  \bibfield  {author} {\bibinfo {author} {\bibfnamefont {P.}~\bibnamefont
  {Lannelongue}}, \bibinfo {author} {\bibfnamefont {R.}~\bibnamefont
  {Bouchal}}, \bibinfo {author} {\bibfnamefont {E.}~\bibnamefont {Mourad}},
  \bibinfo {author} {\bibfnamefont {C.}~\bibnamefont {Bodin}}, \bibinfo
  {author} {\bibfnamefont {M.}~\bibnamefont {Olarte}}, \bibinfo {author}
  {\bibfnamefont {S.}~\bibnamefont {le~Vot}}, \bibinfo {author} {\bibfnamefont
  {F.}~\bibnamefont {Favier}}, \ and\ \bibinfo {author} {\bibfnamefont
  {O.}~\bibnamefont {Fontaine}},\ }\href {\doibase 10.1149/2.0951803jes}
  {\bibfield  {journal} {\bibinfo  {journal} {J. Electrochem. Soc.}\ }\textbf
  {\bibinfo {volume} {165}},\ \bibinfo {pages} {A657} (\bibinfo {year}
  {2018})}\BibitemShut {NoStop}%
\bibitem [{\citenamefont {Chen}\ \emph {et~al.}(2020)\citenamefont {Chen},
  \citenamefont {Feng},\ and\ \citenamefont {Qiao}}]{chen2020water}%
  \BibitemOpen
  \bibfield  {author} {\bibinfo {author} {\bibfnamefont {M.}~\bibnamefont
  {Chen}}, \bibinfo {author} {\bibfnamefont {G.}~\bibnamefont {Feng}}, \ and\
  \bibinfo {author} {\bibfnamefont {R.}~\bibnamefont {Qiao}},\ }\href@noop {}
  {\bibfield  {journal} {\bibinfo  {journal} {Curr. Opin. Colloid Interface
  Sci}\ } (\bibinfo {year} {2020})}\BibitemShut {NoStop}%
\bibitem [{\citenamefont {McEldrew}\ \emph {et~al.}(2018)\citenamefont
  {McEldrew}, \citenamefont {Goodwin}, \citenamefont {Kornyshev},\ and\
  \citenamefont {Bazant}}]{mceldrew2018}%
  \BibitemOpen
  \bibfield  {author} {\bibinfo {author} {\bibfnamefont {M.}~\bibnamefont
  {McEldrew}}, \bibinfo {author} {\bibfnamefont {Z.~A.}\ \bibnamefont
  {Goodwin}}, \bibinfo {author} {\bibfnamefont {A.~A.}\ \bibnamefont
  {Kornyshev}}, \ and\ \bibinfo {author} {\bibfnamefont {M.~Z.}\ \bibnamefont
  {Bazant}},\ }\href@noop {} {\bibfield  {journal} {\bibinfo  {journal} {J.
  Phys. Chem. Lett.}\ }\textbf {\bibinfo {volume} {9}},\ \bibinfo {pages}
  {5840} (\bibinfo {year} {2018})}\BibitemShut {NoStop}%
\bibitem [{\citenamefont {Han}\ \emph {et~al.}(2021)\citenamefont {Han},
  \citenamefont {Zhang}, \citenamefont {Gewirth},\ and\ \citenamefont
  {Espinosa-Marzal}}]{Han2021WiSE}%
  \BibitemOpen
  \bibfield  {author} {\bibinfo {author} {\bibfnamefont {M.}~\bibnamefont
  {Han}}, \bibinfo {author} {\bibfnamefont {R.}~\bibnamefont {Zhang}}, \bibinfo
  {author} {\bibfnamefont {A.~A.}\ \bibnamefont {Gewirth}}, \ and\ \bibinfo
  {author} {\bibfnamefont {R.~M.}\ \bibnamefont {Espinosa-Marzal}},\
  }\href@noop {} {\bibfield  {journal} {\bibinfo  {journal} {Nano Lett.}\
  }\textbf {\bibinfo {volume} {21}},\ \bibinfo {pages} {2304} (\bibinfo {year}
  {2021})}\BibitemShut {NoStop}%
\bibitem [{\citenamefont {Groves}\ \emph {et~al.}(2021)\citenamefont {Groves},
  \citenamefont {Perez-Martinez}, \citenamefont {Lhermerout},\ and\
  \citenamefont {Perkin}}]{Groves2021wise}%
  \BibitemOpen
  \bibfield  {author} {\bibinfo {author} {\bibfnamefont {T.~S.}\ \bibnamefont
  {Groves}}, \bibinfo {author} {\bibfnamefont {C.~S.}\ \bibnamefont
  {Perez-Martinez}}, \bibinfo {author} {\bibfnamefont {R.}~\bibnamefont
  {Lhermerout}}, \ and\ \bibinfo {author} {\bibfnamefont {S.}~\bibnamefont
  {Perkin}},\ }\href@noop {} {\bibfield  {journal} {\bibinfo  {journal} {J.
  Phys. Chem. Lett.}\ }\textbf {\bibinfo {volume} {12}},\ \bibinfo {pages}
  {1702} (\bibinfo {year} {2021})}\BibitemShut {NoStop}%
\bibitem [{\citenamefont {Patsahan}\ and\ \citenamefont
  {Ciach}(2022)}]{Patsahan2022}%
  \BibitemOpen
  \bibfield  {author} {\bibinfo {author} {\bibfnamefont {O.}~\bibnamefont
  {Patsahan}}\ and\ \bibinfo {author} {\bibfnamefont {A.}~\bibnamefont
  {Ciach}},\ }\href@noop {} {\bibfield  {journal} {\bibinfo  {journal} {ACS
  Omega}\ }\textbf {\bibinfo {volume} {7}},\ \bibinfo {pages} {6655} (\bibinfo
  {year} {2022})}\BibitemShut {NoStop}%
\bibitem [{\citenamefont {Yu}\ \emph {et~al.}(2022)\citenamefont {Yu},
  \citenamefont {Balsara}, \citenamefont {Borodin}, \citenamefont {Gewirth},
  \citenamefont {Hahn}, \citenamefont {Maginn}, \citenamefont {Persson},
  \citenamefont {Srinivasan}, \citenamefont {Toney}, \citenamefont {Xu},
  \citenamefont {Zavadil}, \citenamefont {Curtiss},\ and\ \citenamefont
  {Cheng}}]{Yu2022Energy}%
  \BibitemOpen
  \bibfield  {author} {\bibinfo {author} {\bibfnamefont {Z.}~\bibnamefont
  {Yu}}, \bibinfo {author} {\bibfnamefont {N.~P.}\ \bibnamefont {Balsara}},
  \bibinfo {author} {\bibfnamefont {O.}~\bibnamefont {Borodin}}, \bibinfo
  {author} {\bibfnamefont {A.~A.}\ \bibnamefont {Gewirth}}, \bibinfo {author}
  {\bibfnamefont {N.~T.}\ \bibnamefont {Hahn}}, \bibinfo {author}
  {\bibfnamefont {E.~J.}\ \bibnamefont {Maginn}}, \bibinfo {author}
  {\bibfnamefont {K.~A.}\ \bibnamefont {Persson}}, \bibinfo {author}
  {\bibfnamefont {V.}~\bibnamefont {Srinivasan}}, \bibinfo {author}
  {\bibfnamefont {M.~F.}\ \bibnamefont {Toney}}, \bibinfo {author}
  {\bibfnamefont {K.}~\bibnamefont {Xu}}, \bibinfo {author} {\bibfnamefont
  {K.~R.}\ \bibnamefont {Zavadil}}, \bibinfo {author} {\bibfnamefont {L.~A.}\
  \bibnamefont {Curtiss}}, \ and\ \bibinfo {author} {\bibfnamefont
  {L.}~\bibnamefont {Cheng}},\ }\href@noop {} {\bibfield  {journal} {\bibinfo
  {journal} {ACS Energy Lett.}\ }\textbf {\bibinfo {volume} {7}},\ \bibinfo
  {pages} {461} (\bibinfo {year} {2022})}\BibitemShut {NoStop}%
\bibitem [{\citenamefont {Zhang}\ \emph
  {et~al.}(2020{\natexlab{a}})\citenamefont {Zhang}, \citenamefont {Han},
  \citenamefont {Ta}, \citenamefont {Madsen}, \citenamefont {Chen},
  \citenamefont {Zhang}, \citenamefont {Espinosa-Marzal},\ and\ \citenamefont
  {Gewirth}}]{Zhang2020wise}%
  \BibitemOpen
  \bibfield  {author} {\bibinfo {author} {\bibfnamefont {R.}~\bibnamefont
  {Zhang}}, \bibinfo {author} {\bibfnamefont {M.}~\bibnamefont {Han}}, \bibinfo
  {author} {\bibfnamefont {K.}~\bibnamefont {Ta}}, \bibinfo {author}
  {\bibfnamefont {K.~E.}\ \bibnamefont {Madsen}}, \bibinfo {author}
  {\bibfnamefont {X.}~\bibnamefont {Chen}}, \bibinfo {author} {\bibfnamefont
  {X.}~\bibnamefont {Zhang}}, \bibinfo {author} {\bibfnamefont {R.~M.}\
  \bibnamefont {Espinosa-Marzal}}, \ and\ \bibinfo {author} {\bibfnamefont
  {A.~A.}\ \bibnamefont {Gewirth}},\ }\href@noop {} {\bibfield  {journal}
  {\bibinfo  {journal} {ACS Appl. Energy Mater.}\ }\textbf {\bibinfo {volume}
  {3}},\ \bibinfo {pages} {8086} (\bibinfo {year}
  {2020}{\natexlab{a}})}\BibitemShut {NoStop}%
\bibitem [{\citenamefont {Dokko}\ \emph {et~al.}(2013)\citenamefont {Dokko},
  \citenamefont {Tachikawa}, \citenamefont {Yamauchi}, \citenamefont
  {Tsuchiya}, \citenamefont {Yamazaki}, \citenamefont {Takashima},
  \citenamefont {Park}, \citenamefont {Ueno}, \citenamefont {Seki},
  \citenamefont {Serizawa} \emph {et~al.}}]{dokko2013solvate}%
  \BibitemOpen
  \bibfield  {author} {\bibinfo {author} {\bibfnamefont {K.}~\bibnamefont
  {Dokko}}, \bibinfo {author} {\bibfnamefont {N.}~\bibnamefont {Tachikawa}},
  \bibinfo {author} {\bibfnamefont {K.}~\bibnamefont {Yamauchi}}, \bibinfo
  {author} {\bibfnamefont {M.}~\bibnamefont {Tsuchiya}}, \bibinfo {author}
  {\bibfnamefont {A.}~\bibnamefont {Yamazaki}}, \bibinfo {author}
  {\bibfnamefont {E.}~\bibnamefont {Takashima}}, \bibinfo {author}
  {\bibfnamefont {J.-W.}\ \bibnamefont {Park}}, \bibinfo {author}
  {\bibfnamefont {K.}~\bibnamefont {Ueno}}, \bibinfo {author} {\bibfnamefont
  {S.}~\bibnamefont {Seki}}, \bibinfo {author} {\bibfnamefont {N.}~\bibnamefont
  {Serizawa}},  \emph {et~al.},\ }\href@noop {} {\bibfield  {journal} {\bibinfo
   {journal} {J. Electrochem. Soc.}\ }\textbf {\bibinfo {volume} {160}},\
  \bibinfo {pages} {A1304} (\bibinfo {year} {2013})}\BibitemShut {NoStop}%
\bibitem [{\citenamefont {Lewandowski}\ and\ \citenamefont
  {{\'S}widerska-Mocek}(2009)}]{lewandowski2009ionic}%
  \BibitemOpen
  \bibfield  {author} {\bibinfo {author} {\bibfnamefont {A.}~\bibnamefont
  {Lewandowski}}\ and\ \bibinfo {author} {\bibfnamefont {A.}~\bibnamefont
  {{\'S}widerska-Mocek}},\ }\href@noop {} {\bibfield  {journal} {\bibinfo
  {journal} {J. Power Sources}\ }\textbf {\bibinfo {volume} {194}},\ \bibinfo
  {pages} {601} (\bibinfo {year} {2009})}\BibitemShut {NoStop}%
\bibitem [{\citenamefont {Molinari}\ \emph
  {et~al.}(2019{\natexlab{a}})\citenamefont {Molinari}, \citenamefont
  {Mailoa},\ and\ \citenamefont {Kozinsky}}]{molinari2019general}%
  \BibitemOpen
  \bibfield  {author} {\bibinfo {author} {\bibfnamefont {N.}~\bibnamefont
  {Molinari}}, \bibinfo {author} {\bibfnamefont {J.~P.}\ \bibnamefont
  {Mailoa}}, \ and\ \bibinfo {author} {\bibfnamefont {B.}~\bibnamefont
  {Kozinsky}},\ }\href@noop {} {\bibfield  {journal} {\bibinfo  {journal} {J.
  Phys. Chem. Lett.}\ }\textbf {\bibinfo {volume} {10}},\ \bibinfo {pages}
  {2313} (\bibinfo {year} {2019}{\natexlab{a}})}\BibitemShut {NoStop}%
\bibitem [{\citenamefont {Molinari}\ \emph
  {et~al.}(2019{\natexlab{b}})\citenamefont {Molinari}, \citenamefont {Mailoa},
  \citenamefont {Craig}, \citenamefont {Christensen},\ and\ \citenamefont
  {Kozinsky}}]{molinari2019transport}%
  \BibitemOpen
  \bibfield  {author} {\bibinfo {author} {\bibfnamefont {N.}~\bibnamefont
  {Molinari}}, \bibinfo {author} {\bibfnamefont {J.~P.}\ \bibnamefont
  {Mailoa}}, \bibinfo {author} {\bibfnamefont {N.}~\bibnamefont {Craig}},
  \bibinfo {author} {\bibfnamefont {J.}~\bibnamefont {Christensen}}, \ and\
  \bibinfo {author} {\bibfnamefont {B.}~\bibnamefont {Kozinsky}},\ }\href@noop
  {} {\bibfield  {journal} {\bibinfo  {journal} {J. Power Sources}\ }\textbf
  {\bibinfo {volume} {428}},\ \bibinfo {pages} {27} (\bibinfo {year}
  {2019}{\natexlab{b}})}\BibitemShut {NoStop}%
\bibitem [{\citenamefont {Zhang}\ \emph {et~al.}(2018)\citenamefont {Zhang},
  \citenamefont {Ye}, \citenamefont {Henkensmeier}, \citenamefont
  {Hempelmann},\ and\ \citenamefont {Chen}}]{Zhang2018}%
  \BibitemOpen
  \bibfield  {author} {\bibinfo {author} {\bibfnamefont {Y.}~\bibnamefont
  {Zhang}}, \bibinfo {author} {\bibfnamefont {R.}~\bibnamefont {Ye}}, \bibinfo
  {author} {\bibfnamefont {D.}~\bibnamefont {Henkensmeier}}, \bibinfo {author}
  {\bibfnamefont {R.}~\bibnamefont {Hempelmann}}, \ and\ \bibinfo {author}
  {\bibfnamefont {R.}~\bibnamefont {Chen}},\ }\href@noop {} {\bibfield
  {journal} {\bibinfo  {journal} {Electrochim. Acta}\ }\textbf {\bibinfo
  {volume} {263}} (\bibinfo {year} {2018})}\BibitemShut {NoStop}%
\bibitem [{\citenamefont {Yoon}\ \emph {et~al.}(2013)\citenamefont {Yoon},
  \citenamefont {Howlett}, \citenamefont {Best}, \citenamefont {Forsyth},\ and\
  \citenamefont {Macfarlane}}]{yoon2013fast}%
  \BibitemOpen
  \bibfield  {author} {\bibinfo {author} {\bibfnamefont {H.}~\bibnamefont
  {Yoon}}, \bibinfo {author} {\bibfnamefont {P.}~\bibnamefont {Howlett}},
  \bibinfo {author} {\bibfnamefont {A.~S.}\ \bibnamefont {Best}}, \bibinfo
  {author} {\bibfnamefont {M.}~\bibnamefont {Forsyth}}, \ and\ \bibinfo
  {author} {\bibfnamefont {D.~R.}\ \bibnamefont {Macfarlane}},\ }\href@noop {}
  {\bibfield  {journal} {\bibinfo  {journal} {J. Electrochem. Soc.}\ }\textbf
  {\bibinfo {volume} {160}},\ \bibinfo {pages} {A1629} (\bibinfo {year}
  {2013})}\BibitemShut {NoStop}%
\bibitem [{\citenamefont {Howlett}\ \emph {et~al.}(2004)\citenamefont
  {Howlett}, \citenamefont {MacFarlane},\ and\ \citenamefont
  {Hollenkamp}}]{howlett2004high}%
  \BibitemOpen
  \bibfield  {author} {\bibinfo {author} {\bibfnamefont {P.~C.}\ \bibnamefont
  {Howlett}}, \bibinfo {author} {\bibfnamefont {D.~R.}\ \bibnamefont
  {MacFarlane}}, \ and\ \bibinfo {author} {\bibfnamefont {A.~F.}\ \bibnamefont
  {Hollenkamp}},\ }\href@noop {} {\bibfield  {journal} {\bibinfo  {journal}
  {Electrochem. Solid-State Lett.}\ }\textbf {\bibinfo {volume} {7}},\ \bibinfo
  {pages} {A97} (\bibinfo {year} {2004})}\BibitemShut {NoStop}%
\bibitem [{\citenamefont {Fedorov}\ and\ \citenamefont
  {Kornyshev}(2008{\natexlab{a}})}]{Fedorov2008a}%
  \BibitemOpen
  \bibfield  {author} {\bibinfo {author} {\bibfnamefont {M.~V.}\ \bibnamefont
  {Fedorov}}\ and\ \bibinfo {author} {\bibfnamefont {A.~A.}\ \bibnamefont
  {Kornyshev}},\ }\href {\doibase 10.1021/jp803440q} {\bibfield  {journal}
  {\bibinfo  {journal} {J. Phys. Chem. B}\ }\textbf {\bibinfo {volume} {112}},\
  \bibinfo {pages} {11868} (\bibinfo {year} {2008}{\natexlab{a}})}\BibitemShut
  {NoStop}%
\bibitem [{\citenamefont {Fedorov}\ and\ \citenamefont
  {Kornyshev}(2008{\natexlab{b}})}]{Fedorov2008}%
  \BibitemOpen
  \bibfield  {author} {\bibinfo {author} {\bibfnamefont {M.~V.}\ \bibnamefont
  {Fedorov}}\ and\ \bibinfo {author} {\bibfnamefont {A.~A.}\ \bibnamefont
  {Kornyshev}},\ }\href@noop {} {\bibfield  {journal} {\bibinfo  {journal}
  {Electrochim. Acta}\ }\textbf {\bibinfo {volume} {53}},\ \bibinfo {pages}
  {6835} (\bibinfo {year} {2008}{\natexlab{b}})}\BibitemShut {NoStop}%
\bibitem [{\citenamefont {Georgi}\ \emph {et~al.}(2010)\citenamefont {Georgi},
  \citenamefont {Kornyshev},\ and\ \citenamefont {Fedorov}}]{Georgi2010}%
  \BibitemOpen
  \bibfield  {author} {\bibinfo {author} {\bibfnamefont {N.}~\bibnamefont
  {Georgi}}, \bibinfo {author} {\bibfnamefont {A.}~\bibnamefont {Kornyshev}}, \
  and\ \bibinfo {author} {\bibfnamefont {M.}~\bibnamefont {Fedorov}},\ }\href
  {\doibase 10.1016/j.jelechem.2010.07.004} {\bibfield  {journal} {\bibinfo
  {journal} {Journal of Electroanalytical Chemistry}\ }\textbf {\bibinfo
  {volume} {649}},\ \bibinfo {pages} {261} (\bibinfo {year}
  {2010})}\BibitemShut {NoStop}%
\bibitem [{\citenamefont {Bazant}\ \emph {et~al.}(2011)\citenamefont {Bazant},
  \citenamefont {Storey},\ and\ \citenamefont {Kornyshev}}]{Bazant2011}%
  \BibitemOpen
  \bibfield  {author} {\bibinfo {author} {\bibfnamefont {M.~Z.}\ \bibnamefont
  {Bazant}}, \bibinfo {author} {\bibfnamefont {B.~D.}\ \bibnamefont {Storey}},
  \ and\ \bibinfo {author} {\bibfnamefont {A.~A.}\ \bibnamefont {Kornyshev}},\
  }\href {\doibase 10.1103/PhysRevLett.106.046102} {\bibfield  {journal}
  {\bibinfo  {journal} {Phys. Rev. Lett.}\ }\textbf {\bibinfo {volume} {106}},\
  \bibinfo {pages} {046102} (\bibinfo {year} {2011})}\BibitemShut {NoStop}%
\bibitem [{\citenamefont {Coles}\ \emph {et~al.}(2020)\citenamefont {Coles},
  \citenamefont {Park}, \citenamefont {Nikam}, \citenamefont {Kanduc},
  \citenamefont {Dzubiella},\ and\ \citenamefont
  {Rotenberg}}]{Coles2020length}%
  \BibitemOpen
  \bibfield  {author} {\bibinfo {author} {\bibfnamefont {S.}~\bibnamefont
  {Coles}}, \bibinfo {author} {\bibfnamefont {C.}~\bibnamefont {Park}},
  \bibinfo {author} {\bibfnamefont {R.}~\bibnamefont {Nikam}}, \bibinfo
  {author} {\bibfnamefont {M.}~\bibnamefont {Kanduc}}, \bibinfo {author}
  {\bibfnamefont {J.}~\bibnamefont {Dzubiella}}, \ and\ \bibinfo {author}
  {\bibfnamefont {B.}~\bibnamefont {Rotenberg}},\ }\href@noop {} {\bibfield
  {journal} {\bibinfo  {journal} {J. Phys. Chem. B}\ }\textbf {\bibinfo
  {volume} {124}},\ \bibinfo {pages} {1778} (\bibinfo {year}
  {2020})}\BibitemShut {NoStop}%
\bibitem [{\citenamefont {de~Souza}\ \emph {et~al.}(2020)\citenamefont
  {de~Souza}, \citenamefont {Goodwin}, \citenamefont {McEldrew}, \citenamefont
  {Kornyshev},\ and\ \citenamefont {Bazant}}]{pedroRTILs}%
  \BibitemOpen
  \bibfield  {author} {\bibinfo {author} {\bibfnamefont {J.~P.}\ \bibnamefont
  {de~Souza}}, \bibinfo {author} {\bibfnamefont {Z.~A.~H.}\ \bibnamefont
  {Goodwin}}, \bibinfo {author} {\bibfnamefont {M.}~\bibnamefont {McEldrew}},
  \bibinfo {author} {\bibfnamefont {A.~A.}\ \bibnamefont {Kornyshev}}, \ and\
  \bibinfo {author} {\bibfnamefont {M.~Z.}\ \bibnamefont {Bazant}},\ }\href
  {\doibase 10.1103/PhysRevLett.125.116001} {\bibfield  {journal} {\bibinfo
  {journal} {Phys. Rev. Lett.}\ }\textbf {\bibinfo {volume} {125}},\ \bibinfo
  {pages} {116001} (\bibinfo {year} {2020})}\BibitemShut {NoStop}%
\bibitem [{\citenamefont {de~Souza}\ \emph {et~al.}(2021)\citenamefont
  {de~Souza}, \citenamefont {Pivnic}, \citenamefont {Bazant}, \citenamefont
  {Urbakh},\ and\ \citenamefont {Kornyshev}}]{pedro2022force}%
  \BibitemOpen
  \bibfield  {author} {\bibinfo {author} {\bibfnamefont {J.~P.}\ \bibnamefont
  {de~Souza}}, \bibinfo {author} {\bibfnamefont {K.}~\bibnamefont {Pivnic}},
  \bibinfo {author} {\bibfnamefont {M.~Z.}\ \bibnamefont {Bazant}}, \bibinfo
  {author} {\bibfnamefont {M.}~\bibnamefont {Urbakh}}, \ and\ \bibinfo {author}
  {\bibfnamefont {A.~A.}\ \bibnamefont {Kornyshev}},\ }\href@noop {} {\bibfield
   {journal} {\bibinfo  {journal} {J. Phys. Chem. B}\ }\textbf {\bibinfo
  {volume} {126}},\ \bibinfo {pages} {1242} (\bibinfo {year}
  {2021})}\BibitemShut {NoStop}%
\bibitem [{\citenamefont {Gavish}\ \emph {et~al.}(2018)\citenamefont {Gavish},
  \citenamefont {Elad},\ and\ \citenamefont {Yochelis}}]{gavish2018solvent}%
  \BibitemOpen
  \bibfield  {author} {\bibinfo {author} {\bibfnamefont {N.}~\bibnamefont
  {Gavish}}, \bibinfo {author} {\bibfnamefont {D.}~\bibnamefont {Elad}}, \ and\
  \bibinfo {author} {\bibfnamefont {A.}~\bibnamefont {Yochelis}},\ }\href@noop
  {} {\bibfield  {journal} {\bibinfo  {journal} {J. Phys. Chem. Lett.}\
  }\textbf {\bibinfo {volume} {9}},\ \bibinfo {pages} {36} (\bibinfo {year}
  {2018})}\BibitemShut {NoStop}%
\bibitem [{\citenamefont {Krucker-Velasquez}\ and\ \citenamefont
  {Swan}(2021)}]{emily2021}%
  \BibitemOpen
  \bibfield  {author} {\bibinfo {author} {\bibfnamefont {E.}~\bibnamefont
  {Krucker-Velasquez}}\ and\ \bibinfo {author} {\bibfnamefont {J.~W.}\
  \bibnamefont {Swan}},\ }\href@noop {} {\bibfield  {journal} {\bibinfo
  {journal} {J. Chem. Phys.}\ }\textbf {\bibinfo {volume} {155}},\ \bibinfo
  {pages} {134903} (\bibinfo {year} {2021})}\BibitemShut {NoStop}%
\bibitem [{\citenamefont {Levy}\ \emph {et~al.}(2019)\citenamefont {Levy},
  \citenamefont {McEldrew},\ and\ \citenamefont {Bazant}}]{levy2019spin}%
  \BibitemOpen
  \bibfield  {author} {\bibinfo {author} {\bibfnamefont {A.}~\bibnamefont
  {Levy}}, \bibinfo {author} {\bibfnamefont {M.}~\bibnamefont {McEldrew}}, \
  and\ \bibinfo {author} {\bibfnamefont {M.~Z.}\ \bibnamefont {Bazant}},\
  }\href@noop {} {\bibfield  {journal} {\bibinfo  {journal} {Phys. Rev.
  Mater.}\ }\textbf {\bibinfo {volume} {3}},\ \bibinfo {pages} {055606}
  (\bibinfo {year} {2019})}\BibitemShut {NoStop}%
\bibitem [{\citenamefont {Smith}\ \emph {et~al.}(2016)\citenamefont {Smith},
  \citenamefont {Lee},\ and\ \citenamefont {Perkin}}]{Smith2016}%
  \BibitemOpen
  \bibfield  {author} {\bibinfo {author} {\bibfnamefont {A.~M.}\ \bibnamefont
  {Smith}}, \bibinfo {author} {\bibfnamefont {A.~A.}\ \bibnamefont {Lee}}, \
  and\ \bibinfo {author} {\bibfnamefont {S.}~\bibnamefont {Perkin}},\ }\href
  {\doibase 10.1021/acs.jpclett.6b00867} {\bibfield  {journal} {\bibinfo
  {journal} {J. Phys. Chem. Lett.}\ }\textbf {\bibinfo {volume} {7}},\ \bibinfo
  {pages} {2157} (\bibinfo {year} {2016})}\BibitemShut {NoStop}%
\bibitem [{\citenamefont {Gebbie}\ \emph {et~al.}(2013)\citenamefont {Gebbie},
  \citenamefont {Valtiner}, \citenamefont {Banquy}, \citenamefont {Fox},
  \citenamefont {Henderson},\ and\ \citenamefont {Israelachvili}}]{Gebbie2013}%
  \BibitemOpen
  \bibfield  {author} {\bibinfo {author} {\bibfnamefont {M.~A.}\ \bibnamefont
  {Gebbie}}, \bibinfo {author} {\bibfnamefont {M.}~\bibnamefont {Valtiner}},
  \bibinfo {author} {\bibfnamefont {X.}~\bibnamefont {Banquy}}, \bibinfo
  {author} {\bibfnamefont {E.~T.}\ \bibnamefont {Fox}}, \bibinfo {author}
  {\bibfnamefont {W.~A.}\ \bibnamefont {Henderson}}, \ and\ \bibinfo {author}
  {\bibfnamefont {J.~N.}\ \bibnamefont {Israelachvili}},\ }\href {\doibase
  10.1073/pnas.1307871110} {\bibfield  {journal} {\bibinfo  {journal} {PNAS}\
  }\textbf {\bibinfo {volume} {110}},\ \bibinfo {pages} {9674} (\bibinfo {year}
  {2013})}\BibitemShut {NoStop}%
\bibitem [{\citenamefont {Gebbie}\ \emph {et~al.}(2015)\citenamefont {Gebbie},
  \citenamefont {Dobes}, \citenamefont {Valtiner},\ and\ \citenamefont
  {Israelachvili}}]{Gebbie2015}%
  \BibitemOpen
  \bibfield  {author} {\bibinfo {author} {\bibfnamefont {M.~A.}\ \bibnamefont
  {Gebbie}}, \bibinfo {author} {\bibfnamefont {H.~A.}\ \bibnamefont {Dobes}},
  \bibinfo {author} {\bibfnamefont {M.}~\bibnamefont {Valtiner}}, \ and\
  \bibinfo {author} {\bibfnamefont {J.~N.}\ \bibnamefont {Israelachvili}},\
  }\href@noop {} {\bibfield  {journal} {\bibinfo  {journal} {PNAS}\ }\textbf
  {\bibinfo {volume} {112}},\ \bibinfo {pages} {7432} (\bibinfo {year}
  {2015})}\BibitemShut {NoStop}%
\bibitem [{\citenamefont {Gebbie}\ \emph {et~al.}(2017)\citenamefont {Gebbie},
  \citenamefont {Smith}, \citenamefont {Dobbs}, \citenamefont {Lee},
  \citenamefont {Warr}, \citenamefont {Banquy}, \citenamefont {Valtiner},
  \citenamefont {Rutland}, \citenamefont {Israelachvili}, \citenamefont
  {Perkin},\ and\ \citenamefont {Atkin}}]{Gebbie2017rev}%
  \BibitemOpen
  \bibfield  {author} {\bibinfo {author} {\bibfnamefont {M.~A.}\ \bibnamefont
  {Gebbie}}, \bibinfo {author} {\bibfnamefont {A.~M.}\ \bibnamefont {Smith}},
  \bibinfo {author} {\bibfnamefont {H.~A.}\ \bibnamefont {Dobbs}}, \bibinfo
  {author} {\bibfnamefont {A.}~\bibnamefont {Lee}}, \bibinfo {author}
  {\bibfnamefont {G.~G.}\ \bibnamefont {Warr}}, \bibinfo {author}
  {\bibfnamefont {X.}~\bibnamefont {Banquy}}, \bibinfo {author} {\bibfnamefont
  {M.}~\bibnamefont {Valtiner}}, \bibinfo {author} {\bibfnamefont {M.~W.}\
  \bibnamefont {Rutland}}, \bibinfo {author} {\bibfnamefont {J.~N.}\
  \bibnamefont {Israelachvili}}, \bibinfo {author} {\bibfnamefont
  {S.}~\bibnamefont {Perkin}}, \ and\ \bibinfo {author} {\bibfnamefont
  {R.}~\bibnamefont {Atkin}},\ }\href@noop {} {\bibfield  {journal} {\bibinfo
  {journal} {Chem. Commun.}\ }\textbf {\bibinfo {volume} {53}},\ \bibinfo
  {pages} {1214} (\bibinfo {year} {2017})}\BibitemShut {NoStop}%
\bibitem [{\citenamefont {Smith}\ \emph {et~al.}(2017)\citenamefont {Smith},
  \citenamefont {Lee},\ and\ \citenamefont {Perkin}}]{smith2017struct}%
  \BibitemOpen
  \bibfield  {author} {\bibinfo {author} {\bibfnamefont {A.~M.}\ \bibnamefont
  {Smith}}, \bibinfo {author} {\bibfnamefont {A.~A.}\ \bibnamefont {Lee}}, \
  and\ \bibinfo {author} {\bibfnamefont {S.}~\bibnamefont {Perkin}},\
  }\href@noop {} {\bibfield  {journal} {\bibinfo  {journal} {Phys. Rev. Lett.}\
  }\textbf {\bibinfo {volume} {118}},\ \bibinfo {pages} {096002} (\bibinfo
  {year} {2017})}\BibitemShut {NoStop}%
\bibitem [{\citenamefont {Han}\ \emph {et~al.}(2020)\citenamefont {Han},
  \citenamefont {Kim}, \citenamefont {Leal}, \citenamefont {Negrito},
  \citenamefont {Batteas},\ and\ \citenamefont {Espinosa-Marzal}}]{Han2020IL}%
  \BibitemOpen
  \bibfield  {author} {\bibinfo {author} {\bibfnamefont {M.}~\bibnamefont
  {Han}}, \bibinfo {author} {\bibfnamefont {H.}~\bibnamefont {Kim}}, \bibinfo
  {author} {\bibfnamefont {C.}~\bibnamefont {Leal}}, \bibinfo {author}
  {\bibfnamefont {M.}~\bibnamefont {Negrito}}, \bibinfo {author} {\bibfnamefont
  {J.~D.}\ \bibnamefont {Batteas}}, \ and\ \bibinfo {author} {\bibfnamefont
  {R.~M.}\ \bibnamefont {Espinosa-Marzal}},\ }\href@noop {} {\bibfield
  {journal} {\bibinfo  {journal} {Adv. Mater.}\ }\textbf {\bibinfo {volume}
  {7}},\ \bibinfo {pages} {2001313} (\bibinfo {year} {2020})}\BibitemShut
  {NoStop}%
\bibitem [{\citenamefont {Jurado}\ \emph {et~al.}(2016)\citenamefont {Jurado},
  \citenamefont {Kim}, \citenamefont {Rossi}, \citenamefont {Arcifa},
  \citenamefont {Schuh}, \citenamefont {Spencer}, \citenamefont {Leal},
  \citenamefont {Ewoldte},\ and\ \citenamefont {Espinosa-Marzal}}]{Jurado2016}%
  \BibitemOpen
  \bibfield  {author} {\bibinfo {author} {\bibfnamefont {L.~A.}\ \bibnamefont
  {Jurado}}, \bibinfo {author} {\bibfnamefont {H.}~\bibnamefont {Kim}},
  \bibinfo {author} {\bibfnamefont {A.}~\bibnamefont {Rossi}}, \bibinfo
  {author} {\bibfnamefont {A.}~\bibnamefont {Arcifa}}, \bibinfo {author}
  {\bibfnamefont {J.~K.}\ \bibnamefont {Schuh}}, \bibinfo {author}
  {\bibfnamefont {N.~D.}\ \bibnamefont {Spencer}}, \bibinfo {author}
  {\bibfnamefont {C.}~\bibnamefont {Leal}}, \bibinfo {author} {\bibfnamefont
  {R.~H.}\ \bibnamefont {Ewoldte}}, \ and\ \bibinfo {author} {\bibfnamefont
  {R.~M.}\ \bibnamefont {Espinosa-Marzal}},\ }\href@noop {} {\bibfield
  {journal} {\bibinfo  {journal} {Phys. Chem. Chem. Phys.}\ }\textbf {\bibinfo
  {volume} {18}},\ \bibinfo {pages} {22719} (\bibinfo {year}
  {2016})}\BibitemShut {NoStop}%
\bibitem [{\citenamefont {Jurado}\ \emph {et~al.}(2015)\citenamefont {Jurado},
  \citenamefont {Kim}, \citenamefont {Arcifa}, \citenamefont {Rossi},
  \citenamefont {Leal}, \citenamefont {Spencerc},\ and\ \citenamefont
  {Espinosa-Marzal}}]{Jurado2015}%
  \BibitemOpen
  \bibfield  {author} {\bibinfo {author} {\bibfnamefont {L.~A.}\ \bibnamefont
  {Jurado}}, \bibinfo {author} {\bibfnamefont {H.}~\bibnamefont {Kim}},
  \bibinfo {author} {\bibfnamefont {A.}~\bibnamefont {Arcifa}}, \bibinfo
  {author} {\bibfnamefont {A.}~\bibnamefont {Rossi}}, \bibinfo {author}
  {\bibfnamefont {C.}~\bibnamefont {Leal}}, \bibinfo {author} {\bibfnamefont
  {N.~D.}\ \bibnamefont {Spencerc}}, \ and\ \bibinfo {author} {\bibfnamefont
  {R.~M.}\ \bibnamefont {Espinosa-Marzal}},\ }\href@noop {} {\bibfield
  {journal} {\bibinfo  {journal} {Phys. Chem. Chem. Phys.}\ }\textbf {\bibinfo
  {volume} {17}},\ \bibinfo {pages} {13613} (\bibinfo {year}
  {2015})}\BibitemShut {NoStop}%
\bibitem [{\citenamefont {Jurado}\ and\ \citenamefont
  {Espinosa-Marzal}(2017)}]{Jurado2017EDL}%
  \BibitemOpen
  \bibfield  {author} {\bibinfo {author} {\bibfnamefont {L.~A.}\ \bibnamefont
  {Jurado}}\ and\ \bibinfo {author} {\bibfnamefont {R.~M.}\ \bibnamefont
  {Espinosa-Marzal}},\ }\href@noop {} {\bibfield  {journal} {\bibinfo
  {journal} {Sci. Rep.}\ }\textbf {\bibinfo {volume} {17}},\ \bibinfo {pages}
  {4225} (\bibinfo {year} {2017})}\BibitemShut {NoStop}%
\bibitem [{\citenamefont {Mao}\ \emph {et~al.}(2019)\citenamefont {Mao},
  \citenamefont {Brown}, \citenamefont {\v{C}ervinka}, \citenamefont {Hazell},
  \citenamefont {Li}, \citenamefont {Ren}, \citenamefont {Chen}, \citenamefont
  {Atkin}, \citenamefont {Eastoe}, \citenamefont {Grillo}, \citenamefont
  {Padua}, \citenamefont {Gomes},\ and\ \citenamefont {Hatton}}]{Mao2019nano}%
  \BibitemOpen
  \bibfield  {author} {\bibinfo {author} {\bibfnamefont {X.}~\bibnamefont
  {Mao}}, \bibinfo {author} {\bibfnamefont {P.}~\bibnamefont {Brown}}, \bibinfo
  {author} {\bibfnamefont {C.}~\bibnamefont {\v{C}ervinka}}, \bibinfo {author}
  {\bibfnamefont {G.}~\bibnamefont {Hazell}}, \bibinfo {author} {\bibfnamefont
  {H.}~\bibnamefont {Li}}, \bibinfo {author} {\bibfnamefont {Y.}~\bibnamefont
  {Ren}}, \bibinfo {author} {\bibfnamefont {D.}~\bibnamefont {Chen}}, \bibinfo
  {author} {\bibfnamefont {R.}~\bibnamefont {Atkin}}, \bibinfo {author}
  {\bibfnamefont {J.}~\bibnamefont {Eastoe}}, \bibinfo {author} {\bibfnamefont
  {I.}~\bibnamefont {Grillo}}, \bibinfo {author} {\bibfnamefont {A.~A.~H.}\
  \bibnamefont {Padua}}, \bibinfo {author} {\bibfnamefont {M.~F.~C.}\
  \bibnamefont {Gomes}}, \ and\ \bibinfo {author} {\bibfnamefont {T.~A.}\
  \bibnamefont {Hatton}},\ }\href@noop {} {\bibfield  {journal} {\bibinfo
  {journal} {Nat. Mater.}\ }\textbf {\bibinfo {volume} {18}},\ \bibinfo {pages}
  {1350} (\bibinfo {year} {2019})}\BibitemShut {NoStop}%
\bibitem [{\citenamefont {Lee}\ \emph {et~al.}(2017{\natexlab{a}})\citenamefont
  {Lee}, \citenamefont {Perez-Martinez}, \citenamefont {Smith},\ and\
  \citenamefont {Perkin}}]{UND}%
  \BibitemOpen
  \bibfield  {author} {\bibinfo {author} {\bibfnamefont {A.~A.}\ \bibnamefont
  {Lee}}, \bibinfo {author} {\bibfnamefont {C.~S.}\ \bibnamefont
  {Perez-Martinez}}, \bibinfo {author} {\bibfnamefont {A.~M.}\ \bibnamefont
  {Smith}}, \ and\ \bibinfo {author} {\bibfnamefont {S.}~\bibnamefont
  {Perkin}},\ }\href@noop {} {\bibfield  {journal} {\bibinfo  {journal}
  {Faraday Discuss.}\ }\textbf {\bibinfo {volume} {199}},\ \bibinfo {pages}
  {239} (\bibinfo {year} {2017}{\natexlab{a}})}\BibitemShut {NoStop}%
\bibitem [{\citenamefont {Lee}\ \emph {et~al.}(2017{\natexlab{b}})\citenamefont
  {Lee}, \citenamefont {Perez-Martinez}, \citenamefont {Smith},\ and\
  \citenamefont {Perkin}}]{underalpha}%
  \BibitemOpen
  \bibfield  {author} {\bibinfo {author} {\bibfnamefont {A.~A.}\ \bibnamefont
  {Lee}}, \bibinfo {author} {\bibfnamefont {C.~S.}\ \bibnamefont
  {Perez-Martinez}}, \bibinfo {author} {\bibfnamefont {A.~M.}\ \bibnamefont
  {Smith}}, \ and\ \bibinfo {author} {\bibfnamefont {S.}~\bibnamefont
  {Perkin}},\ }\href@noop {} {\bibfield  {journal} {\bibinfo  {journal} {Phys.
  Rev. Lett.}\ }\textbf {\bibinfo {volume} {119}},\ \bibinfo {pages} {026002}
  (\bibinfo {year} {2017}{\natexlab{b}})}\BibitemShut {NoStop}%
\bibitem [{\citenamefont {Wang}\ and\ \citenamefont {Voth}(2005)}]{Wang2005}%
  \BibitemOpen
  \bibfield  {author} {\bibinfo {author} {\bibfnamefont {Y.}~\bibnamefont
  {Wang}}\ and\ \bibinfo {author} {\bibfnamefont {G.~A.}\ \bibnamefont
  {Voth}},\ }\href@noop {} {\bibfield  {journal} {\bibinfo  {journal} {J. Am.
  Chem. Soc.}\ }\textbf {\bibinfo {volume} {35}},\ \bibinfo {pages} {12192}
  (\bibinfo {year} {2005})}\BibitemShut {NoStop}%
\bibitem [{\citenamefont {Bernardes}\ \emph {et~al.}(2011)\citenamefont
  {Bernardes}, \citenamefont {da~Piedade},\ and\ \citenamefont
  {Lopes}}]{Bernardes2011}%
  \BibitemOpen
  \bibfield  {author} {\bibinfo {author} {\bibfnamefont {C.~E.~S.}\
  \bibnamefont {Bernardes}}, \bibinfo {author} {\bibfnamefont {M.~E.~M.}\
  \bibnamefont {da~Piedade}}, \ and\ \bibinfo {author} {\bibfnamefont
  {J.~N.~C.}\ \bibnamefont {Lopes}},\ }\href@noop {} {\bibfield  {journal}
  {\bibinfo  {journal} {J. Phys. Chem. B}\ }\textbf {\bibinfo {volume} {115}},\
  \bibinfo {pages} {2067} (\bibinfo {year} {2011})}\BibitemShut {NoStop}%
\bibitem [{\citenamefont {McEldrew}\ \emph
  {et~al.}(2021{\natexlab{a}})\citenamefont {McEldrew}, \citenamefont
  {Goodwin}, \citenamefont {Zhao}, \citenamefont {Bazant},\ and\ \citenamefont
  {Kornyshev}}]{mceldrew2020correlated}%
  \BibitemOpen
  \bibfield  {author} {\bibinfo {author} {\bibfnamefont {M.}~\bibnamefont
  {McEldrew}}, \bibinfo {author} {\bibfnamefont {Z.~A.~H.}\ \bibnamefont
  {Goodwin}}, \bibinfo {author} {\bibfnamefont {H.}~\bibnamefont {Zhao}},
  \bibinfo {author} {\bibfnamefont {M.~Z.}\ \bibnamefont {Bazant}}, \ and\
  \bibinfo {author} {\bibfnamefont {A.~A.}\ \bibnamefont {Kornyshev}},\
  }\href@noop {} {\bibfield  {journal} {\bibinfo  {journal} {J. Phys. Chem. B}\
  }\textbf {\bibinfo {volume} {125}},\ \bibinfo {pages} {2677} (\bibinfo {year}
  {2021}{\natexlab{a}})}\BibitemShut {NoStop}%
\bibitem [{\citenamefont {Lopes}\ and\ \citenamefont
  {P\'adua}(2006)}]{Lopes2006}%
  \BibitemOpen
  \bibfield  {author} {\bibinfo {author} {\bibfnamefont {J.~N. A.~C.}\
  \bibnamefont {Lopes}}\ and\ \bibinfo {author} {\bibfnamefont {A.~A.~H.}\
  \bibnamefont {P\'adua}},\ }\href@noop {} {\bibfield  {journal} {\bibinfo
  {journal} {J. Phys. Chem. B}\ }\textbf {\bibinfo {volume} {110}},\ \bibinfo
  {pages} {3330} (\bibinfo {year} {2006})}\BibitemShut {NoStop}%
\bibitem [{\citenamefont {Ma}\ \emph {et~al.}(2015)\citenamefont {Ma},
  \citenamefont {Forsman},\ and\ \citenamefont {Woodward}}]{Ma2015}%
  \BibitemOpen
  \bibfield  {author} {\bibinfo {author} {\bibfnamefont {K.}~\bibnamefont
  {Ma}}, \bibinfo {author} {\bibfnamefont {J.}~\bibnamefont {Forsman}}, \ and\
  \bibinfo {author} {\bibfnamefont {C.~E.}\ \bibnamefont {Woodward}},\
  }\href@noop {} {\bibfield  {journal} {\bibinfo  {journal} {J. Chem. Phys.}\
  }\textbf {\bibinfo {volume} {142}},\ \bibinfo {pages} {174704} (\bibinfo
  {year} {2015})}\BibitemShut {NoStop}%
\bibitem [{\citenamefont {Lee}\ \emph {et~al.}(2015{\natexlab{a}})\citenamefont
  {Lee}, \citenamefont {Vella}, \citenamefont {Perkin},\ and\ \citenamefont
  {Goriely}}]{Lee2015}%
  \BibitemOpen
  \bibfield  {author} {\bibinfo {author} {\bibfnamefont {A.~A.}\ \bibnamefont
  {Lee}}, \bibinfo {author} {\bibfnamefont {D.}~\bibnamefont {Vella}}, \bibinfo
  {author} {\bibfnamefont {S.}~\bibnamefont {Perkin}}, \ and\ \bibinfo {author}
  {\bibfnamefont {A.}~\bibnamefont {Goriely}},\ }\href {\doibase
  10.1021/jz502250z} {\bibfield  {journal} {\bibinfo  {journal} {J. Phys. Chem.
  Lett.}\ }\textbf {\bibinfo {volume} {6}},\ \bibinfo {pages} {159} (\bibinfo
  {year} {2015}{\natexlab{a}})}\BibitemShut {NoStop}%
\bibitem [{\citenamefont {Zhang}\ and\ \citenamefont
  {Maginn}(2015)}]{Zhang2015}%
  \BibitemOpen
  \bibfield  {author} {\bibinfo {author} {\bibfnamefont {Y.}~\bibnamefont
  {Zhang}}\ and\ \bibinfo {author} {\bibfnamefont {E.~J.}\ \bibnamefont
  {Maginn}},\ }\href@noop {} {\bibfield  {journal} {\bibinfo  {journal} {J.
  Phys. Chem. Lett.}\ }\textbf {\bibinfo {volume} {6}},\ \bibinfo {pages} {700}
  (\bibinfo {year} {2015})}\BibitemShut {NoStop}%
\bibitem [{\citenamefont {MacFarlane}\ \emph {et~al.}(2009)\citenamefont
  {MacFarlane}, \citenamefont {Forsyth}, \citenamefont {Izgorodina},
  \citenamefont {Abbott}, \citenamefont {Annata},\ and\ \citenamefont
  {Fraser}}]{MacFarlane2009}%
  \BibitemOpen
  \bibfield  {author} {\bibinfo {author} {\bibfnamefont {D.~R.}\ \bibnamefont
  {MacFarlane}}, \bibinfo {author} {\bibfnamefont {M.}~\bibnamefont {Forsyth}},
  \bibinfo {author} {\bibfnamefont {E.~I.}\ \bibnamefont {Izgorodina}},
  \bibinfo {author} {\bibfnamefont {A.~P.}\ \bibnamefont {Abbott}}, \bibinfo
  {author} {\bibfnamefont {G.}~\bibnamefont {Annata}}, \ and\ \bibinfo {author}
  {\bibfnamefont {K.}~\bibnamefont {Fraser}},\ }\href@noop {} {\bibfield
  {journal} {\bibinfo  {journal} {Phys. Chem. Chem. Phys.}\ }\textbf {\bibinfo
  {volume} {11}},\ \bibinfo {pages} {4962} (\bibinfo {year}
  {2009})}\BibitemShut {NoStop}%
\bibitem [{\citenamefont {Holl{\'o}czki}\ \emph {et~al.}(2014)\citenamefont
  {Holl{\'o}czki}, \citenamefont {Malberg}, \citenamefont {Welton},\ and\
  \citenamefont {Kirchner}}]{Kirchner2014}%
  \BibitemOpen
  \bibfield  {author} {\bibinfo {author} {\bibfnamefont {O.}~\bibnamefont
  {Holl{\'o}czki}}, \bibinfo {author} {\bibfnamefont {F.}~\bibnamefont
  {Malberg}}, \bibinfo {author} {\bibfnamefont {T.}~\bibnamefont {Welton}}, \
  and\ \bibinfo {author} {\bibfnamefont {B.}~\bibnamefont {Kirchner}},\
  }\href@noop {} {\bibfield  {journal} {\bibinfo  {journal} {Phys. Chem. Chem.
  Phys.}\ }\textbf {\bibinfo {volume} {16}},\ \bibinfo {pages} {16880}
  (\bibinfo {year} {2014})}\BibitemShut {NoStop}%
\bibitem [{\citenamefont {Adar}\ \emph {et~al.}(2017)\citenamefont {Adar},
  \citenamefont {Markovich},\ and\ \citenamefont {Andelman}}]{adar2017bjerrum}%
  \BibitemOpen
  \bibfield  {author} {\bibinfo {author} {\bibfnamefont {R.~M.}\ \bibnamefont
  {Adar}}, \bibinfo {author} {\bibfnamefont {T.}~\bibnamefont {Markovich}}, \
  and\ \bibinfo {author} {\bibfnamefont {D.}~\bibnamefont {Andelman}},\
  }\href@noop {} {\bibfield  {journal} {\bibinfo  {journal} {J. Chem. Phys.}\
  }\textbf {\bibinfo {volume} {146}},\ \bibinfo {pages} {194904} (\bibinfo
  {year} {2017})}\BibitemShut {NoStop}%
\bibitem [{\citenamefont {Araque}\ \emph {et~al.}(2015)\citenamefont {Araque},
  \citenamefont {Yadav}, \citenamefont {Shadeck}, \citenamefont {Maroncelli},\
  and\ \citenamefont {Margulis}}]{Araque2015}%
  \BibitemOpen
  \bibfield  {author} {\bibinfo {author} {\bibfnamefont {J.~C.}\ \bibnamefont
  {Araque}}, \bibinfo {author} {\bibfnamefont {S.~K.}\ \bibnamefont {Yadav}},
  \bibinfo {author} {\bibfnamefont {M.}~\bibnamefont {Shadeck}}, \bibinfo
  {author} {\bibfnamefont {M.}~\bibnamefont {Maroncelli}}, \ and\ \bibinfo
  {author} {\bibfnamefont {C.~J.}\ \bibnamefont {Margulis}},\ }\href@noop {}
  {\bibfield  {journal} {\bibinfo  {journal} {J. Phys. Chem. B}\ }\textbf
  {\bibinfo {volume} {119}},\ \bibinfo {pages} {7015} (\bibinfo {year}
  {2015})}\BibitemShut {NoStop}%
\bibitem [{\citenamefont {Avni}\ \emph {et~al.}(2020)\citenamefont {Avni},
  \citenamefont {Adar},\ and\ \citenamefont {Andelman}}]{avni2020charge}%
  \BibitemOpen
  \bibfield  {author} {\bibinfo {author} {\bibfnamefont {Y.}~\bibnamefont
  {Avni}}, \bibinfo {author} {\bibfnamefont {R.~M.}\ \bibnamefont {Adar}}, \
  and\ \bibinfo {author} {\bibfnamefont {D.}~\bibnamefont {Andelman}},\
  }\href@noop {} {\bibfield  {journal} {\bibinfo  {journal} {Phys. Rev. E}\
  }\textbf {\bibinfo {volume} {101}},\ \bibinfo {pages} {010601} (\bibinfo
  {year} {2020})}\BibitemShut {NoStop}%
\bibitem [{\citenamefont {Goodwin}\ \emph {et~al.}(2017)\citenamefont
  {Goodwin}, \citenamefont {Feng},\ and\ \citenamefont
  {Kornyshev}}]{goodwin2017mean}%
  \BibitemOpen
  \bibfield  {author} {\bibinfo {author} {\bibfnamefont {Z.~A.}\ \bibnamefont
  {Goodwin}}, \bibinfo {author} {\bibfnamefont {G.}~\bibnamefont {Feng}}, \
  and\ \bibinfo {author} {\bibfnamefont {A.~A.}\ \bibnamefont {Kornyshev}},\
  }\href@noop {} {\bibfield  {journal} {\bibinfo  {journal} {Electrochim.
  Acta}\ }\textbf {\bibinfo {volume} {225}},\ \bibinfo {pages} {190} (\bibinfo
  {year} {2017})}\BibitemShut {NoStop}%
\bibitem [{\citenamefont {Chen}\ \emph {et~al.}(2018)\citenamefont {Chen},
  \citenamefont {Goodwin}, \citenamefont {Feng},\ and\ \citenamefont
  {Kornyshev}}]{Chen2017}%
  \BibitemOpen
  \bibfield  {author} {\bibinfo {author} {\bibfnamefont {M.}~\bibnamefont
  {Chen}}, \bibinfo {author} {\bibfnamefont {Z.~A.~H.}\ \bibnamefont
  {Goodwin}}, \bibinfo {author} {\bibfnamefont {G.}~\bibnamefont {Feng}}, \
  and\ \bibinfo {author} {\bibfnamefont {A.~A.}\ \bibnamefont {Kornyshev}},\
  }\href {\doibase 10.1016/j.jelechem.2017.11.005} {\bibfield  {journal}
  {\bibinfo  {journal} {J. Electroanal. Chem.}\ }\textbf {\bibinfo {volume}
  {819}},\ \bibinfo {pages} {347} (\bibinfo {year} {2018})}\BibitemShut
  {NoStop}%
\bibitem [{\citenamefont {Goodwin}\ and\ \citenamefont
  {Kornyshev}(2017)}]{goodwin2017underscreening}%
  \BibitemOpen
  \bibfield  {author} {\bibinfo {author} {\bibfnamefont {Z.~A.}\ \bibnamefont
  {Goodwin}}\ and\ \bibinfo {author} {\bibfnamefont {A.~A.}\ \bibnamefont
  {Kornyshev}},\ }\href@noop {} {\bibfield  {journal} {\bibinfo  {journal}
  {Electrochem. commun.}\ }\textbf {\bibinfo {volume} {82}},\ \bibinfo {pages}
  {129} (\bibinfo {year} {2017})}\BibitemShut {NoStop}%
\bibitem [{\citenamefont {Feng}\ \emph {et~al.}(2019)\citenamefont {Feng},
  \citenamefont {Chen}, \citenamefont {Bi}, \citenamefont {Goodwin},
  \citenamefont {Postnikov}, \citenamefont {Brilliantov}, \citenamefont
  {Urbakh},\ and\ \citenamefont {Kornyshev}}]{feng2019free}%
  \BibitemOpen
  \bibfield  {author} {\bibinfo {author} {\bibfnamefont {G.}~\bibnamefont
  {Feng}}, \bibinfo {author} {\bibfnamefont {M.}~\bibnamefont {Chen}}, \bibinfo
  {author} {\bibfnamefont {S.}~\bibnamefont {Bi}}, \bibinfo {author}
  {\bibfnamefont {Z.~A.}\ \bibnamefont {Goodwin}}, \bibinfo {author}
  {\bibfnamefont {E.~B.}\ \bibnamefont {Postnikov}}, \bibinfo {author}
  {\bibfnamefont {N.}~\bibnamefont {Brilliantov}}, \bibinfo {author}
  {\bibfnamefont {M.}~\bibnamefont {Urbakh}}, \ and\ \bibinfo {author}
  {\bibfnamefont {A.~A.}\ \bibnamefont {Kornyshev}},\ }\href@noop {} {\bibfield
   {journal} {\bibinfo  {journal} {Phys. Rev. X}\ }\textbf {\bibinfo {volume}
  {9}},\ \bibinfo {pages} {021024} (\bibinfo {year} {2019})}\BibitemShut
  {NoStop}%
\bibitem [{\citenamefont {Dupont}(2004)}]{Dupont2004}%
  \BibitemOpen
  \bibfield  {author} {\bibinfo {author} {\bibfnamefont {J.}~\bibnamefont
  {Dupont}},\ }\href@noop {} {\bibfield  {journal} {\bibinfo  {journal} {J.
  Braz. Chem. Soc.}\ }\textbf {\bibinfo {volume} {3}},\ \bibinfo {pages} {341}
  (\bibinfo {year} {2004})}\BibitemShut {NoStop}%
\bibitem [{\citenamefont {McEldrew}\ \emph {et~al.}(2020)\citenamefont
  {McEldrew}, \citenamefont {Goodwin}, \citenamefont {Bi}, \citenamefont
  {Bazant},\ and\ \citenamefont {Kornyshev}}]{mceldrew2020theory}%
  \BibitemOpen
  \bibfield  {author} {\bibinfo {author} {\bibfnamefont {M.}~\bibnamefont
  {McEldrew}}, \bibinfo {author} {\bibfnamefont {Z.~A.}\ \bibnamefont
  {Goodwin}}, \bibinfo {author} {\bibfnamefont {S.}~\bibnamefont {Bi}},
  \bibinfo {author} {\bibfnamefont {M.~Z.}\ \bibnamefont {Bazant}}, \ and\
  \bibinfo {author} {\bibfnamefont {A.~A.}\ \bibnamefont {Kornyshev}},\
  }\href@noop {} {\bibfield  {journal} {\bibinfo  {journal} {J. Chem. Phys.}\
  }\textbf {\bibinfo {volume} {152}},\ \bibinfo {pages} {234506} (\bibinfo
  {year} {2020})}\BibitemShut {NoStop}%
\bibitem [{\citenamefont {McEldrew}\ \emph
  {et~al.}(2021{\natexlab{b}})\citenamefont {McEldrew}, \citenamefont
  {Goodwin}, \citenamefont {Bi}, \citenamefont {Korvnyshev},\ and\
  \citenamefont {Bazant}}]{mceldrew2021wise}%
  \BibitemOpen
  \bibfield  {author} {\bibinfo {author} {\bibfnamefont {M.}~\bibnamefont
  {McEldrew}}, \bibinfo {author} {\bibfnamefont {Z.~A.}\ \bibnamefont
  {Goodwin}}, \bibinfo {author} {\bibfnamefont {S.}~\bibnamefont {Bi}},
  \bibinfo {author} {\bibfnamefont {A.}~\bibnamefont {Korvnyshev}}, \ and\
  \bibinfo {author} {\bibfnamefont {M.~Z.}\ \bibnamefont {Bazant}},\
  }\href@noop {} {\bibfield  {journal} {\bibinfo  {journal} {J. Electrochem.
  Soc.}\ }\textbf {\bibinfo {volume} {168}},\ \bibinfo {pages} {050514}
  (\bibinfo {year} {2021}{\natexlab{b}})}\BibitemShut {NoStop}%
\bibitem [{\citenamefont {McEldrew}\ \emph
  {et~al.}(2021{\natexlab{c}})\citenamefont {McEldrew}, \citenamefont
  {Goodwin}, \citenamefont {Molinari}, \citenamefont {Kozinsky}, \citenamefont
  {Kornyshev},\ and\ \citenamefont {Bazant}}]{mceldrew2021salt}%
  \BibitemOpen
  \bibfield  {author} {\bibinfo {author} {\bibfnamefont {M.}~\bibnamefont
  {McEldrew}}, \bibinfo {author} {\bibfnamefont {Z.~A.}\ \bibnamefont
  {Goodwin}}, \bibinfo {author} {\bibfnamefont {N.}~\bibnamefont {Molinari}},
  \bibinfo {author} {\bibfnamefont {B.}~\bibnamefont {Kozinsky}}, \bibinfo
  {author} {\bibfnamefont {A.~A.}\ \bibnamefont {Kornyshev}}, \ and\ \bibinfo
  {author} {\bibfnamefont {M.~Z.}\ \bibnamefont {Bazant}},\ }\href@noop {}
  {\bibfield  {journal} {\bibinfo  {journal} {J. Phys. Chem. B}\ }\textbf
  {\bibinfo {volume} {125}},\ \bibinfo {pages} {13752} (\bibinfo {year}
  {2021}{\natexlab{c}})}\BibitemShut {NoStop}%
\bibitem [{\citenamefont {Flory}(1942)}]{flory1942thermodynamics}%
  \BibitemOpen
  \bibfield  {author} {\bibinfo {author} {\bibfnamefont {P.~J.}\ \bibnamefont
  {Flory}},\ }\href@noop {} {\bibfield  {journal} {\bibinfo  {journal} {J.
  Chem. Phys.}\ }\textbf {\bibinfo {volume} {10}},\ \bibinfo {pages} {51}
  (\bibinfo {year} {1942})}\BibitemShut {NoStop}%
\bibitem [{\citenamefont {Tanaka}(1989)}]{tanaka1989}%
  \BibitemOpen
  \bibfield  {author} {\bibinfo {author} {\bibfnamefont {F.}~\bibnamefont
  {Tanaka}},\ }\href@noop {} {\bibfield  {journal} {\bibinfo  {journal}
  {Macromolecules}\ }\textbf {\bibinfo {volume} {22}},\ \bibinfo {pages} {1988}
  (\bibinfo {year} {1989})}\BibitemShut {NoStop}%
\bibitem [{\citenamefont {Tanaka}(1990)}]{tanaka1990thermodynamic}%
  \BibitemOpen
  \bibfield  {author} {\bibinfo {author} {\bibfnamefont {F.}~\bibnamefont
  {Tanaka}},\ }\href@noop {} {\bibfield  {journal} {\bibinfo  {journal}
  {Macromolecules}\ }\textbf {\bibinfo {volume} {23}},\ \bibinfo {pages} {3784}
  (\bibinfo {year} {1990})}\BibitemShut {NoStop}%
\bibitem [{\citenamefont {Tanaka}\ and\ \citenamefont
  {Stockmayer}(1994)}]{tanaka1994}%
  \BibitemOpen
  \bibfield  {author} {\bibinfo {author} {\bibfnamefont {F.}~\bibnamefont
  {Tanaka}}\ and\ \bibinfo {author} {\bibfnamefont {W.~H.}\ \bibnamefont
  {Stockmayer}},\ }\href@noop {} {\bibfield  {journal} {\bibinfo  {journal}
  {Macromolecules}\ }\textbf {\bibinfo {volume} {27}},\ \bibinfo {pages} {3943}
  (\bibinfo {year} {1994})}\BibitemShut {NoStop}%
\bibitem [{\citenamefont {Tanaka}\ and\ \citenamefont
  {Ishida}(1995)}]{tanaka1995}%
  \BibitemOpen
  \bibfield  {author} {\bibinfo {author} {\bibfnamefont {F.}~\bibnamefont
  {Tanaka}}\ and\ \bibinfo {author} {\bibfnamefont {M.}~\bibnamefont
  {Ishida}},\ }\href@noop {} {\bibfield  {journal} {\bibinfo  {journal} {J.
  Chem. Soc. Faraday Trans.}\ }\textbf {\bibinfo {volume} {91}},\ \bibinfo
  {pages} {2663} (\bibinfo {year} {1995})}\BibitemShut {NoStop}%
\bibitem [{\citenamefont {Ishida}\ and\ \citenamefont
  {Tanaka}(1997)}]{ishida1997}%
  \BibitemOpen
  \bibfield  {author} {\bibinfo {author} {\bibfnamefont {M.}~\bibnamefont
  {Ishida}}\ and\ \bibinfo {author} {\bibfnamefont {F.}~\bibnamefont
  {Tanaka}},\ }\href@noop {} {\bibfield  {journal} {\bibinfo  {journal}
  {Macromolecules}\ }\textbf {\bibinfo {volume} {30}},\ \bibinfo {pages} {3900}
  (\bibinfo {year} {1997})}\BibitemShut {NoStop}%
\bibitem [{\citenamefont {Tanaka}(1998)}]{tanaka1998}%
  \BibitemOpen
  \bibfield  {author} {\bibinfo {author} {\bibfnamefont {F.}~\bibnamefont
  {Tanaka}},\ }\href@noop {} {\bibfield  {journal} {\bibinfo  {journal}
  {Physica A: Statistical Mechanics and its Applications}\ }\textbf {\bibinfo
  {volume} {257}},\ \bibinfo {pages} {245} (\bibinfo {year}
  {1998})}\BibitemShut {NoStop}%
\bibitem [{\citenamefont {Tanaka}\ and\ \citenamefont
  {Ishida}(1999)}]{tanaka1999}%
  \BibitemOpen
  \bibfield  {author} {\bibinfo {author} {\bibfnamefont {F.}~\bibnamefont
  {Tanaka}}\ and\ \bibinfo {author} {\bibfnamefont {M.}~\bibnamefont
  {Ishida}},\ }\href@noop {} {\bibfield  {journal} {\bibinfo  {journal}
  {Macromolecules}\ }\textbf {\bibinfo {volume} {32}},\ \bibinfo {pages} {1271}
  (\bibinfo {year} {1999})}\BibitemShut {NoStop}%
\bibitem [{\citenamefont {Tanaka}(2002)}]{tanaka2002}%
  \BibitemOpen
  \bibfield  {author} {\bibinfo {author} {\bibfnamefont {F.}~\bibnamefont
  {Tanaka}},\ }\href@noop {} {\bibfield  {journal} {\bibinfo  {journal} {Polym.
  J.}\ }\textbf {\bibinfo {volume} {34}},\ \bibinfo {pages} {479} (\bibinfo
  {year} {2002})}\BibitemShut {NoStop}%
\bibitem [{\citenamefont {Borodin}\ \emph {et~al.}(2017)\citenamefont
  {Borodin}, \citenamefont {Suo}, \citenamefont {Gobet}, \citenamefont {Ren},
  \citenamefont {Wang}, \citenamefont {Faraone}, \citenamefont {Peng},
  \citenamefont {Olguin}, \citenamefont {Schroeder}, \citenamefont {Ding} \emph
  {et~al.}}]{borodin2017liquid}%
  \BibitemOpen
  \bibfield  {author} {\bibinfo {author} {\bibfnamefont {O.}~\bibnamefont
  {Borodin}}, \bibinfo {author} {\bibfnamefont {L.}~\bibnamefont {Suo}},
  \bibinfo {author} {\bibfnamefont {M.}~\bibnamefont {Gobet}}, \bibinfo
  {author} {\bibfnamefont {X.}~\bibnamefont {Ren}}, \bibinfo {author}
  {\bibfnamefont {F.}~\bibnamefont {Wang}}, \bibinfo {author} {\bibfnamefont
  {A.}~\bibnamefont {Faraone}}, \bibinfo {author} {\bibfnamefont
  {J.}~\bibnamefont {Peng}}, \bibinfo {author} {\bibfnamefont {M.}~\bibnamefont
  {Olguin}}, \bibinfo {author} {\bibfnamefont {M.}~\bibnamefont {Schroeder}},
  \bibinfo {author} {\bibfnamefont {M.~S.}\ \bibnamefont {Ding}},  \emph
  {et~al.},\ }\href@noop {} {\bibfield  {journal} {\bibinfo  {journal} {ACS
  nano}\ }\textbf {\bibinfo {volume} {11}},\ \bibinfo {pages} {10462} (\bibinfo
  {year} {2017})}\BibitemShut {NoStop}%
\bibitem [{\citenamefont {Choi}\ \emph {et~al.}(2018)\citenamefont {Choi},
  \citenamefont {Lee}, \citenamefont {Choi},\ and\ \citenamefont
  {Cho}}]{choi2018graph}%
  \BibitemOpen
  \bibfield  {author} {\bibinfo {author} {\bibfnamefont {J.-H.}\ \bibnamefont
  {Choi}}, \bibinfo {author} {\bibfnamefont {H.}~\bibnamefont {Lee}}, \bibinfo
  {author} {\bibfnamefont {H.~R.}\ \bibnamefont {Choi}}, \ and\ \bibinfo
  {author} {\bibfnamefont {M.}~\bibnamefont {Cho}},\ }\href@noop {} {\bibfield
  {journal} {\bibinfo  {journal} {Annu. Rev. Phys. Chem.}\ }\textbf {\bibinfo
  {volume} {69}},\ \bibinfo {pages} {125} (\bibinfo {year} {2018})}\BibitemShut
  {NoStop}%
\bibitem [{\citenamefont {Jeon}\ \emph {et~al.}(2020)\citenamefont {Jeon},
  \citenamefont {Lee}, \citenamefont {Choi},\ and\ \citenamefont
  {Cho}}]{jeon2020modeling}%
  \BibitemOpen
  \bibfield  {author} {\bibinfo {author} {\bibfnamefont {J.}~\bibnamefont
  {Jeon}}, \bibinfo {author} {\bibfnamefont {H.}~\bibnamefont {Lee}}, \bibinfo
  {author} {\bibfnamefont {J.-H.}\ \bibnamefont {Choi}}, \ and\ \bibinfo
  {author} {\bibfnamefont {M.}~\bibnamefont {Cho}},\ }\href@noop {} {\bibfield
  {journal} {\bibinfo  {journal} {J. Phys. Chem. C}\ } (\bibinfo {year}
  {2020})}\BibitemShut {NoStop}%
\bibitem [{\citenamefont {France-Lanord}\ and\ \citenamefont
  {Grossman}(2019)}]{france2019}%
  \BibitemOpen
  \bibfield  {author} {\bibinfo {author} {\bibfnamefont {A.}~\bibnamefont
  {France-Lanord}}\ and\ \bibinfo {author} {\bibfnamefont {J.~C.}\ \bibnamefont
  {Grossman}},\ }\href@noop {} {\bibfield  {journal} {\bibinfo  {journal}
  {Phys. Rev. Lett.}\ }\textbf {\bibinfo {volume} {122}},\ \bibinfo {pages}
  {136001} (\bibinfo {year} {2019})}\BibitemShut {NoStop}%
\bibitem [{\citenamefont {Goodwin}\ \emph {et~al.}(2021)\citenamefont
  {Goodwin}, \citenamefont {de~Souza}, \citenamefont {Bazant},\ and\
  \citenamefont {Kornyshev}}]{goodwin2021review}%
  \BibitemOpen
  \bibfield  {author} {\bibinfo {author} {\bibfnamefont {Z.~A.~H.}\
  \bibnamefont {Goodwin}}, \bibinfo {author} {\bibfnamefont {J.~P.}\
  \bibnamefont {de~Souza}}, \bibinfo {author} {\bibfnamefont {M.~Z.}\
  \bibnamefont {Bazant}}, \ and\ \bibinfo {author} {\bibfnamefont {A.~A.}\
  \bibnamefont {Kornyshev}},\ }\href
  {https://doi.org/10.1007/978-981-10-6739-6_62-1} {\emph {\bibinfo {title}
  {Mean-Field Theory of the Electrical Double Layer in Ionic Liquids}}}\
  (\bibinfo  {publisher} {In: Zhang S. (eds) Encyclopedia of Ionic Liquids.
  Springer, Singapore.},\ \bibinfo {year} {2021})\BibitemShut {NoStop}%
\bibitem [{\citenamefont {Kornyshev}(2007)}]{Kornyshev2007}%
  \BibitemOpen
  \bibfield  {author} {\bibinfo {author} {\bibfnamefont {A.~A.}\ \bibnamefont
  {Kornyshev}},\ }\href {\doibase 10.1021/jp067857o} {\bibfield  {journal}
  {\bibinfo  {journal} {J. Phys. Chem. B}\ }\textbf {\bibinfo {volume} {111}},\
  \bibinfo {pages} {5545} (\bibinfo {year} {2007})}\BibitemShut {NoStop}%
\bibitem [{\citenamefont {Kilic}\ \emph {et~al.}(2007)\citenamefont {Kilic},
  \citenamefont {Bazant},\ and\ \citenamefont {Ajdari}}]{kilic2007a}%
  \BibitemOpen
  \bibfield  {author} {\bibinfo {author} {\bibfnamefont {M.~S.}\ \bibnamefont
  {Kilic}}, \bibinfo {author} {\bibfnamefont {M.~Z.}\ \bibnamefont {Bazant}}, \
  and\ \bibinfo {author} {\bibfnamefont {A.}~\bibnamefont {Ajdari}},\
  }\href@noop {} {\bibfield  {journal} {\bibinfo  {journal} {Phys. Rev. E}\
  }\textbf {\bibinfo {volume} {75}},\ \bibinfo {pages} {021502} (\bibinfo
  {year} {2007})}\BibitemShut {NoStop}%
\bibitem [{\citenamefont {Bazant}\ \emph {et~al.}(2009)\citenamefont {Bazant},
  \citenamefont {Kilic}, \citenamefont {Storey},\ and\ \citenamefont
  {Ajdari}}]{Bazant2009a}%
  \BibitemOpen
  \bibfield  {author} {\bibinfo {author} {\bibfnamefont {M.~Z.}\ \bibnamefont
  {Bazant}}, \bibinfo {author} {\bibfnamefont {M.~S.}\ \bibnamefont {Kilic}},
  \bibinfo {author} {\bibfnamefont {B.}~\bibnamefont {Storey}}, \ and\ \bibinfo
  {author} {\bibfnamefont {A.}~\bibnamefont {Ajdari}},\ }\href@noop {}
  {\bibfield  {journal} {\bibinfo  {journal} {Adv. Colloid Interface Sci.}\
  }\textbf {\bibinfo {volume} {152}},\ \bibinfo {pages} {48} (\bibinfo {year}
  {2009})}\BibitemShut {NoStop}%
\bibitem [{\citenamefont {A.A.Kornyshev}\ and\ \citenamefont
  {M.A.Vorotyntsev}(1981)}]{Kornyshev1981}%
  \BibitemOpen
  \bibfield  {author} {\bibinfo {author} {\bibnamefont {A.A.Kornyshev}}\ and\
  \bibinfo {author} {\bibnamefont {M.A.Vorotyntsev}},\ }\href@noop {}
  {\bibfield  {journal} {\bibinfo  {journal} {Electrochim. Acta.}\ }\textbf
  {\bibinfo {volume} {26}},\ \bibinfo {pages} {303} (\bibinfo {year}
  {1981})}\BibitemShut {NoStop}%
\bibitem [{\citenamefont {Lee}\ \emph {et~al.}(2014)\citenamefont {Lee},
  \citenamefont {Vella}, \citenamefont {Perkin},\ and\ \citenamefont
  {Goriely}}]{lee2014room}%
  \BibitemOpen
  \bibfield  {author} {\bibinfo {author} {\bibfnamefont {A.~A.}\ \bibnamefont
  {Lee}}, \bibinfo {author} {\bibfnamefont {D.}~\bibnamefont {Vella}}, \bibinfo
  {author} {\bibfnamefont {S.}~\bibnamefont {Perkin}}, \ and\ \bibinfo {author}
  {\bibfnamefont {A.}~\bibnamefont {Goriely}},\ }\href@noop {} {\bibfield
  {journal} {\bibinfo  {journal} {J. Phys. Chem. Lett.}\ }\textbf {\bibinfo
  {volume} {6}},\ \bibinfo {pages} {159} (\bibinfo {year} {2014})}\BibitemShut
  {NoStop}%
\bibitem [{\citenamefont {Budkov}\ \emph {et~al.}(2018)\citenamefont {Budkov},
  \citenamefont {Kolesnikov}, \citenamefont {Goodwin}, \citenamefont
  {Kiselev},\ and\ \citenamefont {Kornyshev}}]{BBKGK}%
  \BibitemOpen
  \bibfield  {author} {\bibinfo {author} {\bibfnamefont {Y.~A.}\ \bibnamefont
  {Budkov}}, \bibinfo {author} {\bibfnamefont {A.~L.}\ \bibnamefont
  {Kolesnikov}}, \bibinfo {author} {\bibfnamefont {Z.~A.~H.}\ \bibnamefont
  {Goodwin}}, \bibinfo {author} {\bibfnamefont {M.}~\bibnamefont {Kiselev}}, \
  and\ \bibinfo {author} {\bibfnamefont {A.~A.}\ \bibnamefont {Kornyshev}},\
  }\href@noop {} {\bibfield  {journal} {\bibinfo  {journal} {Electrochim.
  Acta}\ }\textbf {\bibinfo {volume} {284}},\ \bibinfo {pages} {346} (\bibinfo
  {year} {2018})}\BibitemShut {NoStop}%
\bibitem [{\citenamefont {Jitvisate}\ and\ \citenamefont
  {Seddon}(2018)}]{Monchai2018}%
  \BibitemOpen
  \bibfield  {author} {\bibinfo {author} {\bibfnamefont {M.}~\bibnamefont
  {Jitvisate}}\ and\ \bibinfo {author} {\bibfnamefont {J.~R.~T.}\ \bibnamefont
  {Seddon}},\ }\href@noop {} {\bibfield  {journal} {\bibinfo  {journal} {J.
  Phys. Chem. Lett.}\ }\textbf {\bibinfo {volume} {9}},\ \bibinfo {pages} {126}
  (\bibinfo {year} {2018})}\BibitemShut {NoStop}%
\bibitem [{\citenamefont {Feng}\ \emph {et~al.}(2014)\citenamefont {Feng},
  \citenamefont {Jiang}, \citenamefont {Qiao},\ and\ \citenamefont
  {Kornyshev}}]{Feng2014}%
  \BibitemOpen
  \bibfield  {author} {\bibinfo {author} {\bibfnamefont {G.}~\bibnamefont
  {Feng}}, \bibinfo {author} {\bibfnamefont {X.}~\bibnamefont {Jiang}},
  \bibinfo {author} {\bibfnamefont {R.}~\bibnamefont {Qiao}}, \ and\ \bibinfo
  {author} {\bibfnamefont {A.~A.}\ \bibnamefont {Kornyshev}},\ }\href {\doibase
  10.1021/nn505017c} {\bibfield  {journal} {\bibinfo  {journal} {ACS Nano}\
  }\textbf {\bibinfo {volume} {8}},\ \bibinfo {pages} {11685} (\bibinfo {year}
  {2014})}\BibitemShut {NoStop}%
\bibitem [{\citenamefont {de~Souza}\ and\ \citenamefont
  {Bazant}(2020)}]{de2020continuum}%
  \BibitemOpen
  \bibfield  {author} {\bibinfo {author} {\bibfnamefont {J.~P.}\ \bibnamefont
  {de~Souza}}\ and\ \bibinfo {author} {\bibfnamefont {M.~Z.}\ \bibnamefont
  {Bazant}},\ }\href@noop {} {\bibfield  {journal} {\bibinfo  {journal} {The
  Journal of Physical Chemistry C}\ }\textbf {\bibinfo {volume} {124}},\
  \bibinfo {pages} {11414} (\bibinfo {year} {2020})}\BibitemShut {NoStop}%
\bibitem [{\citenamefont {Chao}\ and\ \citenamefont {Wang}(2020)}]{Chao2020}%
  \BibitemOpen
  \bibfield  {author} {\bibinfo {author} {\bibfnamefont {H.}~\bibnamefont
  {Chao}}\ and\ \bibinfo {author} {\bibfnamefont {Z.-G.}\ \bibnamefont
  {Wang}},\ }\href@noop {} {\bibfield  {journal} {\bibinfo  {journal} {J. Phys.
  Chem. Lett.}\ }\textbf {\bibinfo {volume} {11}},\ \bibinfo {pages} {1767}
  (\bibinfo {year} {2020})}\BibitemShut {NoStop}%
\bibitem [{\citenamefont {Lee}\ \emph {et~al.}(2015{\natexlab{b}})\citenamefont
  {Lee}, \citenamefont {Kondrat}, \citenamefont {Vella},\ and\ \citenamefont
  {Goriely}}]{Leedynamics2015}%
  \BibitemOpen
  \bibfield  {author} {\bibinfo {author} {\bibfnamefont {A.~A.}\ \bibnamefont
  {Lee}}, \bibinfo {author} {\bibfnamefont {S.}~\bibnamefont {Kondrat}},
  \bibinfo {author} {\bibfnamefont {D.}~\bibnamefont {Vella}}, \ and\ \bibinfo
  {author} {\bibfnamefont {A.}~\bibnamefont {Goriely}},\ }\href@noop {}
  {\bibfield  {journal} {\bibinfo  {journal} {Phys. Rev. Lett.}\ }\textbf
  {\bibinfo {volume} {115}},\ \bibinfo {pages} {106101} (\bibinfo {year}
  {2015}{\natexlab{b}})}\BibitemShut {NoStop}%
\bibitem [{\citenamefont {Damaskin}\ and\ \citenamefont
  {Frumkin}(1974)}]{Damaskin1974}%
  \BibitemOpen
  \bibfield  {author} {\bibinfo {author} {\bibfnamefont {B.~B.}\ \bibnamefont
  {Damaskin}}\ and\ \bibinfo {author} {\bibfnamefont {A.~N.}\ \bibnamefont
  {Frumkin}},\ }\href@noop {} {\bibfield  {journal} {\bibinfo  {journal}
  {Electrochim. Acta}\ }\textbf {\bibinfo {volume} {19}},\ \bibinfo {pages}
  {173} (\bibinfo {year} {1974})}\BibitemShut {NoStop}%
\bibitem [{\citenamefont {Parsons}(1975)}]{Parson1975}%
  \BibitemOpen
  \bibfield  {author} {\bibinfo {author} {\bibfnamefont {R.}~\bibnamefont
  {Parsons}},\ }\href@noop {} {\bibfield  {journal} {\bibinfo  {journal} {J.
  Electroanal. Chem. Interfacial Electrochem.}\ }\textbf {\bibinfo {volume}
  {59}},\ \bibinfo {pages} {229} (\bibinfo {year} {1975})}\BibitemShut
  {NoStop}%
\bibitem [{\citenamefont {Zhang}\ \emph
  {et~al.}(2020{\natexlab{b}})\citenamefont {Zhang}, \citenamefont {Ye},
  \citenamefont {Chen}, \citenamefont {Goodwin}, \citenamefont {Feng},
  \citenamefont {Huang},\ and\ \citenamefont {Kornyshev}}]{yufan2020}%
  \BibitemOpen
  \bibfield  {author} {\bibinfo {author} {\bibfnamefont {Y.}~\bibnamefont
  {Zhang}}, \bibinfo {author} {\bibfnamefont {T.}~\bibnamefont {Ye}}, \bibinfo
  {author} {\bibfnamefont {M.}~\bibnamefont {Chen}}, \bibinfo {author}
  {\bibfnamefont {Z.~A.~H.}\ \bibnamefont {Goodwin}}, \bibinfo {author}
  {\bibfnamefont {G.}~\bibnamefont {Feng}}, \bibinfo {author} {\bibfnamefont
  {J.}~\bibnamefont {Huang}}, \ and\ \bibinfo {author} {\bibfnamefont {A.~A.}\
  \bibnamefont {Kornyshev}},\ }\href@noop {} {\bibfield  {journal} {\bibinfo
  {journal} {Energy Environ. Mater.}\ }\textbf {\bibinfo {volume} {3}},\
  \bibinfo {pages} {414} (\bibinfo {year} {2020}{\natexlab{b}})}\BibitemShut
  {NoStop}%
\end{thebibliography}%

\end{document}